\documentclass[aps,pra,twocolumn,showpacs,floatfix,superscriptaddress,floatfix]{revtex4-2}
\usepackage{natbib}
\usepackage{bm}
\usepackage[normalem]{ulem}
\usepackage{lipsum}

\PassOptionsToPackage{table,dvipsnames}{xcolor}
\usepackage{graphicx}
\usepackage{color}
\usepackage{xcolor}
\usepackage{amsmath}
\usepackage{amssymb}
\usepackage[utf8]{inputenc}
\usepackage{comment}
\usepackage{dcolumn}
\usepackage{bm}
\usepackage{hyperref}
\usepackage{soul}
\usepackage{footmisc}
\usepackage{xfrac}

\usepackage{ulem}
\usepackage{amsfonts}
\usepackage{verbatim}
\usepackage{stackrel}
\usepackage[capitalize]{cleveref}
\usepackage{tikz}

\usetikzlibrary{shapes.geometric, arrows}
\usetikzlibrary{decorations.markings}
\usetikzlibrary{decorations.pathmorphing}
\usetikzlibrary{decorations.pathreplacing}

\nocite{BreathingBonitz2009,BeyondGP2010}

\graphicspath{{/Images}}
\usepackage{float}

\usepackage{placeins}
\newcolumntype{P}[1]{>{\centering\arraybackslash}p{#1}}
\definecolor{tdgreen}{rgb}{0,0.4,0}
\definecolor{tlgreen}{rgb}{0,0.7,0}
\definecolor{tpurple}{RGB}{103, 0, 31}
\definecolor{tbrown}{RGB}{127, 39, 4}
\definecolor{tviolet}{RGB}{73, 0, 106}
\definecolor{tgreen}{RGB}{1, 70, 54}
\definecolor{clblue}{rgb}{0.050, 0.35, 0.90}

\newcommand{\Rmnum}[1]{\expandafter\@slowromancap\romannumeral  #1@}

\usepackage{bbold}
\usepackage{color}

\newcommand{\rd} {\color{black}}

\makeatletter
\allowdisplaybreaks
\makeatother

\begin{document}

\title{Few-Photon Fluorescence from Ultracold Bosons in an Optical Cavity}

\author{Megha Gopalakrishna}
\email{meghagopalakrishna@rnsit.ac.in}
\affiliation{Department of Physics, Division of Mathematical Physics, Lund University, 22100  Lund, Sweden}
\affiliation{Department of Physics, RNS Institute of Technology (Affilated to Visvesvaraya Technological University), Bengaluru 560098, Karnataka, India}

\author{Emil Vi\~nas Bostr\"om} 
\email{emil.bostrom@mpsd.mpg.de}
\affiliation{Max Planck Institute for the Structure and Dynamics of Matter, Luruper Chaussee 149, 22761 Hamburg, Germany}
\affiliation{Nano-Bio Spectroscopy Group, Departamento de Fisica de Materiales, Universidad del Pais Vasco, 20018 San Sebastian, Spain}

\author{Claudio Verdozzi}
\email{claudio.verdozzi@fysik.lu.se (corresponding author)}
\affiliation{Department of Physics, Division of Mathematical Physics and ETSF, Lund University, 22100  Lund, Sweden}
\date{\today}

\begin{abstract}
Within a cavity quantum electrodynamics framework, we analyse the fluorescent spectrum of ultracold bosons in resonant and second-harmonic generation (SHG) regimes. Two settings are considered: (i) (few) bosons in an optical lattice and (ii) a trapped two-component dilute Bose–Einstein condensate (2BEC) within the Bogoliubov scheme. Atom–photon dynamics are treated exactly via time-dependent configuration interaction, while cavity leakage is modelled via classical oscillator baths.
For optical lattices, bosons in short Bose–Hubbard chains with two levels per site show spectral intensity that increases with atom number at weak interactions but decreases and redshifts at strong interactions once atoms outnumber sites. In the 2BEC case, spectra largely scale with particle number when 2BEC–cavity and inter-particle couplings are rescaled, while strong correlations induce SHG redshifts.
These results reveal some generic features of fluorescence in ultracold bosons in optical cavities, and hint at possible experiments and further theoretical developments.
\end{abstract} 

%%%%%%%%%%%%%%%%%%%%%%%%%%%%%%%%%%%%%
%%%%%%%%%%%%%%%%%%%%%%%%%%%%%%%%%%%%%
%%%%%%%%%%%%%%%%%%%%%%%%%%%%%%%%%%%%%
%%%%%%%%%%%%%%%%%%%%%%%%%%%%%%%%%%%%%
%%%%%%%%%%%%%%%%%%%%%%%%%%%%%%%%%%%%%
%%%%%%%%%%%%%%%%%%%%%%%%%%%%%%%%%%%%%

\maketitle
\section{Introduction}
Ultracold boson systems \citep{zhai_2021} provide highly tuneable platforms for exploring fundamental and applied physics across atomic, molecular and condensed matter fields \citep{Molecule1999,Lewenstein2007}, including light–matter interaction, thermalization and localization \citep{Goldman_2014,Thermal_2019,Souza2023}, and quantum information and metrology \citep{Greiner_QI_2015,RevModPhys.90.035005}. 

Two paradigmatic settings are free atoms in a trapping potential \citep{pethick_smith_2008} and atoms in standing-wave, laser-generated periodic potentials \citep{optical} (optical lattices). The first enabled the original Bose–Einstein condensate (BEC) detection \citep{Wieman1995,Ketterle1995} and fostered areas such as atom chips \citep{atom_chips}, topological magnetism \citep{Topolog_magn2001}, supersolidity \citep{Ferlaino2021}, and analogues of early-universe expansion \citep{cosmic_inflation}. Particularly relevant are two-component BECs \citep{binary_mixture}, realisable by combining different internal states \citep{Javanainen1996} or spatially separating components, for instance in double-well traps \citep{Double_well1997,Dalton}.

Optical lattices greatly broaden ultracold boson physics by enabling controlled realizations of condensed-matter lattice models \citep{Zoller1998,Bloch2005,Lewenstein2007,Gross2017,Schafer2020}. By tuning lattice depth and interactions, one can investigate superfluid–Mott transitions \citep{Troyer2002}, chaotic dynamics \citep{Fromhold2000}, entanglement \citep{Pichler2016}, artificial gauge fields \citep{Galitski2019}, and precision measurements such as atomic and molecular clocks \citep{Takamoto2005,Borkowski2018}.

Electromagnetic fields, besides creating trapping and optical lattice landscapes, also probe ultracold atoms in the quantum electrodynamics regime. In optical cavities \citep{Zoubi2009,Pohl2012,Essingler2013} they allow photon-mode selection, control of photon number \citep{schleich01}, and tuneable cavity–matter coupling, enabling studies of light-induced phases,
gravitational effects \cite{photonics11030205} and theoretical methods \citep{Ritsch2021}. Dissipation \citep{Ghasemian2017,PhysRevLett.134.073604}, spontaneous emission \citep{Ghasemian18} and the Stark effect \citep{Ghasemian2021} have been analysed in this context (for a recent review of
cavity QED setups with ultracold boson atoms, see \cite{Chanda_2025}). Furthermore,
cavity Bose–Hubbard models have been used to describe superfluid–Mott transitions \citep{Zoubi2009},  two-photon spectroscopy \citep{TwophotonSpec2011}, atom–photon  entanglement \citep{Maschler2005},  and the interplay of temporal bistability and dissipation \citep{DickBoseHubbard_bistable}.

Fluorescence is a common optical de-excitation process, and second-harmonic generation (SHG)—which produces photons at integer multiples of the incident frequency \citep{BloeRMP}—is one of its prominent manifestations. SHG underlies nonlinear optics and many spectroscopic applications \citep{physi,engine,chemist,biolog,medicine}, and is well studied in atoms, molecules, and solids \citep{SHGmaterials}, but far less in ultracold atoms, particularly ultracold bosons in cavities operating in the low-photon regime. This is precisely the regime where, as shown by previous work in other contexts \citep{SHG93,SHG95}, the quantal nature of the optical field is expected to play a significant role in shaping the fluorescence spectral line (see e.g. the discussion in \cite{Emil2020} of the Mollow and SHG spectra for a two level system).

Here, we take a step in this less explored direction by analysing fluorescence from interacting ultracold bosons at low photon number, treating the cavity mode fully quantum mechanically. Starting from the quantum field theory of two-component bosons, we focus on two widely used regimes of broad relevance: (i) the Bose–Hubbard/tight-binding limit, which captures strongly correlated lattice physics, and (ii) the fully quantum two-mode BEC limit, which describes internal-state dynamics in dilute condensates. A third important regime—the Gross–Pitaevskii description retaining spatial condensate fluctuations—is left to future work.

As commonly done, we consider zero-range interactions describing the BEC \citep{pethick_smith_2008}, while Hubbard-type terms model the lattice bosons interactions \citep{Zoller1998,Bloch2005,Lewenstein2007,Gross2017,Schafer2020,Landig2016,Nagy2018}. Both types of systems couple to two quantized photon modes (cavity and fluorescence) and to oscillator baths representing cavity leakage. The dynamics is solved exactly via time-dependent configuration interaction, without the rotating-wave approximation (RWA), whilst bath degrees of freedom are evolved classically.  This confines us to small samples, but at the same time makes it possible to adopt a full quantum description. The time-dependent approach to fluorescence employed here was introduced in recent work \cite{Emil2020,Scipost}, but is applied here for the first time in the context of fluorescence from ultracold atom systems.

For optical lattices we find that SHG grows with atom number at weak interactions but is quenched once atoms outnumber lattice sites. Resonant fluorescence shows a non-monotonic dependence on particle number under strong interactions, and Mollow sidebands weaken when incident photons enter the cavity at a finite rate. Leakage reduces both SHG and resonant intensities and shifts the SHG peak. For BECs, SHG spectra remain similar at weak atom–cavity coupling but become strongly dependent on atom number and interaction strength at strong coupling. Resonant spectra are largely insensitive to particle number, with the Mollow triplet replaced by multiple peaks merging into a nearly flat profile around resonance.

The paper is organized as follows: Sec.~\ref{sec.model} introduces the different systems considered
and the approach used; Sec.~\ref{HB-chains} presents resonant and SHG fluorescence spectra, also with the inclusion of cavity leakage effects; Sec.~\ref{sec.BEC_GP_limit} describes and analyses the two-mode BEC (2BEC); conclusions are given in Sect.~\ref{sec.conclude}.

This paper is dedicated to the 65th birthday of Michael Bonitz and highlights the intersection of three areas within his rich and diverse scientific work: interacting bosonic systems \cite{BreathingBonitz2009,BeyondGP2010}, light–matter interaction phenomena \cite{LaserBonitz2010,Laser_Bonitz_Graphene2025}, and nonequilibrium physics 
\citep{MBonitz_1996,PhysRevB.93.035107,PhysRevLett.124.076601}, the latter standing as a cornerstone of his distinguished career.

%%%%%%%%%%%%%%%%%%%%%%%%%%%%%%%% 
%%%%%%%%%%%%%%%%%%%%%%%%%%%%%%%% 
%%%%%%%%%%%%%%%%%%%%%%%%%%%%%%%% 
%%%%%%%%%%%%%%%%%%%%%%%%%%%%%%%% 
%%%%%%%%%%%%%%%%%%%%%%%%%%%%%%%% 

\section{Systems, cavity leakage, theoretical approach, and observables \label{sec.model}}
\begin{figure}
 {\includegraphics[width=0.99\columnwidth]{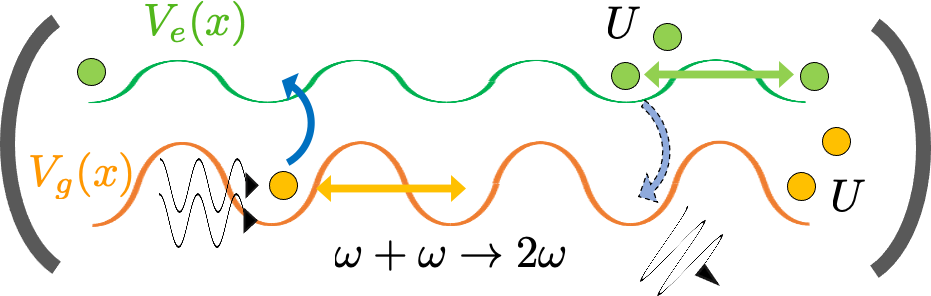}}
 \caption{Bosonic atoms in an optical lattice experience different potentials $V_{e/g}$ depending on their internal state.
 The bare transition $\omega_{eg}$ is the mean separation of these potentials. Atoms hop within each potential
 and interact onsite [Eq.~(\ref{Eq_prima})]. Photon absorption or emission couples the two manifolds.}
 \label{fig0}
\end{figure}
To characterise fluorescence from ultracold bosonic atoms in a cavity, we consider two system typologies: 1) 1D repulsive bosons in an optical lattice in the tight-binding/Hubbard-like regime and 2) a two-state 1D Bose-Einstein condensate (2BEC).
We assume that each atom can be in either of two internal states (see e.g. \citep{Zoubi2007}), namely a ground (excited) state $g$ ($e$), thus experiencing the optical potential $\mathcal{V}_g(x)$ ($\mathcal{V}_e(x))$ (see Fig.~\ref{fig0} for the case of an optical lattice). For both cases, the starting point will be
a 1D many-body Hamiltonian for two interacting bosonic fields in external potentials (this resembles an experimental setup with strong transverse atom confinement): 
\begin{align}\label{H_start}
&\!\!\!\!\!\hat{H}=\sum_{\kappa\in\{g,e\}} \int d x~\hat{\psi}_\kappa^\dagger(x) 
\big[-\frac{\hbar^2}{2m}\frac{d^2}{dx^2}+\mathcal{V}_\kappa(x)\big] \hat{\psi}_\kappa(x)\nonumber\\
&\!\!\!\!\!+\!\!\!\!\!\sum_{\kappa,\kappa'\in\{e,g\}}\int dx dy~ \hat{\psi}_\kappa^\dagger(x)\hat{\psi}
_{\kappa'}^\dagger(y) \mathcal{U}_{\kappa \kappa'}(x,y) \hat{\psi}_{\kappa'} (y)\hat{\psi}_{\kappa}(x),
\end{align}
where  the specific form of $\mathcal{V}_\kappa(x)$ and $\mathcal{U}_{\kappa \kappa'}(x,y)$,
together with the approximations introduced, define the two physical systems under consideration.
Moreover, in both scenarios the systems couple to two optical modes, 
reflecting the fact that the Bose–Hubbard lattice and the 2BEC reside inside a quantum optical cavity.
Most features of the boson systems and Hamiltonians presented next are standard and we refer to the literature
for details. Here, we concisely sketch how to arrive at the expressions (and associated notation)
used in the numerical calculations.
%%%%%%%%%%%%%%%
\subsection{A Bose-Hubbard System in an Optical Lattice}
\noindent In this case, in Eq.~(\ref{H_start}) $\mathcal{V}_g(x)=V_g(x)$ and $\mathcal{V}_e(x)=\Omega_{ef}+V_e(x)$, with $V_{e/g}$ optical periodic potentials, and $\Omega_{ef}$ the bare (i.e. with no photon dressing) atom excitation energy. Then,
i) moving to a representation in terms of ground and excited Wannier functions, ii) including  only the two lowest Wannier $g/e$ bands 
(i.e. $\hat{\Psi}^\dagger_g(x)=\sum_i \Omega_g(x-x_i)\hat{\alpha}^\dagger_i, 
\hat{\Psi}^\dagger_e(x)=\sum_i \Omega_e(x-x_i)\hat{\beta}^\dagger_i$,  a quite reasonable assumption in the low-temperature regime ~\citep{Maschler2008,Zoubi2013}) 
and iii) employing a tight-binding approximation with only on-site and nearest-neighbour terms, the effective two-mode boson Hamiltonian in the presence of optical cavity modes reads
\begin{equation}\label{Eq_prima}
\hat{H}_{2BH}(t)=\hat{H}^{0}_A+\hat{H}^{I}_A+\hat{H}_R(t)+\hat{H}_{AR}(t).
\end{equation}
Explicitly, the different terms are:
\begin{subequations}
\begin{align}
&\!\!\!\!\hat{H}^{(0)}_A = \sum_i [\epsilon_g\hat{\alpha}^\dagger_i\hat{\alpha}_i+\epsilon_e\hat{\beta}^\dagger_i\hat{\beta}_i]
 + \sum_{\langle ij\rangle} [t_g\hat{\alpha}^\dagger_i\hat{\alpha}_j+t_e\hat{\beta}^\dagger_i\hat{\beta}_j] \label{lab_a} \\
&\!\!\!\!\hat{H}^I_A =\!\!\sum_{i}
\!\left[\frac{U_{gg}}{2}\hat{n}_i^g(\hat{n}_i^g\!-\!1)
\!+\!\frac{U_{ee}}{2}\hat{n}_i^e(\hat{n}_i^e\!-\!1)\!
+\!U_{eg}\hat{n}_i^g\hat{n}_i^e\!\right] \label{lab_b} \\
&\!\!\!\!\hat{H}_R(t)=\omega_0\hat{b}^\dagger\hat{b}+\omega\hat{b}_f^\dagger\hat{b}_f\!+\!\hat{H}_{\rm Drive}(t)\label{lab_c}\\
&\!\!\!\!\hat{H}_{AR}(t)=[\gamma_0(\hat{b}^\dagger\!+\!\hat{b})\!+\!\gamma_f\!(t)(\hat{b}_f^\dagger\!+\!\hat{b}_f)]\!
\sum_i(\hat{\alpha}^\dagger_i\hat{\beta}_i\!+\!\mathrm{h.c.})\label{lab_d}.
\end{align}
\end{subequations}
In Eqs.~(\ref{lab_a},\ref{lab_c},\ref{lab_d}), $\hat{\alpha}^\dagger_i$ ($\hat{\beta}^\dagger_i$) creates an atom in the $g$ ($e$) Wannier state at site $i$ with energies $\epsilon_g$ and $\epsilon_e$. Similarly, the hopping matrix elements between adjacent sites for the ground and excited Wannier bands are denoted by $t_g$ and $t_e$. Furthermore, in Eq.~(\ref{lab_c}), describing boson-boson interactions,
$\hat{n}_i^g=\hat{\alpha}^\dagger_i\hat{\alpha}_i$, and $\hat{n}_i^e=\hat{\beta}^\dagger_i\hat{\beta}_i$ are site occupation operators, and $U_{gg}, U_{ee}$ and $U_{eg}$ are respectively the onsite interaction strengths for bosons in $gg, ee$ and $eg$ pairs of Wannier states at the same optical lattice site $i$. Considering now Eqs.~(\ref{lab_c},d) for the radiation modes, $b^\dagger$ ($b^\dagger_f$) creates a cavity (fluorescent) photon of frequency $\omega_0$ ($\omega$). As indicated  in Eq.~(\ref{lab_d}), the light-matter interaction associated with these photons occurs with coupling strength $\gamma_0$ ($\gamma_f $) and induces local `vertical' transitions at each site between the $e$ and $g$ states.

The  basic energy unit in our system is $\Omega_R/2$ (set equal to 1), where $\Omega_R= \epsilon_e - \epsilon_g$. All energy parameters in the model are expressed in this unit.
The system features several interconnected scales: the ratio 
$\omega_0/\Omega_R$, the effective Rabi-like couplings, and the interaction and hopping ratios 
$U/t$ and $\gamma/t$ which quantify intra-band and inter-band processes, respectively.
Finally, the coupling to the fluorescent field is taken as decaying in time \cite{Emil2020,Scipost,Dicke24}, i.e.
$\gamma_f (t) = \tilde{\gamma} e^{-\Gamma t}$, and the drive term
$\hat{H}_{\rm Drive}(t)$ of Eq.~(\ref{lab_c}), to be specified later,
permits to pump $\omega_0$-type photons in the cavity. A similar single-mode Hamiltonian under RWA was analysed in~\citep{Zoubi2009}.

%%%%%%%%%%%%%%%%%%%%%%

\subsection{A Two-component, Two-mode BEC}\label{sec.BEC_GP_limit}
For the 2BEC regime, where $\mathcal{V}_{g/e}$ are trapping potentials, 
we start again from Eq.~(\ref{H_start}) and assume 
 i) the large-particle-number (large-$N$) limit \citep{currently} and ii) weak, dilute interactions and low temperature, such that thermal excitations and condensate depletion are small. We further iii) model intra- and inter- components interactions between atoms by effective contact potentials 
 $\overline{\mathcal{U}}_{\kappa \kappa'}$ characterised by $s$-wave scattering lengths, as done in \citep{Dalton,Kuang}.
Under the same assumptions, we approximate the field operators as
 $\hat{\psi}^\dagger_g({\bf x})\approx \phi_g({\bf x})\hat{\alpha}^\dagger$,
$\hat{\psi}^\dagger_e({\bf x})\approx \phi_e({\bf x})\hat{\beta}^\dagger$, where $\phi_{\kappa=g/e}$ is the 
the lowest energy eigenstate of
\begin{equation}\label{1body_BEC}
\big[-\frac{\hbar^2\nabla^2}{2m}+\mathcal{V}_{g/e}({\bf x})\big]\phi_{g/e}({\bf x})=E_{g/e} \phi_{g/e}({\bf x}),
\end{equation}
with $E_e > E_g$. 
Making use of the assumptions  i-iii) in  Eq.~(\ref{H_start}), performing the space integrations, and then absorbing inherent parameters (e.g., the scattering lengths) into the definition of the contact interaction parameters (re-labeled as $ U_{\kappa \kappa'}$), we arrive at
\begin{align}
& \hat{H}_{\rm 2BEC} = E_g \hat{\alpha}^\dagger\hat{\alpha} + E_e \hat{\beta}^\dagger \hat{\beta}\nonumber\\
&+\frac{1}{2}U_{gg}(\hat{\alpha}^\dagger)^2 \hat{\alpha}^2
+\frac{1}{2}U_{ee}(\hat{\beta}^\dagger)^2 \hat{\beta}^2 
+U_{eg} \hat{\alpha}^\dagger \hat{\alpha} \hat{\beta}^\dagger \hat{\beta}.
\end{align}
The light–2BEC coupling, in the dipole approximation and for two photon modes of frequencies $\omega_0$ and $\omega$, is described by \citep{Dalton,Kuang,Huang08,Ghasemian18}
\begin{align}\label{coupling_BEC}
\hat{H}' = \big[\gamma_0(\hat{b}^\dagger+\hat{b})+\gamma_f(t)(\hat{b}_f^\dagger+\hat{b}_f)\big]  (\hat{\alpha}^\dagger\hat{\beta}+\hat{\beta}^\dagger\hat{\alpha}),
\end{align}
where the fields spatial dependencies are absorbed into the coupling constants,
and $\gamma_f(t)=\tilde{\gamma}e^{-\Gamma t}$ is again a damped fluorescent coupling.
Thus, the total two-mode Hamiltonian is
%%%%
\begin{align}\label{2BECav}
\!\hat{H}&_{\rm 2BEC-cav}(t)= E_g \hat{\alpha}^\dagger\hat{\alpha} + E_e \hat{\beta}^\dagger \hat{\beta}
+\omega_0\hat{b}^\dagger\hat{b} +\omega \hat{b}_f^\dagger\hat{b}_f \nonumber\\
&+\frac{1}{2}U_{gg}(\hat{\alpha}^\dagger)^2 \hat{\alpha}^2
+\frac{1}{2}U_{ee}(\hat{\beta}^\dagger)^2 \hat{\beta}^2 
+U_{eg} \hat{\alpha}^\dagger \hat{\alpha} \hat{\beta}^\dagger \hat{\beta}\nonumber\\
&+
(\hat{\alpha}^\dagger \hat{\beta}+\hat{\beta}^\dagger \hat{\alpha}) \big[\gamma_0 (\hat{b}^\dagger+\hat{b})+
\gamma_f (t)(\hat{b}_f^\dagger+\hat{b}_f)\big].
\end{align}

%%%%%

\subsection{The Role of Dissipation and Leakage}
A rigorous account of radiative and non-radiative damping and leakage is attainable via Lindblad dynamics or non-equilibrium Green’s functions (or their combination \citep{Stefanucci_dissip}), thus guaranteeing convergence to a steady-state fluorescence spectrum.  Here, to stay within a wavefunction/CI framework, we adopt a phenomenological prescription based on a time dependent coupling $\gamma_f(t)$ that is exponentially damped in time [Eqs.~(\ref{lab_d}, \ref{coupling_BEC})], the plausibility of which has been addressed previously~\citep{SHG93,SHG95,Emil2020,Scipost,Dicke24}

At the same time, for some of the results presented, we also consider a more microscopic dissipation/leakage mechanism, based on coupling the $\omega_0$ and $\omega$ modes to 
external oscillator baths, i.e. a Caldeira–Leggett–type description. 
\citep{Caldeira1983,Venkataraman2014,Scipost}.
As in \citep{Ghasemian2017}, we introduce one bath for each of the two modes. This  augments 
Eqs.~(\ref{Eq_prima}) and (\ref{2BECav}) with the term
\begin{align}
&\hat{H}_{\rm Loss} = \frac{1}{2}\sum_{k=1}^{\mathcal{N}_K} \left\{ \hat{p}^2_k+\omega^2_k  \big[\hat{x}_k-(C_k/\omega^2_k)(\hat{b}^\dagger+\hat{b})\big[^2\right\} \nonumber \\
&+\frac{1}{2} \sum_{j=1}^{\mathcal{N}_J} \left\{ \hat{p}^2_j+\omega^2_j  \big[\hat{x}_j-(C_j/\omega^2_j)(\hat{b}_f^\dagger+\hat{b}_f)\big[^2\right\}.
\end{align}
where \(\mathcal{N}_{K/J}\) are the numbers of bath oscillators with unit masses, coordinates \((\hat{p}_{k/j}, \hat{x}_{k/j})\), and frequencies \(\omega_{k/j}\). The couplings \(\{C_{k/j}\}\) set the bath–photon strength for incident/fluorescent fields. The treatment of \(\hat{H}_{\rm Loss}\) should be quantum mechanical; however, to maintain a unitary wavefunction description for atoms and cavity, we employ \emph{Ehrenfest dynamics} in which the system–bath dynamics is treated within a mixed quantum–classical scheme, with classical bath coordinates:
\(\{\hat{x}_{k/j},\hat{p}_{k/j}\}\!\to\!\{x_{k/j},p_{k/j}\}\). 

{\rd In general, the Ehrenfest scheme does not enforce detailed balance \citep{DetBal1,DetBal2,DetBal3,DetBal4,DetBal5}. As a consequence, Ehrenfest dynamics (ED) may yield slightly overheated steady-state populations, particularly in anharmonic systems. This limitation originates from its mean-field nature, which suppresses stochastic transitions and tends to preserve coherences for longer than they should persist physically.

This should be contrasted with Lindblad dynamics (LD), arguably the most widely used framework for describing open quantum systems and dissipative dynamics. While LD guarantees positivity and Hermiticity of the density matrix, it relies on Markovian approximations and may itself become inaccurate in the presence of structured environments, strong system--bath correlations, or strongly dressed states. In such situations, it may also fail to reproduce the correct steady state.

For the present problem, the conditions are particularly favourable to Ehrenfest dynamics.
The bath consists of $10^3$ harmonic oscillators (i.e., a large number of degrees of freedom; see below), linearly coupled to the cavity mode and serving primarily as a channel for energy dissipation. This places the system close to the conditions under which mixed quantum--classical approaches are generally expected to perform best. We therefore expect ED to provide a qualitatively correct description of cavity damping and energy flow, with deviations from the exact thermal steady state remaining predominantly quantitative rather than qualitative. 

Finally, because the method propagates the system wavefunction rather than its density matrix, it is computationally very efficient. Moreover, although the bath is treated in terms of classical variables, the approach naturally retains non-Markovian memory effects in the system dynamics.}

\subsection{Observables and Theoretical Approach}
To characterize the fluorescence spectrum of the ultracold atom system, we use the probability $ \mathcal{S}(t,\omega)$ of observing at a given time $t$ the emission of one or more photons \citep{Emil2020,Scipost}:
\begin{equation}\label{spectrum_def}
\mathcal{S}(t,\omega)=\sum_{\lambda n} \sum_{m>0}\rvert \langle \lambda  nm|\mathcal{T}\big[e^{-i\int_0^t\hat{H}(t')dt'}]\big|\Phi_0\rangle \rvert^2,
\end{equation}
where $\lambda$, $n$, and $m$ respectively label the states corresponding to the atoms, the incident field, and the fluorescent field, and $|\Phi_0\rangle$ is the initial state of the system (in the actual calculations, different types of initial states will be considered). In presenting
the numerical  results, we will mostly focus on the long-time, steady-state limit $\mathcal{S}(t \rightarrow \infty,\omega)$ of Eq.~(\ref{spectrum_def}), denoted as $\mathcal{S}(\omega)$. While $\mathcal{S}(t, \omega)$, $\mathcal{S}(\omega)$ are the quantities of primary interest, we will also  consider the time dependent populations of specific single particle states of the  ultracold atom system.

Solving for $\mathcal{S}(t, \omega)$ and $\mathcal{S}(\omega)$ requires to perform the exact time evolution of the system. For this,
we employ the short iterated Lanczos algorithm. The associated problem size in the space of configurations is $M=M_B M_c M_f$, with $M_B$ representing the size of the cold bosonic atom system, and $M_c$ and $M_f$ respectively denoting the maximum number of incident and fluorescent photons included in the light Fock space. In the presence of the Caldeira-Leggett baths,
each classical oscillator evolves in time according to
\begin{align}
\ddot{x}_{\mu}(t)=-\omega_{\mu}^2x_{\mu}(t)+C_{\mu} \langle\hat{b}_\mu^\dagger + \hat{b}_\mu \rangle_{\bar{x},t}~,
\end{align}
with \(\bar{x}=\{x_k\}\cup\{x_j\}\) and $\hat{b}_\mu \equiv \hat{b}$ for \(\mu\in[1,\mathcal{N}_K]\) and
$\hat{b}_\mu \equiv \hat{b}_f$ for \(\mu\in[1,\mathcal{N}_J]\).

%
%%%%%%%%%%%%%%%%%%%%%%%%%%%%%%%%%%%%
%%%%%%%%%%%%%%%%%%%%%%%%%%%%%%%%%%%%
%%%%%%%%%%%%%%%%%%%%%%%%%%%%%%%%%%%%
%%%%%%%%%%%%%%%%%%%%%%%%%%%%%%%%%%%%
%%%%%%%%%%%%%%%%%%%%%%%%%%%%%%%%%%%%
%%%%%%%%%%%%%%%%%%%%%%%%%%%%%%%%%%%%
\begin{figure}%[H]
	\begin{tikzpicture}
	\node at (-2.3, 2.3) {\includegraphics[width=0.48\columnwidth]{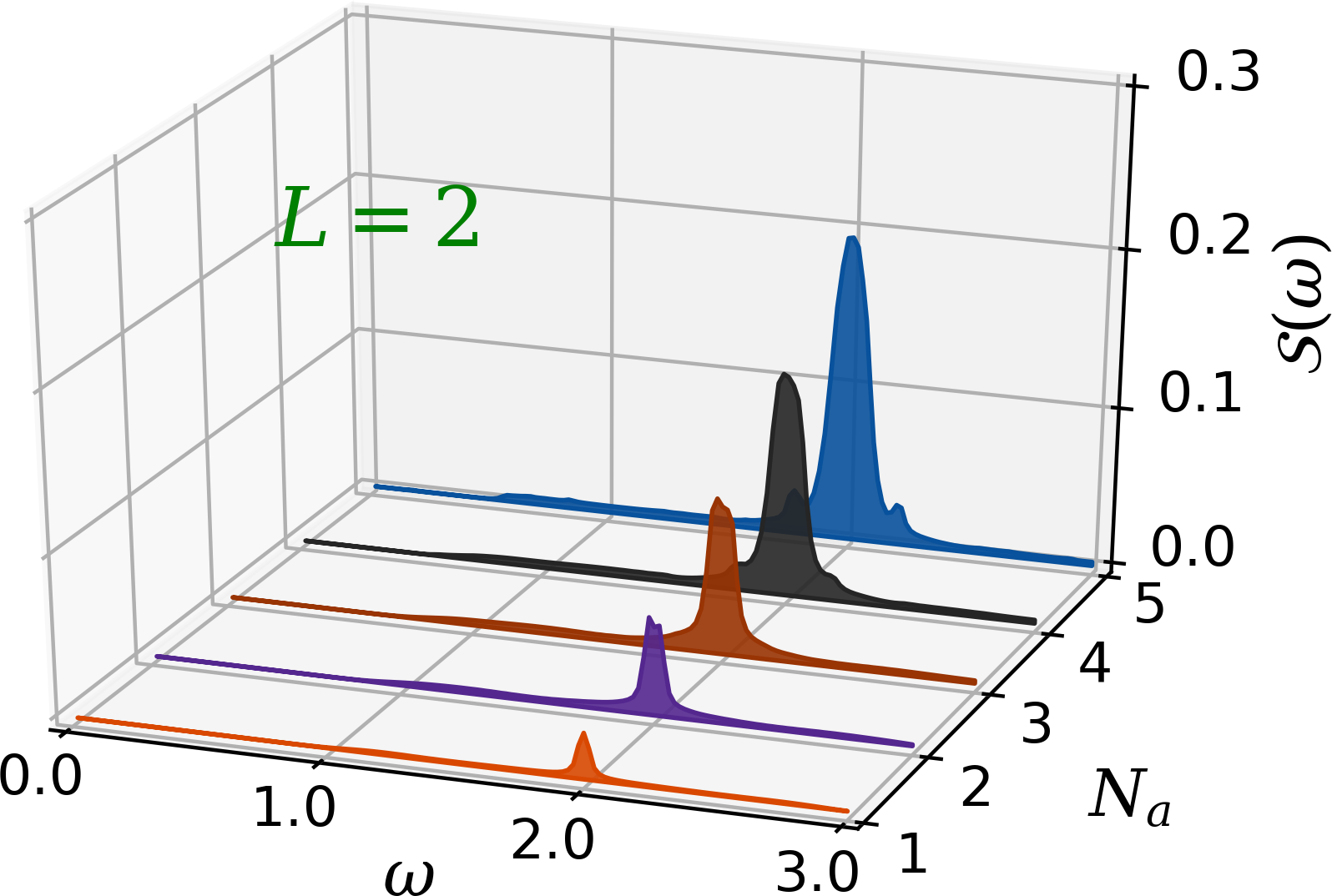}};
	\node at (2.0,2.3){\includegraphics[width=0.48\columnwidth]{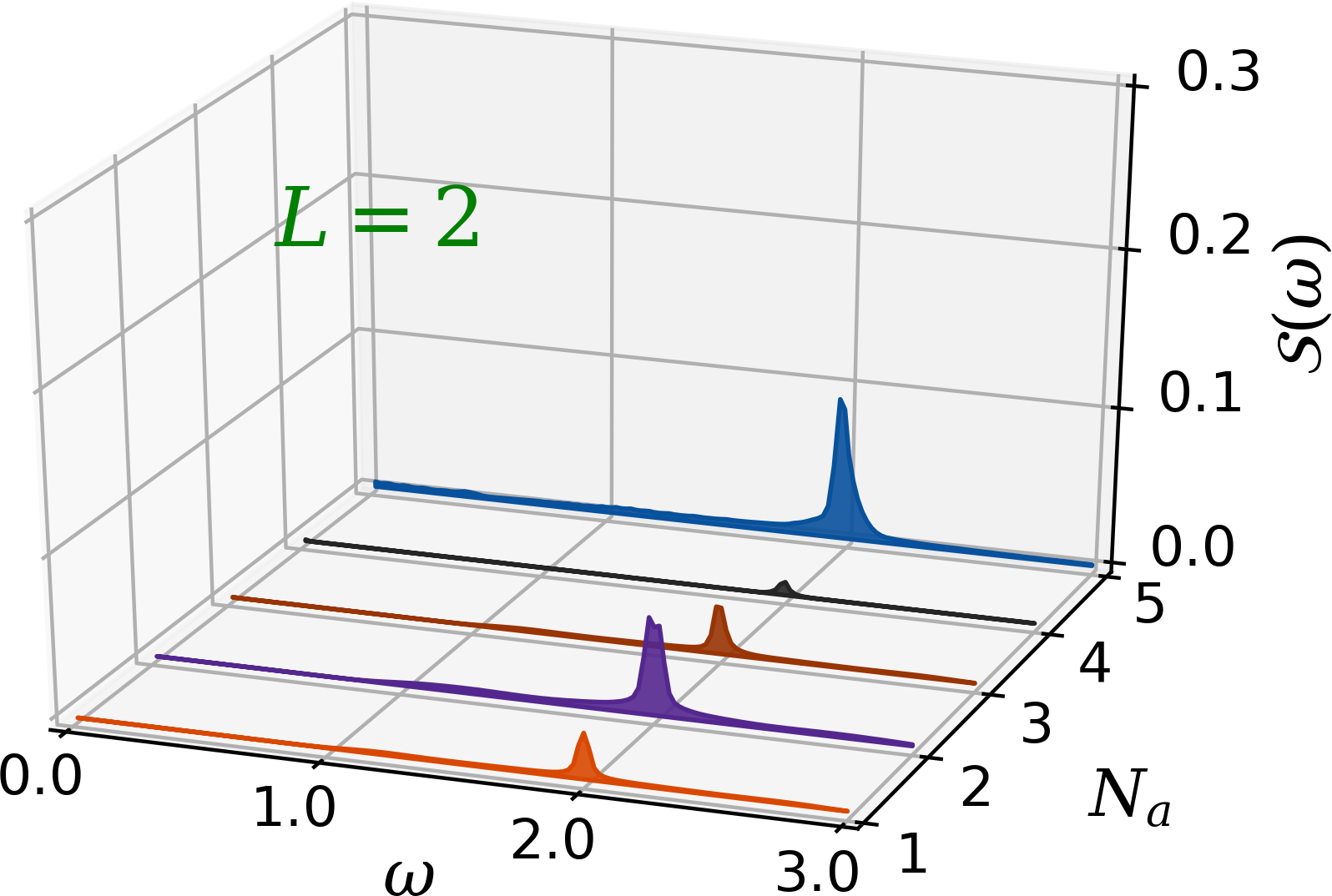}};
	\node[black] at (-1.30,3.0) {$\bm{(a)}$};
     	\node[black] at (3.0,3.0) {$\bm{(b)}$};
	\end{tikzpicture}
	\caption{Resonant fluorescence spectra in a dimer ($L=2$) for varying $N_a$ and interaction $U$ ($U_{eg}=U/2$). Panels $\bm{(a)}$ and $\bm{(b)}$ refer to $U=0.0001$ and $U=5.0$, respectively. The parameters are $\omega_0=2.0$, $\tau_1=\tau_2=2.0$, $t_1=\frac6\pi/\omega_0$ and $t_2=31\pi/\omega_0$, and the initial state is $|\Phi_0^{\rm Res}\rangle$. $\gamma_d$ is set such that $\langle \hat{b}^\dagger \hat{b}\rangle\approx 16.0$ at the end of the drive. It is very hard to see the spectral features here, especially the side bands.}
	\label{fig5}
\end{figure}

\section{Small Bose-Hubbard chains}\label{HB-chains}
We consider a small cluster with $L$ sites and a number $N_a$ of bosonic atoms.
For all the calculations of this section, the hopping parameters are $t_g=t_e=t=-0.01$, and the energies $\epsilon_g=0$, and $\epsilon_e=2$, giving a resonant atomic transition of $\Omega_R=\epsilon_e-\epsilon_g=2.0$. We use $\Gamma=0.02$, atom–photon couplings $\gamma_0=0.1$ and $\gamma_f=0.01$ for the incident and fluorescent fields, and atom–atom interactions $U_{gg}=U_{ee}=U$ and $U_{eg}=U/2$, with different values of $U$ considered (going to the large $U$ regime permits to gain insight into the Dicke limit of the system, where the inter-site hopping effectively vanishes).

The values chosen for
the optical lattice parameters and the incident photon
couplings correspond to those used in Ref.~\citep{Zoubi2009} (after a
rescaling of units by a factor 10$^2$); an average cavity photon number of $\langle \hat{b}^\dagger \hat{b} \rangle \approx 16$ is chosen to reveal multiphoton features while keeping computations tractable.
Our aim is to illustrate qualitative fluorescence features in ultracold atoms. The chosen parameters serve solely this purpose and bear no relation to the applicability of RWA (used in Ref.~\citep{Zoubi2009}). Our approach is RWA-free, which is less important near resonance but essential for SHG.
 
Convergence requirements on the cavity and fluorescent photon numbers — set mainly by
$\omega_0$, $\omega$ and $\gamma_0$, $\gamma_f$ and $\gamma_d$ — limit  the value of $N_a$ and
lattice size $L$, since the overall size of the  full Hilbert space is ${2L+N_a-1 \choose N_a} N_0 N_f$, 
and the spectra $\mathcal{S}(t,\omega)$ are computed on a grid of $N_\omega=150$ frequency points. We deliberately remain in a regime that is computationally comfortable
rather than maximally exhaustive; for tractability we use $L = 2\!-\!4$ and 
$N_a = 1\!-\!5$, and focus on $L = 2$, which already captures the essential trends. 
Representative larger-$L$ examples appear in the Appendixes \ref{AdditionalReso} and \ref{AdditionalSHG}.
\subsection{Resonant Regime \label{sec.res}}
At resonance, i.e. $\omega_0=\Omega_R$, the fluorescence intensity is not extremely sensitive to the photon ramping speed \citep{Emil2020,Scipost}, while the pumping rate noticeably influences the spectral weight distribution (the situation is different for SHG, see Sect.~\ref{sec.SHG}).  We therefore drive the cavity at a finite rate, which also enables characterization of the transient response (some results with instantaneous drive are shown in Appendix \ref{AdditionalReso}). 

We will consider different particle-particle interaction regimes, and start with 
an initial state, $|\Phi_0\rangle \equiv |\Phi_0^{\rm Res}\rangle$, which is the exact ground state of $H(t=0)$, with atoms and cavity in equilibrium (in this state, $\langle \hat{b}^\dagger \hat{b}\rangle \approx 0$). The dynamics is then triggered by injecting photons via an external laser described by the driving term
\begin{equation}
\hat{H}_{\rm Drive}(t)=\gamma_d(\hat{b}+\hat{b}^\dagger) f(t)\sin(\omega_0 t),
\label{eqHR2}
\end{equation}
where $f(t)$ is a smoothed rectangular envelope in the interval $(t_1,t_2)$, explicitly given by
$f(t)=[1-\mathcal{F}_1(t)]\mathcal{F}_2(t)$ with $\mathcal{F}_i(t)=[\exp((t-t_i)/\tau_i)+1]^{-1}$,
$\tau_1 = \tau_2=2.0$, $t_1 = 6\pi/\omega_0$ and $t_2=31\pi/\omega_0$. The value of $\gamma_d$ is fixed so that the pulse ends with $\langle \hat{b}^\dagger \hat{b}\rangle\approx 16$. The spectra $\mathcal{S}(t,\omega)$ are computed for $N_\omega=150$ frequency points.

Figure~\ref{fig5} shows resonant fluorescence for $L=2$ and various $N_a$ at weak and strong interactions. For weak interactions (Panel {$\bm a$), increasing $N_a$ enhances the spectral intensity, while at strong interactions (Panel $\bm b$) the dependence becomes non-monotonic. Moreover, the Mollow sidebands are considerably suppressed, an effect induced by the finite-time photon injection and also observed for other $L$ and $N_a$ (Fig.~\ref{A2fig1}, Appendix~\ref{AdditionalReso}). Finite ramping also hinders the low-frequency shoulder of the spectrum. To illustrate this, Appendix~\ref{AdditionalReso} reports data for $L=2$ and $N_a=3,4$ and $5$, with the cavity initially in a coherent state: in this “instantaneous’’ pumping case, low-frequency peaks reappear (Fig.~\ref{A2fig2}), unlike in Fig.~\ref{fig5}.

To see how the above trends reflect the interplay of interaction strength and atom number, we use as indicator
$\mathcal{D}_e=\sum_i ^L \langle\hat{n}_i^e\rangle$ (Figure~\ref{fig6}), corresponding to the total occupation of the excited states. For $N_a \le 3$, $\mathcal{D}_e(t)$ is nearly interaction–independent. For larger $N_a$, strong interactions promote excited-state occupation already at $t=0$, hindering further excitation. It is also instructive to look at the behavior of $\gamma_d$. For weak interactions, achieving $\langle \hat{b}^\dagger \hat{b} \rangle\approx 16$ requires larger $\gamma_d$ as $N_a$ increases (Fig.~\ref{fig6}a), reflecting greater absorption. Conversely, for strong interactions $\gamma_d$ does not necessarily increase with $N_a$ (e.g., low for $N_a=3$ and $4$), which is indicative of reduced absorption, and consistent with the weaker fluorescence in Fig.~\ref{fig5}b.
\begin{figure}
	\begin{tikzpicture}
	\node at (-2.3, 2.3) {\includegraphics[width=0.98\columnwidth]{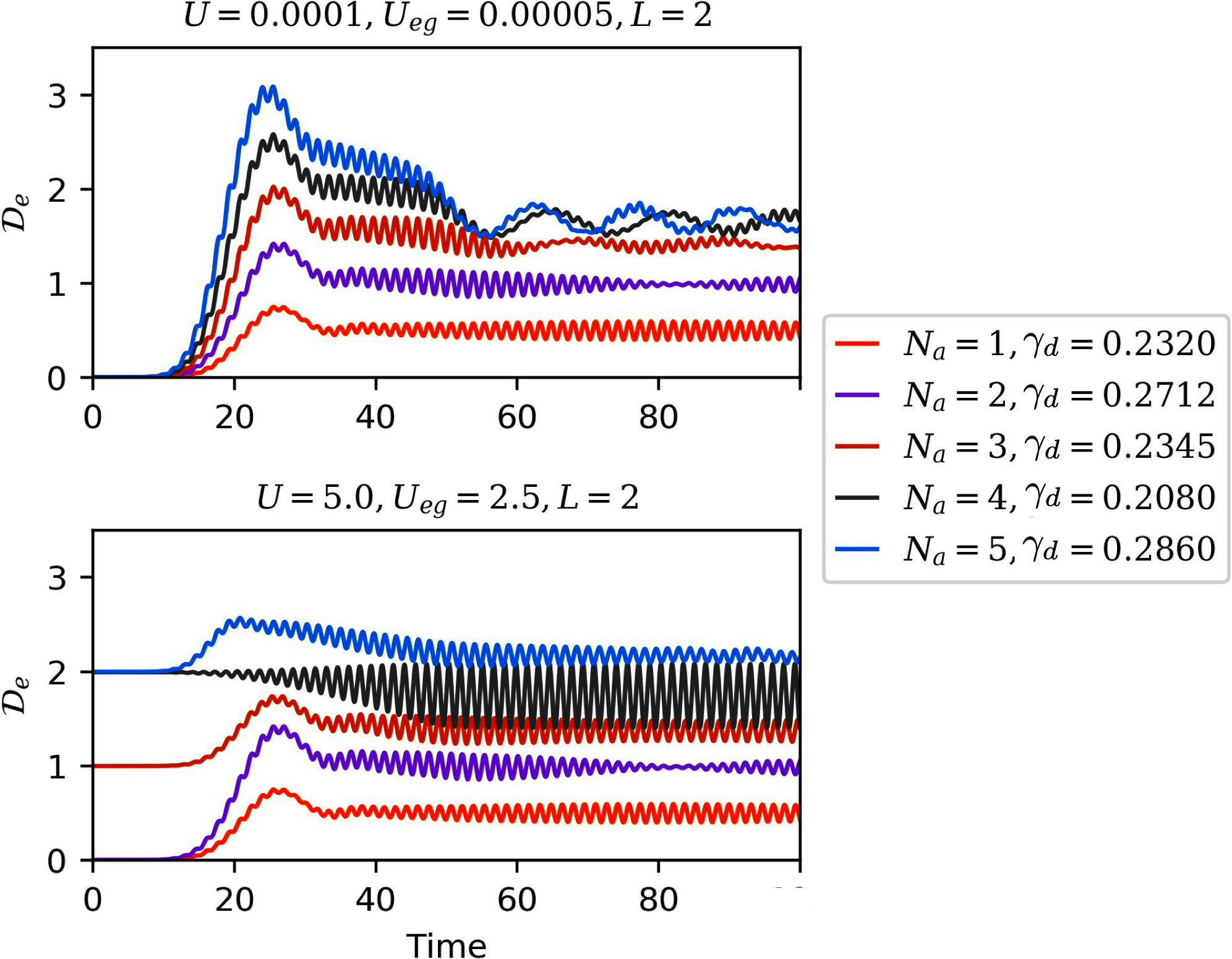}};
	\node[black] at (-5.50,4.70) {$\bm{(a)}$};
     	\node[black] at (-5.50,1.60) {$\bm{(b)}$};
	\end{tikzpicture}
	\caption{Total number of excited atoms $\mathcal{D}_e$ as a function of time, for resonant fluorescence. The initial state is $|\Phi_0^{\rm Res}\rangle$, $U_{eg}=U/2$, $L=2$, $\omega_0=2.0$, $\tau_1=\tau_2=2.0$, $t_1 = 6\pi/\omega_0$, $t_2 = 31\pi/\omega_0$. In panel $\bm{(a)}$, $U=0.0001$, and in panel $\bm{(b)}$, $U=5$. $\gamma_d$ is set such that $\langle \hat{b}^\dagger \hat{b}\rangle\approx 16.0$.}
	\label{fig6}
\end{figure}
The situation changes at intermediate coupling. The fluorescence intensity now decreases monotonically with increasing $U$ (Fig.~\ref{fig7}a). In  contrast, for $U=3$ (Fig.~\ref{fig7}b), $\mathcal{D}_e$ exhibits a plateau in time, indicative of a blockade effect due to already excited atoms, which prevents further excitations.

{\rd
In general, interactions make the optical excitation energy in our small Bose--Hubbard chains dependent on the local occupation, causing successive $g\rightarrow e$ transitions to become increasingly detuned. Because of the discrete many-body spectrum, this can suppress further population transfer, resulting in the observed blockade-like mechanism. To consider the behavior in larger systems, we note that in the noninteracting limit each momentum mode behaves as an independent optical two-level system. Interactions, however, couple different momenta and embed optical excitations in a dense many-body continuum. This broadening of the available final states is expected to weaken the blockade as the system size increases. Some degree of blockade may nevertheless survive when the interaction scale remains sufficiently large compared with the kinetic bandwidth and the many-body continuum. We therefore expect the blockade effect to be reduced, though not necessarily eliminated, in extended systems.}

{\rd As a final remark, the dependence of the Mollow-like sidebands on the shape and duration of the driving pulse has not been systematically investigated here. This feature of the fluorescence spectrum was already noted in Ref.~\cite{Scipost}, and associated to the transient formation of dressed states,  Similar evidence was later reported for Mollow-like absorption spectra induced by classical light \cite{PhysRevLett.133.063202}, where sidebands emerged for flat-top or supra-Gaussian pulses but were suppressed for Gaussian driving. Although absorption and emission, as well as classical and quantum driving fields, correspond to distinct physical situations, both cases point to the same underlying mechanism: pronounced sidebands are favoured by driving protocols with sufficiently sharp temporal features to induce nonadiabatic coherences and mixing among dressed-state manifolds. We therefore expect the Mollow-like sidebands in the present cold-atom--cavity system to depend sensitively on the temporal profile of the driving pulse.}

\begin{figure}
	\begin{tikzpicture}
	\node at (-3.3, 2.3) {\includegraphics[width=0.8\columnwidth]{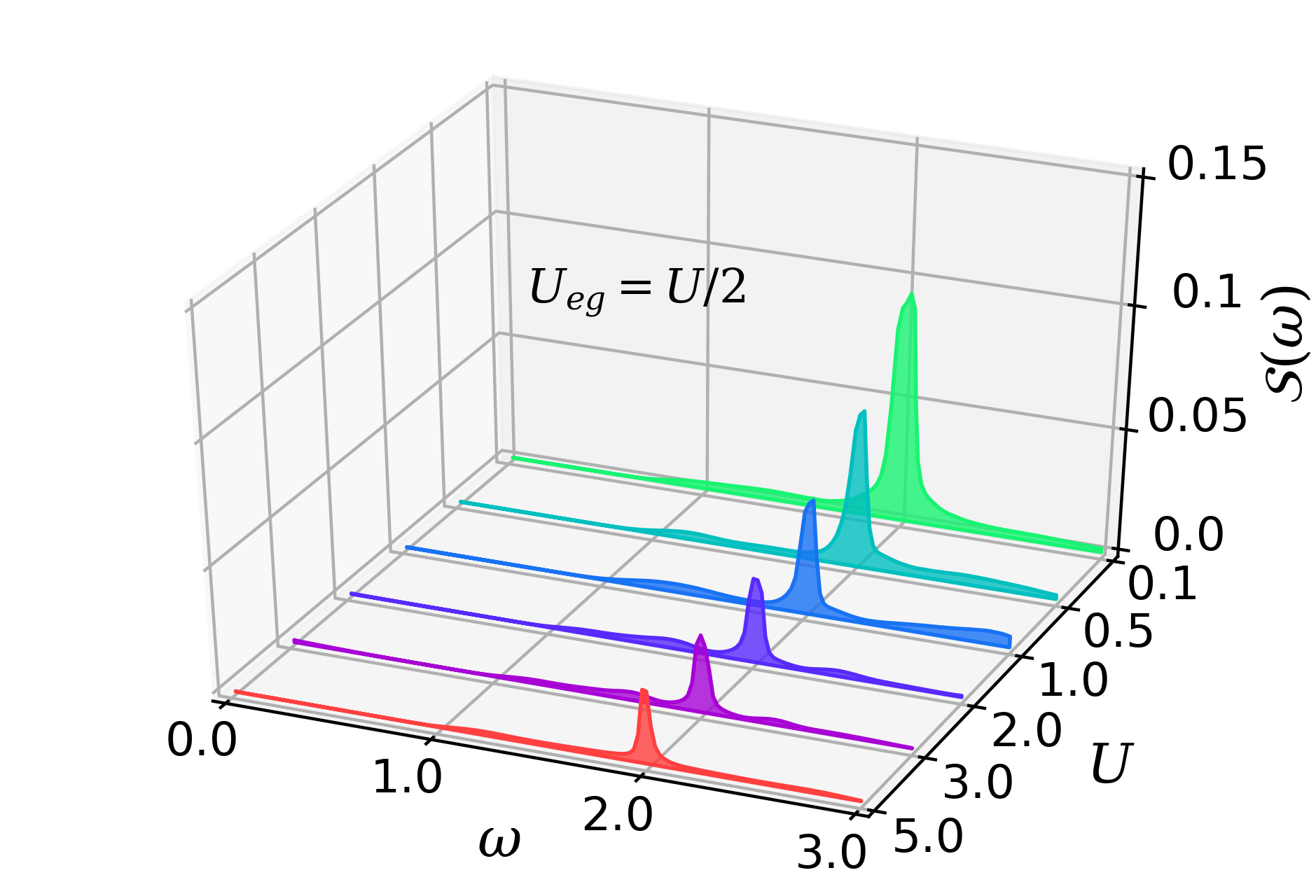}};
	\node at (-2.5, -2.50) {\includegraphics[width=1.0\columnwidth]{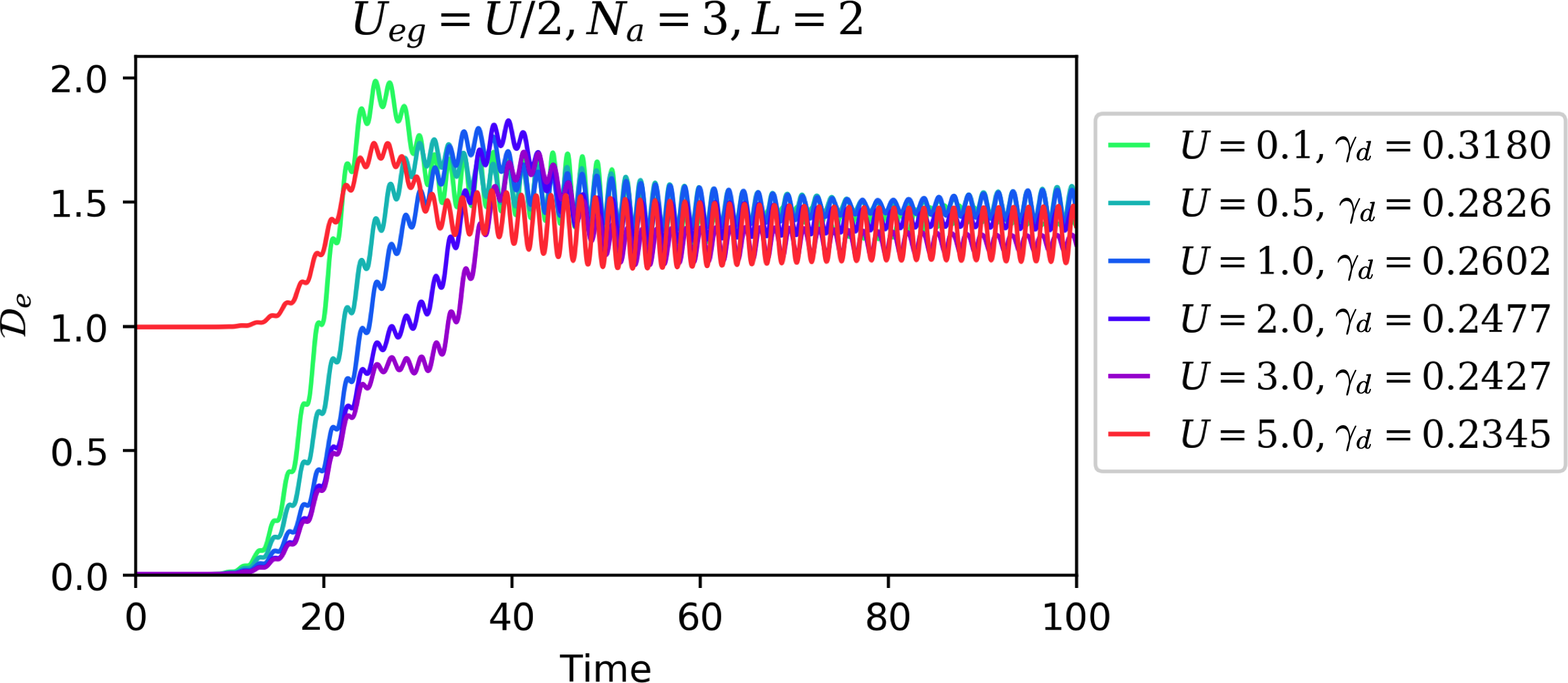}};
	    \node[black] at (-5.0,2.60) {$\bm{(a)}$};
     	\node[black] at (-5.50,-1.30) {$\bm{(b)}$};
	\end{tikzpicture}
	\caption{Resonant fluorescence at intermediate interactions. Panel $\bm{(a)}$ shows the spectrum $\mathcal{S}(\omega)$ while panel $\bm{(b)}$ shows the excited state occupation $\mathcal{D}_e(t)$. Parameters are $L=2$, $N_a=3$, $U_{eg}=U/2$, $\omega_0=2.0$, $\tau_1=\tau_2=2.0$, $t_1 = 6\pi/\omega_0$, $t_2 = 31\pi/\omega_0$, and the initial state is $|\Phi_0^{\rm Res}\rangle$. $\gamma_d$ is set such that $\langle \hat{b}^\dagger \hat{b}\rangle\approx 16.0$.}
	\label{fig7}
\end{figure}
%%%%%%%%%%%%%%%%%%%%%%%%
 
\subsection{Second-Harmonic Generation (SHG)  \label{sec.SHG}}
\subsubsection{Preliminaries}
In Ref.~\citep{Scipost} it was reported that if the material–cavity system starts in its ground state and photons of frequency $\Omega_R/2$ are injected at a finite rate, the SHG response is strongly suppressed. A sizeable signal is recovered only in the ultrafast pumping limit, effectively equivalent to preparing the cavity in a coherent state $|\eta\rangle$ at $t=0$ \citep{Scipost}.

Accordingly, for the SHG calculations, the initial state is taken as $|\Phi_0^{\mathrm{SHG}}\rangle = |\phi_A\rangle|\eta\rangle|0_f\rangle$, where $|\phi_A\rangle$ is the ground state of $H_A$, $|\eta\rangle$ is a coherent state representing the incident field, and $|0_f\rangle$ is the fluorescent vacuum. 
This mimics injecting $\omega_0$ photons via $\hat{H}_{\rm Drive}(t)$ [Eq.~(\ref{lab_c})] over a time interval $[0,T]$ into a cavity initially void of photons, with the atoms in their ground state and {\it no} atom--light coupling: that is, in [Eq.~(\ref{lab_d})]
$\hat{H}_{AR}(t) \rightarrow \hat{H}_{AR}(t)\theta(t-T)$.
This approximation may not be unwarranted: simulations in Ref.~\citep{Scipost} showed that, with the system-photon coupling active during the drive and for moderate coupling and a ramp time faster than the system’s response, the final state approaches a product of a coherent photon state and the ground state of the material system.

To exemplify, consider the driving Hamiltonian
$
\hat{H}_{\rm Drive}(t)=\gamma_d(\hat{b}^\dagger+\hat{b})\sin(\omega_0 t)
$
(the same form used in the resonant case). For an initially empty cavity without atoms, after a time interval $T$ this drive produces a coherent photon state with 
$
n_b(T) \equiv \langle \hat{b}^\dagger \hat{b}\rangle_T =
(\gamma_d^2/4) |\sin(\omega_0 T)/\omega_0 - T\,e^{-i\omega_0 T}|^2
$,
obtained from the dynamics of the driven quantum harmonic oscillator. To simplify this result and for sake of illustration we choose  $T=\pi/\omega_0$ and a target $n_T=n_b(T)$; this gives
$
\gamma_d=\frac{2\omega_0}{\pi}\sqrt{n_T}.
$
Then, if $T$ is small compared to the system’s response timescale (the latter can in two-level systems can be related to the inverse of the Rabi frequency), preparing a factorised initial state or explicitly ramping photons into the fully interacting cavity  will produce essentially equivalent physical initial conditions. The incident field to generate the SHG has frequency $\omega_0 = 1$, and in all simulations $\langle \hat{b}^\dagger \hat{b}\rangle = \eta^2 = 16$.

\subsubsection{SHG vs Interactions, Atom number,  and Lattice Size}
We consider short Bose–Hubbard chains with $L=2$, $3$ or $4$ sites, two orbitals per site, and $N_a =1,\dots,5$. We include up to $N_0=59$ incident and $N_f=4$ fluorescent photons, and to induce SHG we use using an incident field with $\omega_0=\Omega_R/2$. As for the  resonant case, most results focus on $L=2$, which already exhibits the qualitative trends observed for larger systems (additional data for larger $L$ appear in Appendix~\ref{AdditionalSHG}). The dependence of SHG on interaction strength, atom number, and lattice size is summarized in Fig.~\ref{fig2}. In the weak interaction regime ($U=10^{-4}$), increasing $N_a$ simply increases the SHG intensity (Fig.~\ref{fig2}a), as independent atoms can be excited without constraint. Similarly, increasing $L$ has negligible effect at small $U$ (Appendix~\ref{AdditionalSHG}, Fig.~\ref{figAPPL34}).

\begin{figure}
	\begin{tikzpicture}
	\node at (-2.3, 2.3) {\includegraphics[width=0.5\columnwidth]{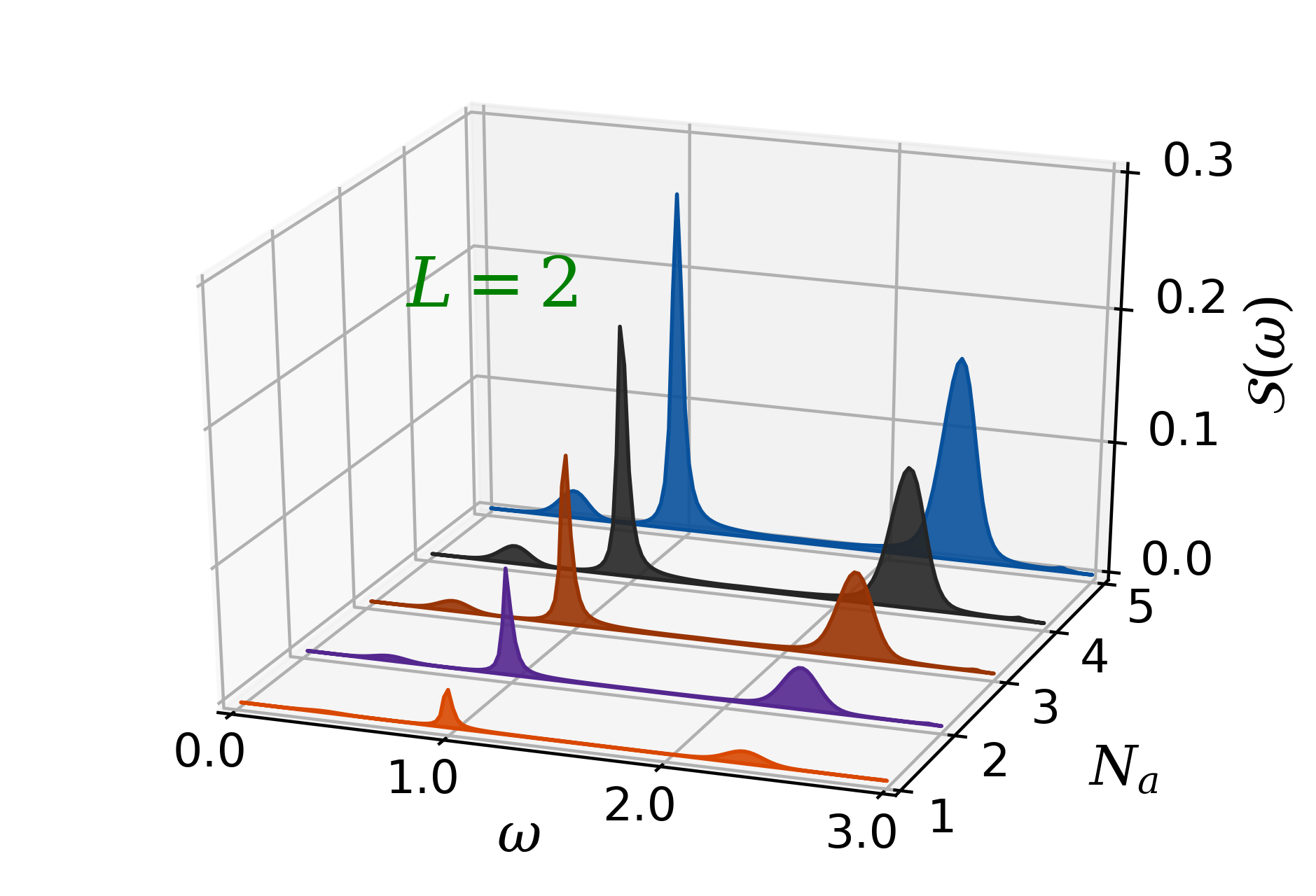}};
	\node at (2.0,2.3){\includegraphics[width=0.5\columnwidth]{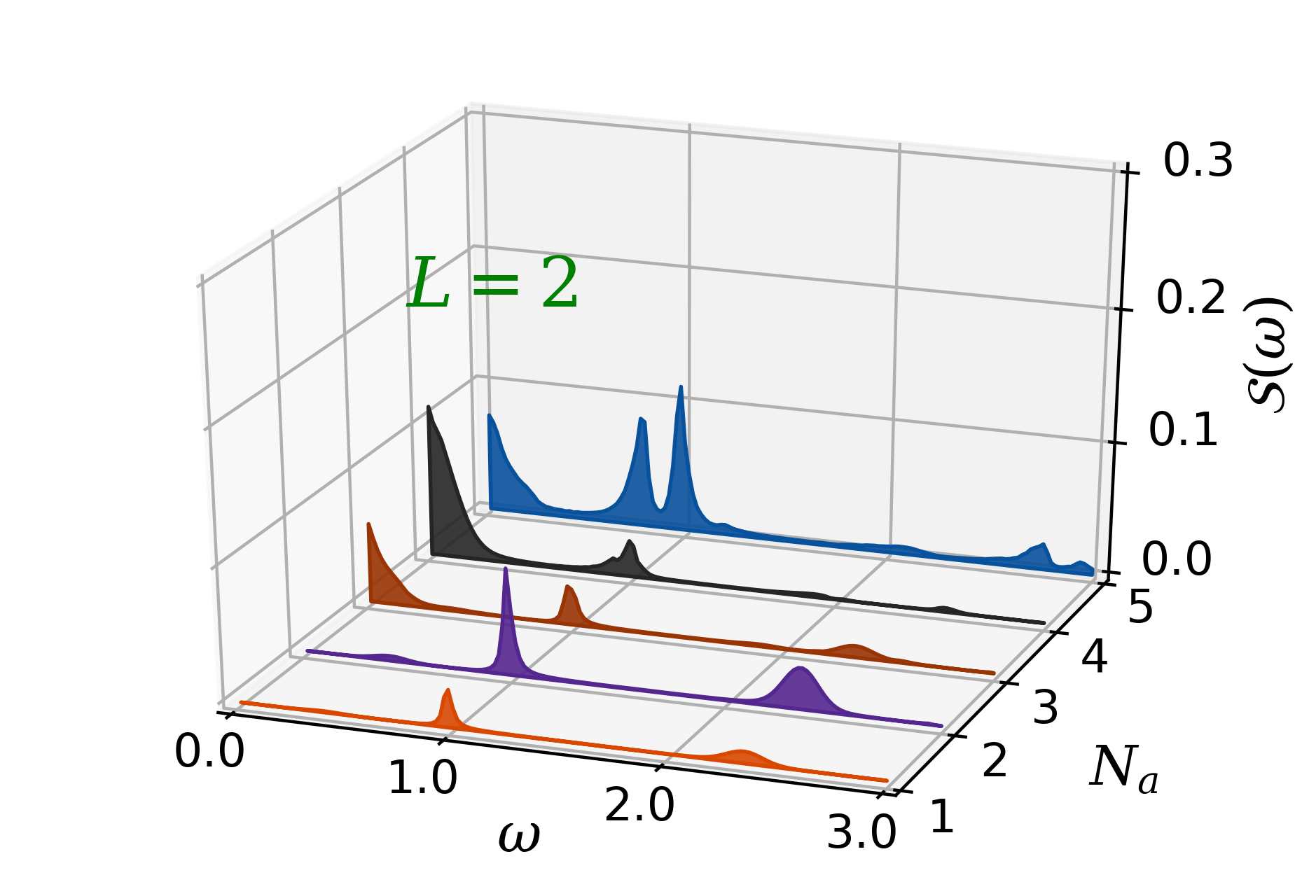}};
	\node at (-2.3, -0.60) {\includegraphics[width=0.5\columnwidth]{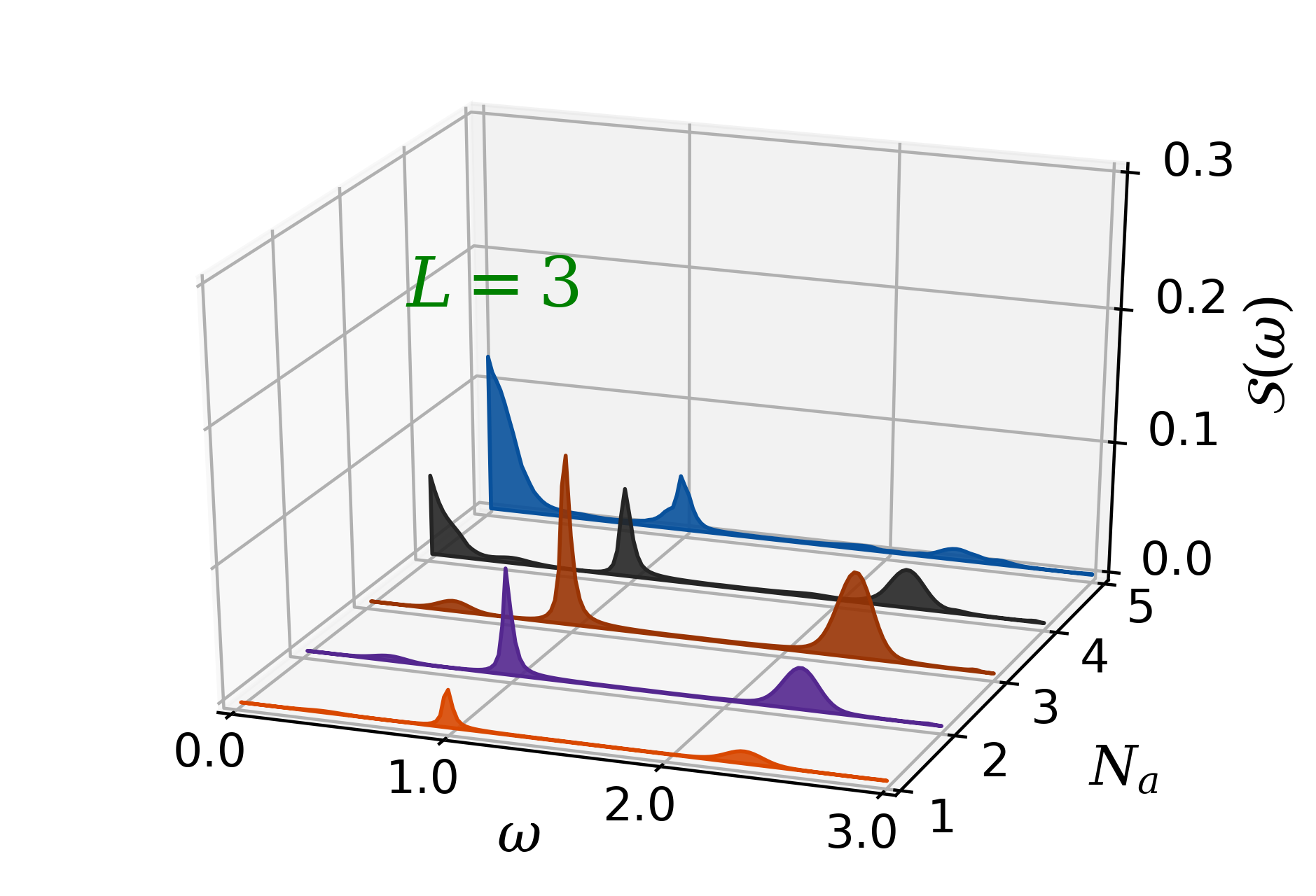}};
	\node at (2.0, -0.60) {\includegraphics[width=0.5\columnwidth]{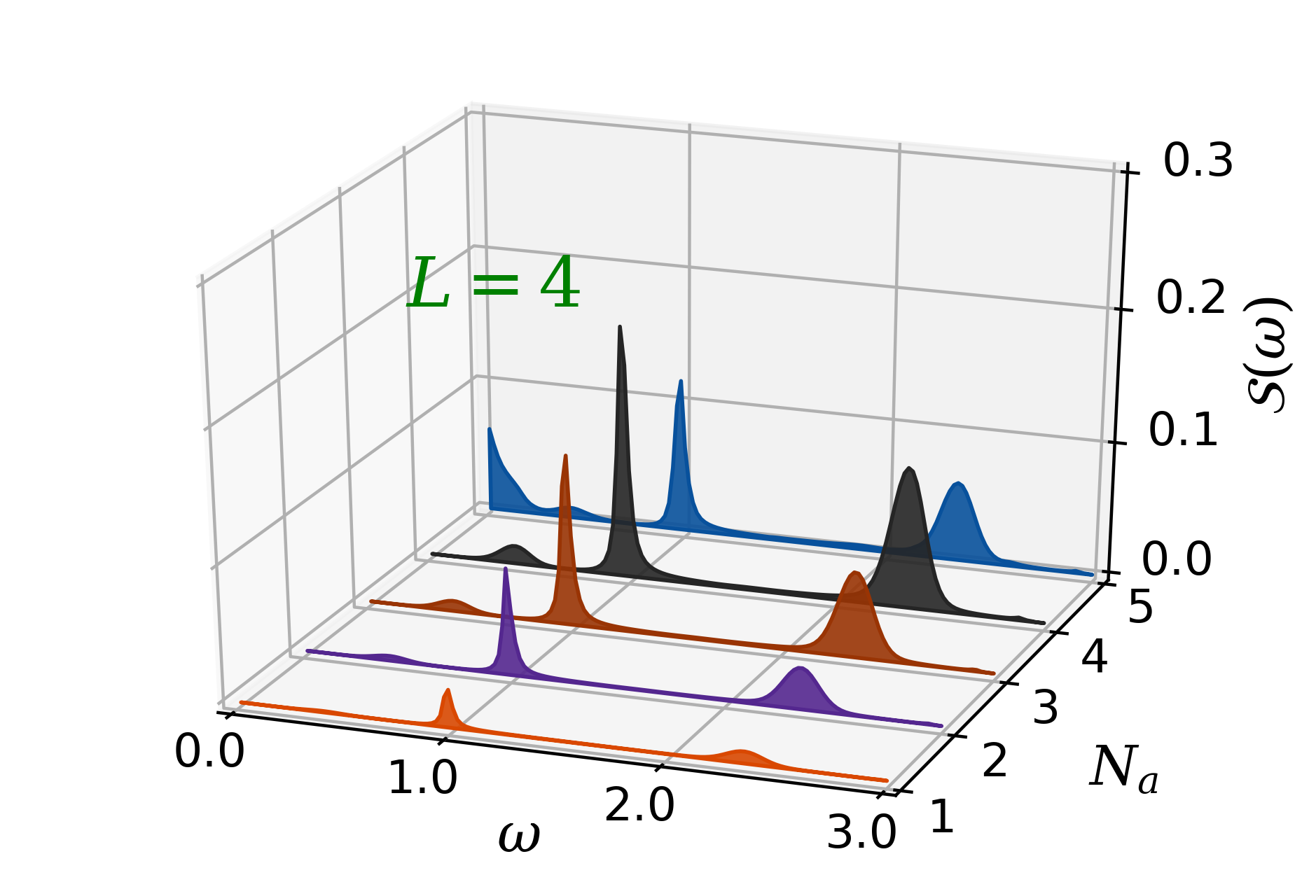}};
	\node[black] at (-1.30,3.0) {$\bm{(a)}$};
	\node[black] at (3.0,3.0) {$\bm{(b)}$};
	\node[black] at (-1.30,-0.0) {$\bm{(c)}$};
	\node[black] at (3.0,-0.0) {$\bm{(d)}$};
	\end{tikzpicture}
	\caption{Long-time SHG signal as a function of chain length $L$, atom number $N_a$, and interaction strength $U$. Initial state $|\Phi_0^{\mathrm{SHG}}\rangle$, with $U_{eg}=U/2$, $\eta=4.0$, and $\omega_0=1.0$. (a) Weak interactions ($U=10^{-4}$). (b–d) Strong interactions ($U=5.0$).}
	\label{fig2}
\end{figure}

The behavior changes markedly at strong interactions ($U=5$), Fig.~\ref{fig2}b–d. Here the SHG intensity grows with $N_a$ only up to $N_a \le L$; for $N_a>L$, the signal decreases and the main SHG peak can be reduced. When more than one excited atom occupies a site on average, interaction penalties suppress additional excitations.

To validate this picture, Fig.~\ref{fig3} shows the total excited-atom number $\mathcal{D}_e(t)=\sum_i\langle \hat{n}_i^e\rangle$. At small $U$ (Fig.~\ref{fig3}a), $\mathcal{D}_e$ varies widely with $N_a$ at early times. At large $U$ and $N_a>L$ (Fig.~\ref{fig3}b), a substantial excited state occupation is present already at $t=0$ and hinders further excitation, resulting in smaller changes in $\mathcal{D}_e$ before pumping stops or fluorescence losses dominate.

\begin{figure}
	\begin{tikzpicture}
	\node at (-2.3, 2.3) {\includegraphics[width=0.99\columnwidth]{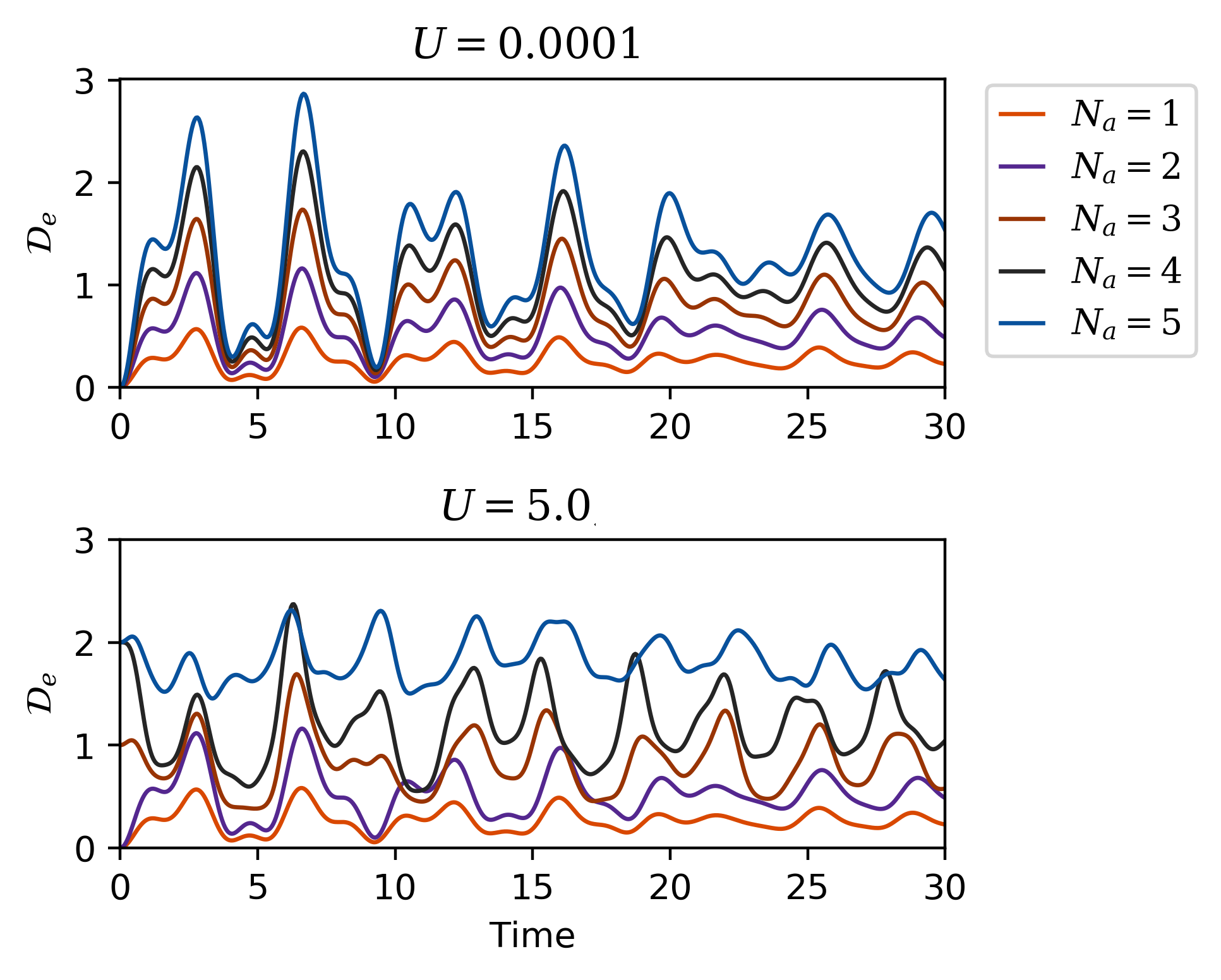}};
	\node[black] at (-0.50,4.70) {$\bm{(a)}$};
	\node[black] at (-0.50,1.60) {$\bm{(b)}$};
	\end{tikzpicture}
	\caption{Total excited-atom number $\mathcal{D}_e(t)$ for weak (a) and strong (b) interactions. Parameters: $L=2$, $U_{eg}=U/2$, $\eta=4.0$, $\omega_0=1.0$, initial state $|\Phi_0^{\mathrm{SHG}}\rangle$.}
	\label{fig3}
\end{figure}

For $U$ not too small and $N_a>L$, another feature emerges: significant spectral weight at frequencies well below 
$\omega_0=1$ (Fig.~\ref{fig2}b–d), plausibly arising from the many-body interactions. 
These, at large \(U\), introduce fine structure in the low-energy excitation spectrum (on the order of the \(t^{2}/U\) scale), yielding multiple nearby states that become optically accessible via non-resonant transitions.
Overall, scanning a broad interaction range (Fig.~\ref{fig4}) we note a gradual suppression of spectral intensity with increasing $U$, accompanied by splitting and shifts of the SHG peak. The corresponding evolution of $\mathcal{D}_e(t)$ confirms that interactions impede transitions to excited states. Notably, after the initial excitation spike, $\mathcal{D}_e$ changes only weakly even when the excited levels remain sparsely populated, indicating persistent interaction-induced hindrance.

\begin{figure}[b]
	\begin{tikzpicture}
	\node at (-2.3, 2.3) {\includegraphics[width=0.8\columnwidth]{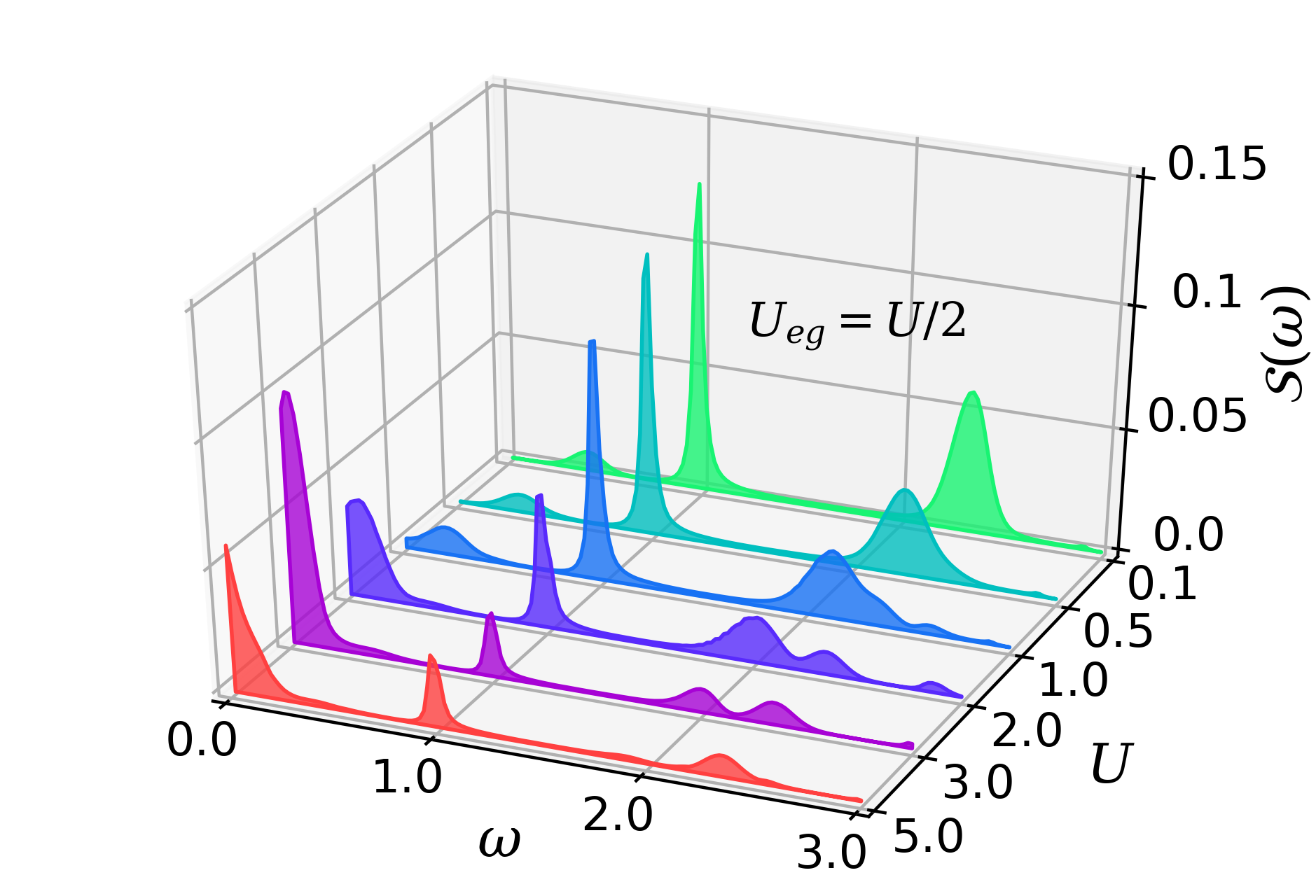}};
	\node at (-1.9, -2.10) {\includegraphics[width=0.85\columnwidth]{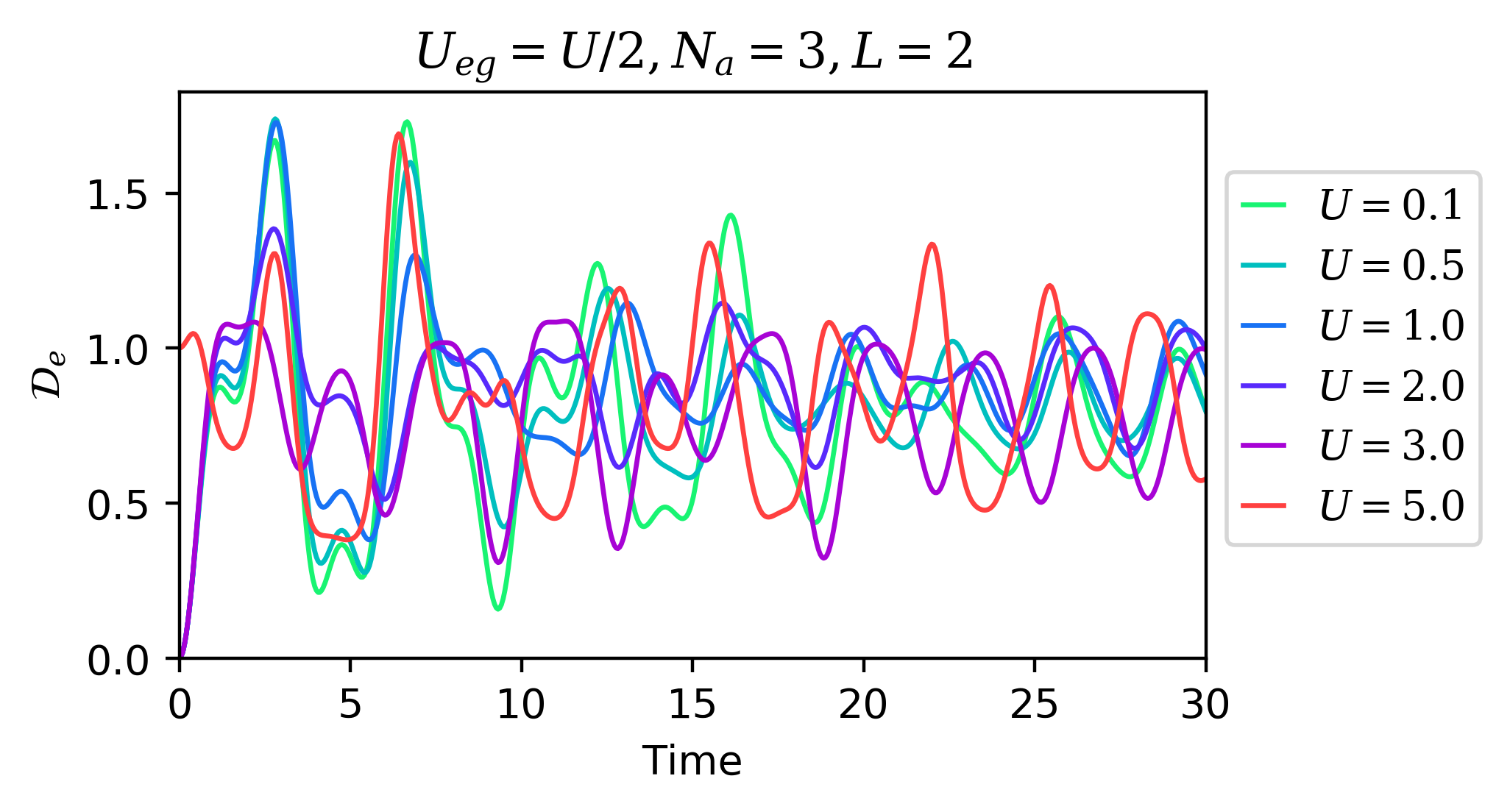}};
	\node[black] at (-3.50,3.0) {$\bm{(a)}$};
	\node[black] at (-0.50,-1.20) {$\bm{(b)}$};
	\end{tikzpicture}
	\caption{SHG for Intermediate interactions. (a) SHG spectra. (b) Total excited-atom number $\mathcal{D}_e(t)$. Parameters: $L=2$, $N_a=3$, $U_{eg}=U/2$, $\eta=4.0$, $\omega_0=1.0$, initial state $|\Phi_0^{\mathrm{SHG}}\rangle$.}
	\label{fig4}
\end{figure}

As general observations to both resonance and SHG results presented so far, we note that at long times the average incident photon number $\langle b^\dagger b \rangle$ stabilises, up to small oscillations, as the coupling to the fluorescent field vanishes. Furthermore, interactions and substantial excited state occupation can hinder further transitions, though this may depend on our chosen interaction strengths. Other regimes, that can  possibly change the picture,  are under investigation. Finally, the SHG region often displays pronounced structure, which we attribute to the non-perturbative boson–photon coupling and the interplay between Autler–Townes renormalization, inter-site hopping  and boson–boson interactions. A detailed analysis and a validation of this conjecture lies beyond the scope of this work.

%%%%%%%%%%%%%%%%%%%%%%%%%%%%%%%%%%%%%
%%%%%%%%%%%%%%%%%%%%%%%%%%%%%%%%%%%%%
%%%%%%%%%%%%%%%%%%%%%%%%%%%%%%%%%%%%%
%%%%%%%%%%%%%%%%%%%%%%%%%%%%%%%%%%%%%
%%%%%%%%%%%%%%%%%%%%%%%%%%%%%%%%%%%%%
%%%%%%%%%%%%%%%%%%%%%%%%%%%%%%%%%%%%%

\subsubsection{Cavity Leakage}
 Finally, we briefly address the role of cavity leakage as accounted for by the classical Caldeira-Leggett oscillators. Resonant spectra are shown in Fig.~\ref{fig8}(a–b) for \(L=2\) and two atom numbers. The time-evolving surfaces plots include leakage; orange curves are long-time spectra without it.  In this regime, we drive the dynamics by ramping the cavity field, with the full system-cavity-bath initially in the ground state (so that $\langle \hat{b}^\dagger \hat{b}\rangle \simeq 0$ at $t=0$). To highlight the impact of leakage, we consider the strongly interacting case ($U=5$).
As seen in Fig.~\ref{fig8}, cavity leakage clearly suppresses the fluorescence intensity, regardless of particle number. Compared to the weak-interaction regime (not shown), this suppression is more pronounced when atom–atom interactions increase. This is consistent with the behavior of $\langle \hat{b}^\dagger \hat{b}\rangle$ in Fig.~\ref{fig8}\textbf{(e)}: stronger Hubbard repulsion limits excitation to higher levels, the cavity photon population is more strongly reduced by leakage, and fewer incident photons are absorbed. Consequently, the fluorescence response is diminished.

This trend is similar for SHG, as shown in Fig.~\ref{fig8}(c–d): for the same system considered in the resonant case, leakage reduces the SHG intensity and produces a slight redshift of the peak. This behavior also occurs at low interaction strengths. The evolution of $\langle \hat{b}^\dagger \hat{b}\rangle$ in Fig.~\ref{fig8}\textbf{(e)} confirms that the photon population decreases over time due to leakage, thereby diminishing the effectiveness of the excitation process.

%%%%%%%% XXXXXXXXXXXXXXXXXXXXX %%%%%%%%%
\begin{figure}
	\begin{tikzpicture}
	\node at (-2.3, 2.3){\includegraphics[width=0.48\columnwidth]{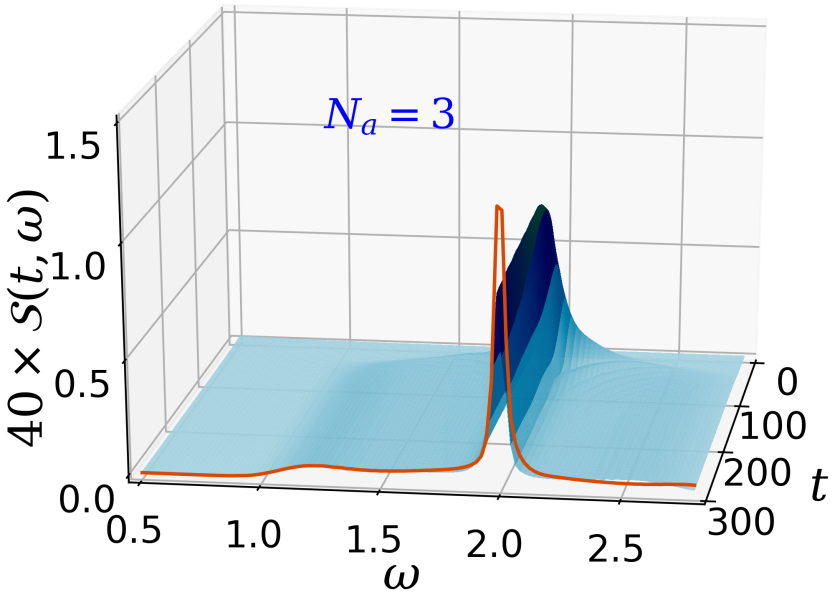}};
	\node at (2.2,2.3) {\includegraphics[width=0.48\columnwidth]{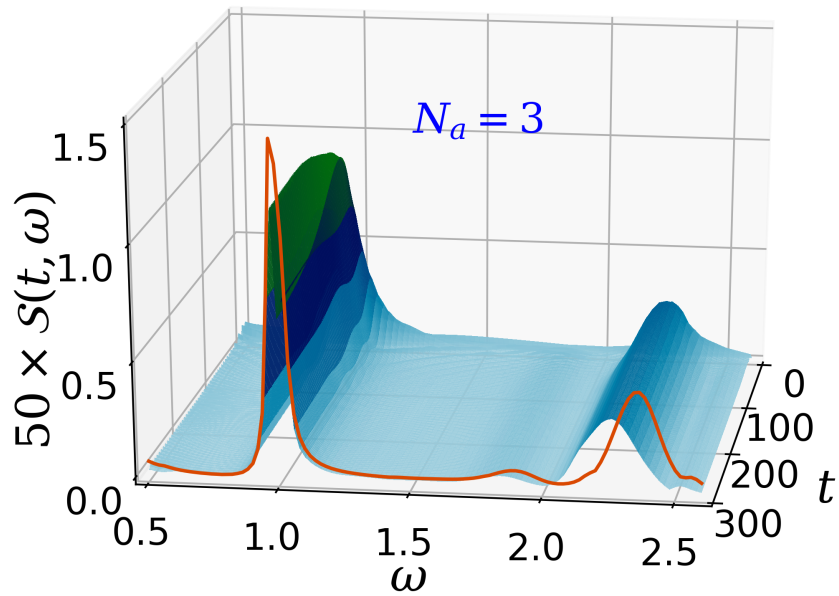}};
	\node at (-2.3, -0.8) {\includegraphics[width=0.48\columnwidth]{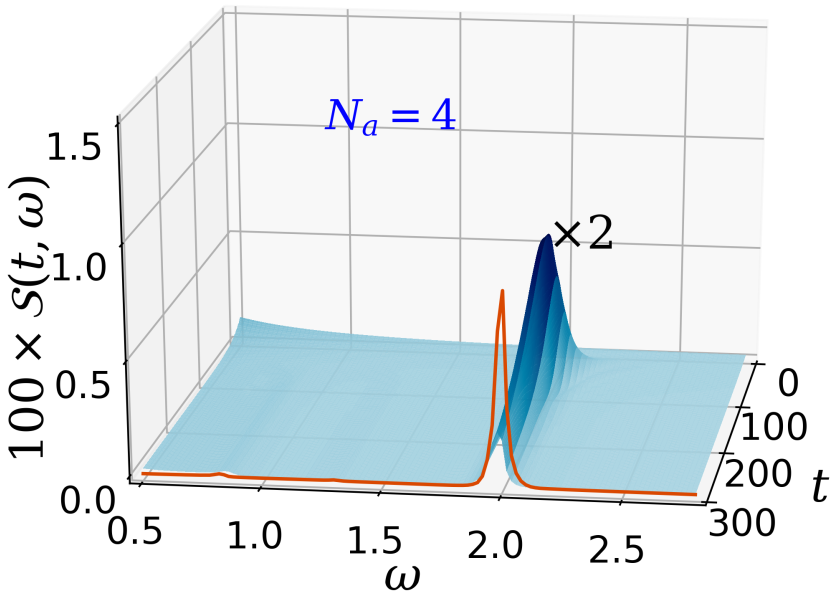}};
	\node at (2.2, -0.8) {\includegraphics[width=0.48\columnwidth]{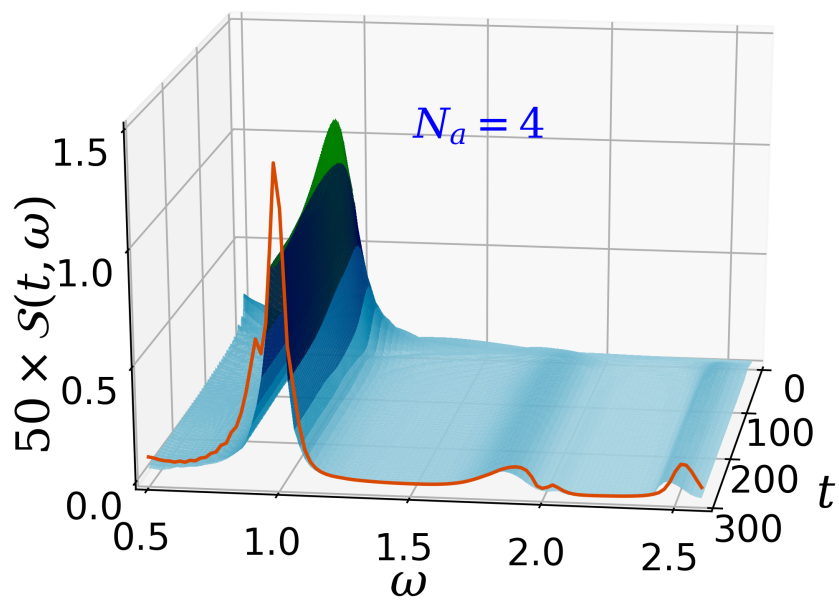}};
	\node at (-0.1, -3.80) {\includegraphics[width=1.0\columnwidth]{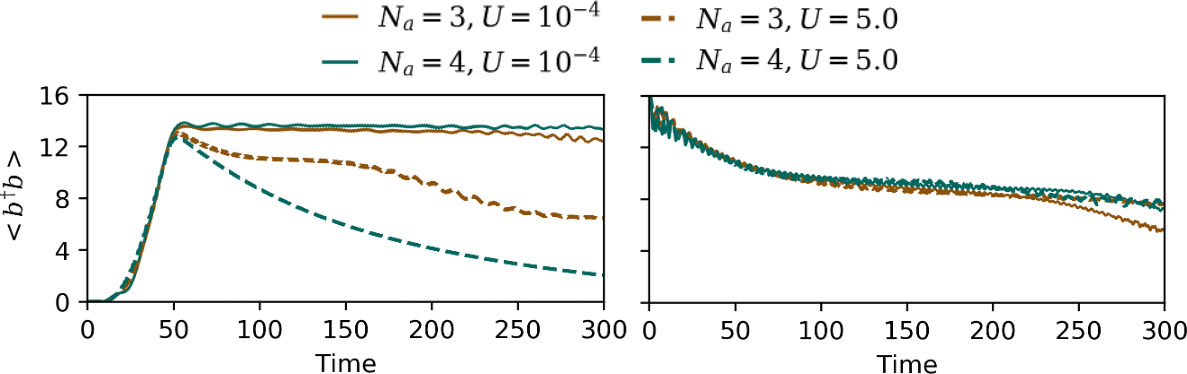}};	
	    \node[black] at (-1.10,3.40) {$\bm{(a)}$};
     	\node[black] at (3.50,3.40) {$\bm{(c)}$};
     	\node[black] at (-1.10,0.28) {$\bm{(b)}$};
     	\node[black] at ( 3.50,0.28) {$\bm{(d)}$};
     	\node[black] at (-2.8,-2.70) {$\bm{(e)}$};
	\end{tikzpicture}
\caption{$(\mathbf{a})$–$(\mathbf{b})$: Time dependent resonance spectra for atom numbers $N_a=3,4$ in the presence of leakage (3D plots) versus long-time-limit spectra without leakage (red curve) for $L=2$ chains with $U=5$, $U_{eg}=U/2$, $\eta=4.0$, 
$\omega_0=\Omega_R=2.0$, and initial state $|\Phi_0^{\mathrm{Res}}\rangle$.
Bath parameters: $C_k = A(\Delta k)^a$ and $C_j = A(\Delta j)^a$, with $A=0.005$, $a=0.6$, $\Delta=0.01$, and $\mathcal{N}_{K/J}=1000$. In this case, a laser drive is applied, with 
drive strength $\gamma_d$ chosen so that, without leakage, $\langle \hat{b}^\dagger \hat{b}\rangle \approx 16$ after the drive. $(\mathbf{c})$–$(\mathbf{d})$: Time dependent SHG spectra with leakage (3D plots) versus  long-time-limit  no-leakage spectra (red), using the same parameters  in $(\mathbf{a})$–$(\mathbf{b})$  but with $\omega_0=\Omega_R/2=1.0$ and initial state $|\Phi_0^{\mathrm{SHG}}\rangle$. 
$(\mathbf{e})$: Incident-field photon number $\langle \hat{b}^\dagger \hat{b}\rangle$ with leakage for the resonance (left) and SHG (right) cases. Results for small $U$ are also shown.} 
\label{fig8}
\end{figure}

%%%%%%%%%%%%%%%%%%%%%%%%%%%%%%%%%%%%%
%%%%%%%%%%%%%%%%%%%%%%%%%%%%%%%%%%%%%
%%%%%%%%%%%%%%%%%%%%%%%%%%%%%%%%%%%%%
%%%%%%%%%%%%%%%%%%%%%%%%%%%%%%%%%%%%%
%%%%%%%%%%%%%%%%%%%%%%%%%%%%%%%%%%%%%
%%%%%%%%%%%%%%%%%%%%%%%%%%%%%%%%%%%%%

\section{Fluorescence from the 2BEC}\label{sec.BEC_GP_limit}
%%%%%%
\noindent In the simulations we take the initial state as
\begin{equation}
|\Psi(0)\rangle = |\eta\rangle\, |0_f\rangle\, |\Phi_{\rm BEC}\rangle ,
\end{equation}
where $|\eta\rangle$ is a coherent  cavity photon state with mean photon number $\eta^2$, and $|0_f\rangle$ is the vacuum of the fluorescent mode. The state $|\Phi_{\rm BEC}\rangle$ is the ground state of the uncoupled 2BEC, and for all results shown we take $U_{ee}=U_{gg}=2U_{eg}$.
To compare results for different atom numbers, we follow the standard Dicke-model convention \citep{Dicke1954,Kirton2019}, where both atom--atom interactions and  light--matter coupling are scaled with $N_a$ \citep{Ghasemian2021} to avoid divergences of the spectral intensity in the thermodynamic limit \citep{Nagy2010}.
We thus rewrite $\hat{H}_{\rm 2BEC-cav}(t)$ of Eq.~(\ref{2BECav}) as
\begin{align}
 &\hat{H}_{\rm 2BEC-cav}(t)=E_g\hat{\alpha}^\dagger\hat{\alpha}+E_e\hat{\beta}^\dagger\hat{\beta}
 +\frac{U_{gg}}{2N_a}\hat{\alpha}^\dagger\hat{\alpha}^\dagger\hat{\alpha}\hat{\alpha}\nonumber
 \\
 &
 +\frac{U_{ee}}{2N_a}\hat{\beta}^\dagger\hat{\beta}^\dagger\hat{\beta}\hat{\beta}\nonumber
 +\frac{U_{eg}}{N_a}\hat{\beta}^\dagger\hat{\alpha}^\dagger\hat{\alpha}\hat{\beta}
 +\omega_0\hat{b}^\dagger\hat{b}+\omega \hat{b}_f^\dagger\hat{b}_f \nonumber\\
 &+(\hat{\alpha}^\dagger\hat{\beta}+\mathrm{h.c.})\big[
 \frac{\gamma_0}{\sqrt{N_a}}(\hat{b}^\dagger+\hat{b})+ \frac{\tilde{\gamma}\, e^{-\Gamma t}}{\sqrt{N_a}}(\hat{b}_f^\dagger+\hat{b}_f)\big]. \label{rescaled}
\end{align}
In these initial calculations, we set $U_{gg}/N_a=U_{ee}/N_a=U$ and $U_{eg}/N_a=U/2$ for illustration, deferring to future
work the case of interactions with other relative strengths, where different physics is likely to emerge. Worth mentioning, the two-mode approach assumes weak, dilute interactions, and in our case the effective interaction strength decreases with increasing 
$N_a$, consistent with \citep{Ghasemian2021}.
Resonance spectra for $N_a=40$ and $N_a=100$ with cavity coupling $\gamma_0=0.3$ are shown in Fig.~\ref{figBEC2}. Although the incident frequency $\omega_0$ matches the nominal atomic resonance, i.e. 
$\omega_0=\Omega_R=2$, the fluorescent spectra do not display the familiar Mollow triplet (the latter is present at low
number of atoms $N_a$, see Fig.~\ref{Mollow_BEC_App}). Instead, they feature several peaks---some broad and merging into a plateau around $\omega=2$. 

This behaviour is rather intriguing and likely due to the quantum nature of photons. Indeed, even for a single two-level system and at low photon number, deviations from the usual Mollow triplet are known to occur \cite{SHG93,SHG95}: Semi-classically, the triplet depends on the effective coupling $g^{SC}_{\mathrm{eff}}=\gamma_0\eta$. Quantum mechanically, however, different pairs $(\gamma_0,\eta)$ can yield the same $g^{SC}_{\mathrm{eff}}$ yet produce qualitatively different fluorescence spectra \cite{SHG93,SHG95}. This factor may contribute to the structure of 
$\mathcal{S}(\omega)$ in Fig.~\ref{figBEC2}, together with the presence of many bosons distributed across photon-dressed ground and excited states, which introduces multiple nearby many-body resonant channels. 
Furthermore, increasing the atom–atom interaction strength (dashed curves) produces a slight redshift of the resonant spectra. However, because of the chosen scaling of the interaction strength with $N_a$, spectra corresponding to different particle numbers but identical interaction strength remain essentially very similar in shape. 

{\rd For further perspective, it can be expedient to consider the exact mapping of 2BEC part Hamiltonian into a collective-spin model,
i.e. $H_{\rm 2BEC} \rightarrow C_{N_a}+\Omega_a \hat{J}_z+\chi \hat{J}_z^2 $.  For $N_a>1$, The $U_{gg},U_{ee},U_{eg}$ interactions give $\chi\neq0$, and this lifts the degeneracy of the atomic transitions, yielding a ladder of frequencies $ \Delta_m=\Omega_a+\chi(2m+1),$
spanning a width $W\sim2|\chi|(N_a-1)$. The cavity coupling
$\hat{J}_x(b+b^\dagger)$
connects neighboring states $m\rightarrow m\pm1$, with nonuniform transition strengths transition strengths determined by the collective-spin ladder-operator matrix elements
$\sqrt{J(J+1)-m(m\pm1)}$ 
which favor transitions near the center of the ladder. Consequently, the cavity field dresses a multilevel collective system rather than a single two-level transition, and the conventional Mollow triplet evolves into a multiplet of overlapping dressed resonances. A numerical analysis of the full interacting 2BEC+cavity spectrum (not reported here) further shows that interactions and cavity dressing shift and strongly broaden and redistribute the many-body eigenenergies. The fluorescence signal probes transitions among these dressed collective states, producing the observed broad envelope and central plateau rather than three well-resolved Mollow peaks.}

The fluorescence spectra for the SHG regime are reported in Figure~\ref{figBEC1}. We considered
systems  with $N_a=40$ and $N_a=400$ and different
strengths of the cavity-BEC coupling $\gamma_0$ and of the particle-particle interaction $U$.
At weak coupling ($\gamma_0=0.1$), spectra for different $N_a$ are essentially identical, both with and without interactions, again an effect of the parameters rescaling via Eq.~(\ref{rescaled}) (as for the resonant case,
the spectra are are redshifted when $U\neq 0$}.
The situation changes at stronger coupling: for $\gamma_0=0.3$ the spectra for $N_a=40$ and $N_a=400$ differ significantly, and interactions lead to clear qualitative changes. The redshift of the SHG peak is also larger than at weak coupling. Moreover, the overall fluorescence intensity increases at large $\gamma_0$, enhancing the SHG peak while leaving the Rayleigh peak height almost unchanged. 

In this initial work, we just provide a preliminary, not exhaustive discussion and interpretation of the 2BEC fluorescence. A more detailed analysis—including a broader parameter scan and the inclusion of a driving laser field that injects photons at a finite rate—is underway. Finally, we note that despite its simplicity, the present cavity-BEC model already produces a wide range of resonant and SHG fluorescence behaviors and non-trivial spectral features, arising from the quantum nature of cavity photons in the low-density regime. This observation represents one of the main outcomes of the present study.

\begin{figure}[t]
    \centering
    \includegraphics*[scale=0.35]{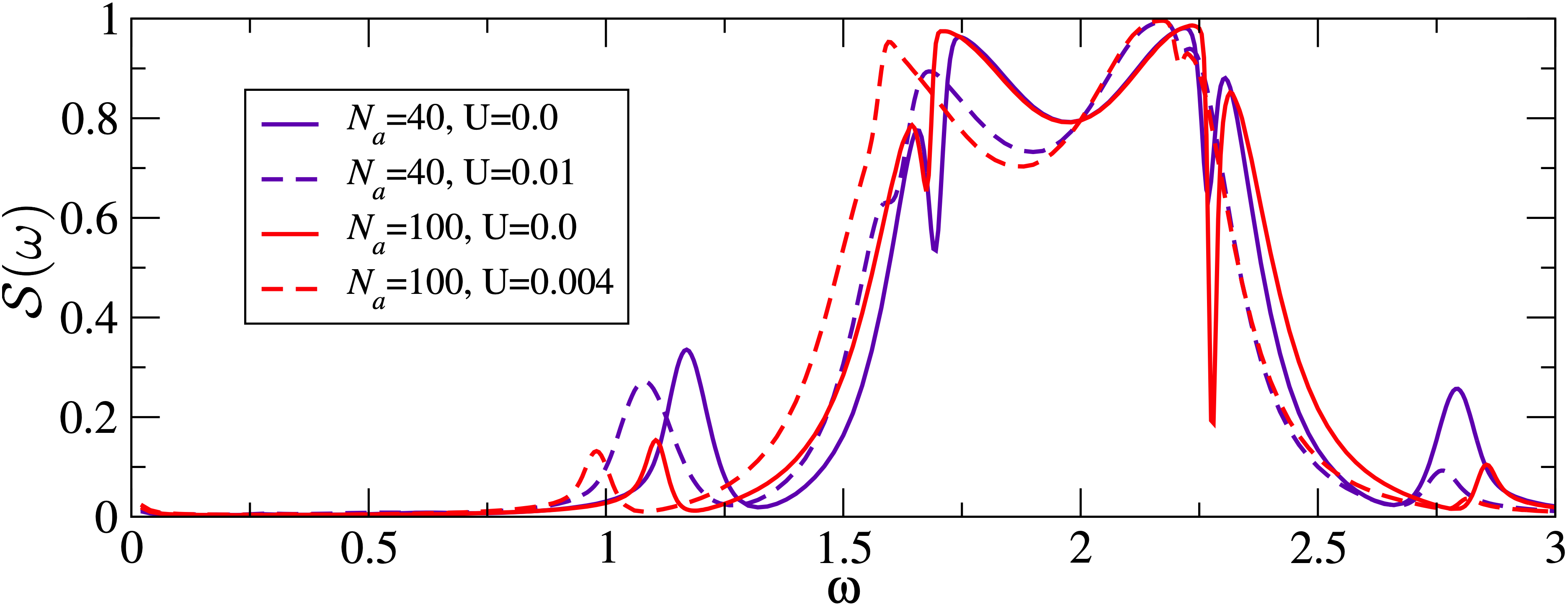}
    \caption{Resonance fluorescence spectra for $N_a=40$ and $N_a=100$ obtained with an initial coherent photon state satisfying $\langle \hat{b}^\dagger \hat{b}\rangle = 16$ and cavity frequency $\omega_0=\Omega_R=2.0$.  
Calculations use $U_{gg}/N_a = U_{ee}/N_a = U$, $U_{eg}/N_a = U/2.0$, $\gamma_0 = 0.3$, $\gamma_f = 0.1$, $\Gamma = 0.02$, $E_g = 0.0$, and $E_e = 2.0$.  
The plots show the long-time fluorescence spectra, highlighting the emergence of multiple broad peaks rather than a standard Mollow-triplet structure.    
    }
    \label{figBEC2}
\end{figure}

\begin{figure}[t]
    \centering
    \includegraphics*[scale=0.15]{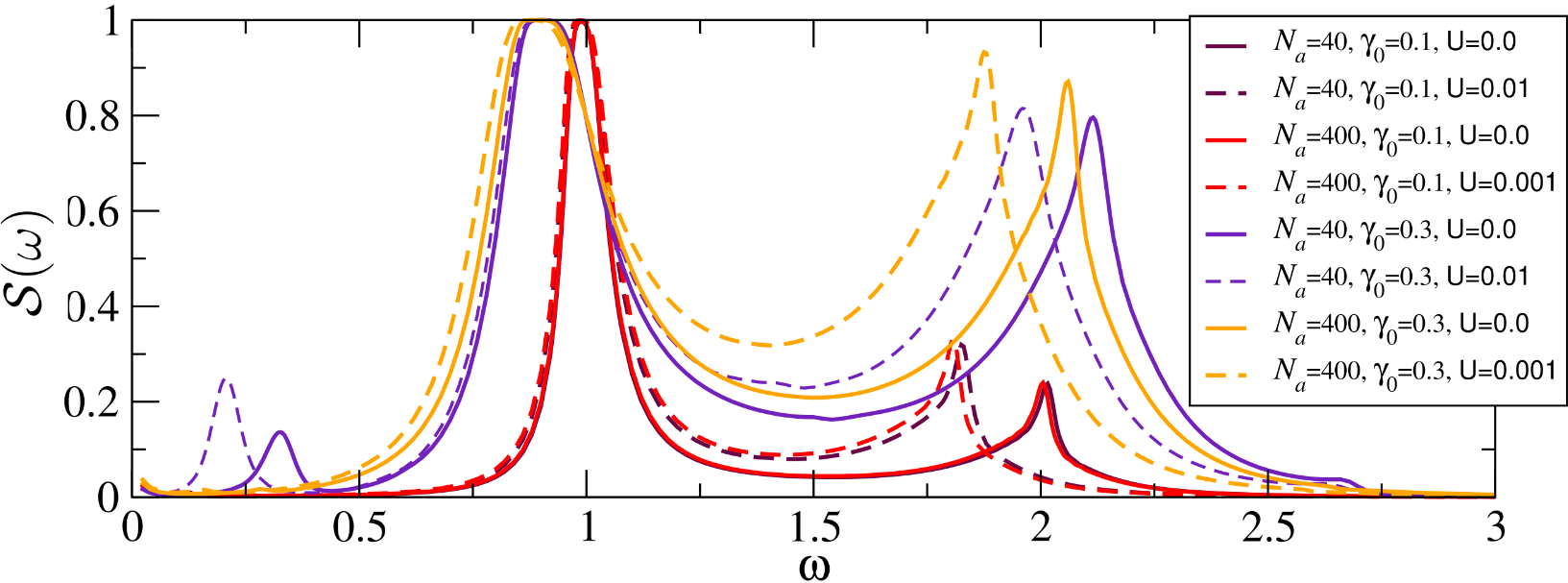}
    \caption{
SHG fluorescence spectra for $N_a=40$ and $N_a=400$ obtained with an initial coherent photon state satisfying $\langle \hat{b}^\dagger \hat{b}\rangle = 16$ and cavity frequency $\omega_0=\Omega_R/2=1.0$.  
The parameters used are $U_{gg}/N_a = U_{ee}/N_a = U$, $U_{eg}/N_a = U/2.0$, $\gamma_0$ as specified in the plots, $\gamma_f = 0.1$, $\Gamma = 0.02$, $E_g = 0.0$, and $E_e = 2.0$.  
Results illustrate the dependence of the SHG peak, Rayleigh response, and overall spectral intensity on atom number, atom–atom interactions, and cavity–BEC coupling. 
    }\label{figBEC1}
\end{figure}%
%
%
%%%%%%%%%%%%%%%%%%%%%%%%%%%%%%%%%%%%%%%%%%%%
%%%%%%%%%%%%%%%%%%%%%%%%%%%%%%%%%%%%%%%%%%%%
%%%%%%%%%%%%%%%%%%%%%%%%%%%%%%%%%%%%%%%%%%%%
%%%%%%%%%%%%%%%%%%%%%%%%%%%%%%%%%%%%%%%%%%%%
%%%%%%%%%%%%%%%%%%%%%%%%%%%%%%%%%%%%%%%%%%%%
%%%%%%%%%%%%%%%%%%%%%%%%%%%%%%%%%%%%%%%%%%%%
\section{Conclusions\label{sec.conclude}}
In this work we investigated fluorescence from ultracold bosons in an optical cavity using an exact time-dependent configuration-interaction approach (ETDCI). We examined both small Bose–Hubbard lattices and two-component Bose–Einstein condensates (2BECs). In the explored parameter range, the fluorescent response depends strongly on atom–atom interactions, particle number, and— for finite lattices—the number of sites.

For optical lattices, SHG intensity at weak interactions grows with atom number and is only weakly influenced by interaction strength or lattice size. At strong interactions the same trend holds while atoms remain fewer than sites, but once atoms outnumber sites the SHG peak is suppressed by the interaction cost of multiple excitations per site. Strong interactions also generate low-frequency spectral weight: many-body states appear that can be optically accessed from the ground state.

Resonant fluorescence shows a comparable pattern but with non-monotonic intensity versus particle number under strong interactions. Mollow sidebands are strongly reduced when photons are pumped into the cavity at a finite rate. Cavity leakage decreases overall spectral intensity and redshifts the SHG peak, especially for strong interactions.
For 2BECs, SHG spectra at weak atom–cavity coupling are nearly particle-number invariant once interactions are suitably rescaled, though interacting and non-interacting cases differ. At strong coupling the spectra change markedly with particle number and interaction strength; strong interactions also redshift SHG. Resonant 2BEC spectra, by contrast, remain largely particle-number independent for different interaction strength, and show no clear Mollow triplet: instead several peaks merge into an almost flat profile near resonance, likely due to multiphoton processes and photon dressing of atomic levels.

Overall, our simple models and the inherent results outline general trends for fluorescence from ultracold bosons in cavities, offering guidance for future experimental and theoretical work, while further parameter exploration and model extensions are expected to reveal additional phenomena {\rd worthy of further investigation. 

The ETDCI treatment used here  provides an essentially exact characterization of the optical response of the boson--cavity models considered, in contrast to many-body approaches commonly used in nonequilibrium quantum theory, which generally rely on some level of approximation. At the same time, it is restricted to relatively small system sizes. Extending the present calculations would naturally call for a finite-size scaling analysis, which, while robust in principle, is computationally demanding for systems combining interacting bosons with cavity and fluorescent photon modes. Tensor-network methods could in principle be employed, but the large local bosonic Hilbert spaces and globally coupled photon fields are expected to generate substantial entanglement growth, while a classical treatment of the radiation fields would modify the nature of the light--matter problem. 

Other approaches, including Holstein--Primakoff mappings, Green-function methods, Gutzwiller treatments, strong-coupling expansions, and Bogoliubov theories, could also be considered, although each relies on assumptions and approximations that affect its range of validity. 

How some of these methods, individually or in combination, can be employed to explore the effects of interactions, particle-number scaling, and cavity-coupling strength on the response of larger boson--cavity systems is currently under investigation and will be reported elsewhere.}

%%%%%
%%%%%
%%%%%
%%%%%
%%%%%
%%%%%%%%%%%%%%%%%%%%%%%%%%%%%%%%%%%%%%%%%%%%
%%%%%%%%%%%%%%%%%%%%%%%%%%%%%%%%%%%%%%%%%%%%
%%%%%%%%%%%%%%%%%%%%%%%%%%%%%%%%%%%%%%%%%%%%
%%%%%%%%%%%%%%%%%%%%%%%%%%%%%%%%%%%%%%%%%%%%
%%%%%%%%%%%%%%%%%%%%%%%%%%%%%%%%%%%%%%%%%%%%
%%%%%%%%%%%%%%%%%%%%%%%%%%%%%%%%%%%%%%%%%%%%
 \subsection*{Acknowledgments}
Megha Gopalakrishna and Claudio Verdozzi acknowledge support from the Swedish Research
Council (grants number 2017-03945 and 2022-04486).
Emil Vi\~nas Bostr\"om acknowledges funding from the European Union's
Horizon Europe research and innovation programme under
the Marie Sk\l odowska-Curie grant agreement No
101106809.

 \subsection*{Funding}
 This work was supported by the Vetenskapsrådet (2017-03945,
 2022-04486) and European Union’s Horizon Europe Research and
 Innovation Programme Under the Marie Skłodowska-Curie Grant
 Agreement (101106809).

 \bibliography{bib}

%apsrev4-2.bst 2019-01-14 (MD) hand-edited version of apsrev4-1.bst
%Control: key (0)
%Control: author (8) initials jnrlst
%Control: editor formatted (1) identically to author
%Control: production of article title (0) allowed
%Control: page (0) single
%Control: year (1) truncated
%Control: production of eprint (0) enabled
\begin{thebibliography}{83}%
\makeatletter
\providecommand \@ifxundefined [1]{%
 \@ifx{#1\undefined}
}%
\providecommand \@ifnum [1]{%
 \ifnum #1\expandafter \@firstoftwo
 \else \expandafter \@secondoftwo
 \fi
}%
\providecommand \@ifx [1]{%
 \ifx #1\expandafter \@firstoftwo
 \else \expandafter \@secondoftwo
 \fi
}%
\providecommand \natexlab [1]{#1}%
\providecommand \enquote  [1]{``#1''}%
\providecommand \bibnamefont  [1]{#1}%
\providecommand \bibfnamefont [1]{#1}%
\providecommand \citenamefont [1]{#1}%
\providecommand \href@noop [0]{\@secondoftwo}%
\providecommand \href [0]{\begingroup \@sanitize@url \@href}%
\providecommand \@href[1]{\@@startlink{#1}\@@href}%
\providecommand \@@href[1]{\endgroup#1\@@endlink}%
\providecommand \@sanitize@url [0]{\catcode `\\12\catcode `\$12\catcode
  `\&12\catcode `\#12\catcode `\^12\catcode `\_12\catcode `\%12\relax}%
\providecommand \@@startlink[1]{}%
\providecommand \@@endlink[0]{}%
\providecommand \url  [0]{\begingroup\@sanitize@url \@url }%
\providecommand \@url [1]{\endgroup\@href {#1}{\urlprefix }}%
\providecommand \urlprefix  [0]{URL }%
\providecommand \Eprint [0]{\href }%
\providecommand \doibase [0]{https://doi.org/}%
\providecommand \selectlanguage [0]{\@gobble}%
\providecommand \bibinfo  [0]{\@secondoftwo}%
\providecommand \bibfield  [0]{\@secondoftwo}%
\providecommand \translation [1]{[#1]}%
\providecommand \BibitemOpen [0]{}%
\providecommand \bibitemStop [0]{}%
\providecommand \bibitemNoStop [0]{.\EOS\space}%
\providecommand \EOS [0]{\spacefactor3000\relax}%
\providecommand \BibitemShut  [1]{\csname bibitem#1\endcsname}%
\let\auto@bib@innerbib\@empty
%</preamble>
\bibitem [{\citenamefont {Zhai}(2021)}]{zhai_2021}%
  \BibitemOpen
  \bibfield  {author} {\bibinfo {author} {\bibfnamefont {H.}~\bibnamefont
  {Zhai}},\ }\href {https://doi.org/10.1017/9781108595216} {\emph {\bibinfo
  {title} {Ultracold Atomic Physics}}}\ (\bibinfo  {publisher} {Cambridge
  University Press},\ \bibinfo {year} {2021})\BibitemShut {NoStop}%
\bibitem [{\citenamefont {Stwalley}\ and\ \citenamefont
  {Wang}(1999)}]{Molecule1999}%
  \BibitemOpen
  \bibfield  {author} {\bibinfo {author} {\bibfnamefont {W.}~\bibnamefont
  {Stwalley}}\ and\ \bibinfo {author} {\bibfnamefont {H.}~\bibnamefont
  {Wang}},\ }\bibfield  {title} {\bibinfo {title} {Special review lecture -
  photoassociation of ultracold atoms: A new spectroscopic technique},\ }\href
  {https://doi.org/10.1006/jmsp.1999.7838} {\bibfield  {journal} {\bibinfo
  {journal} {{J}ournal of {M}olecular {S}pectroscopy}\ }\textbf {\bibinfo
  {volume} {195}},\ \bibinfo {pages} {194} (\bibinfo {year}
  {1999})}\BibitemShut {NoStop}%
\bibitem [{\citenamefont {Lewenstein}\ \emph {et~al.}(2007)\citenamefont
  {Lewenstein}, \citenamefont {Sanpera}, \citenamefont {Ahufinger},
  \citenamefont {Damski}, \citenamefont {Sen(De)},\ and\ \citenamefont
  {Sen}}]{Lewenstein2007}%
  \BibitemOpen
  \bibfield  {author} {\bibinfo {author} {\bibfnamefont {M.}~\bibnamefont
  {Lewenstein}}, \bibinfo {author} {\bibfnamefont {A.}~\bibnamefont {Sanpera}},
  \bibinfo {author} {\bibfnamefont {V.}~\bibnamefont {Ahufinger}}, \bibinfo
  {author} {\bibfnamefont {B.}~\bibnamefont {Damski}}, \bibinfo {author}
  {\bibfnamefont {B.}~\bibnamefont {Sen(De)}},\ and\ \bibinfo {author}
  {\bibfnamefont {U.}~\bibnamefont {Sen}},\ }\bibfield  {title} {\bibinfo
  {title} {Ultracold atomic gases in optical lattices: mimicking condensed
  matter physics and beyond},\ }\href
  {https://doi.org/10.1080/00018730701223200} {\bibfield  {journal} {\bibinfo
  {journal} {Advances in Physics}\ }\textbf {\bibinfo {volume} {56}},\ \bibinfo
  {pages} {243} (\bibinfo {year} {2007})},\ \Eprint
  {https://arxiv.org/abs/https://doi.org/10.1080/00018730701223200}
  {https://doi.org/10.1080/00018730701223200} \BibitemShut {NoStop}%
\bibitem [{\citenamefont {Goldman}\ \emph {et~al.}(2014)\citenamefont
  {Goldman}, \citenamefont {Juzeliūnas}, \citenamefont {Öhberg},\ and\
  \citenamefont {Spielman}}]{Goldman_2014}%
  \BibitemOpen
  \bibfield  {author} {\bibinfo {author} {\bibfnamefont {N.}~\bibnamefont
  {Goldman}}, \bibinfo {author} {\bibfnamefont {G.}~\bibnamefont
  {Juzeliūnas}}, \bibinfo {author} {\bibfnamefont {P.}~\bibnamefont
  {Öhberg}},\ and\ \bibinfo {author} {\bibfnamefont {I.~B.}\ \bibnamefont
  {Spielman}},\ }\bibfield  {title} {\bibinfo {title} {Light-induced gauge
  fields for ultracold atoms},\ }\href
  {https://doi.org/10.1088/0034-4885/77/12/126401} {\bibfield  {journal}
  {\bibinfo  {journal} {Reports on Progress in Physics}\ }\textbf {\bibinfo
  {volume} {77}},\ \bibinfo {pages} {126401} (\bibinfo {year}
  {2014})}\BibitemShut {NoStop}%
\bibitem [{\citenamefont {Abanin}\ \emph {et~al.}(2019)\citenamefont {Abanin},
  \citenamefont {Altman}, \citenamefont {Bloch},\ and\ \citenamefont
  {Serbyn}}]{Thermal_2019}%
  \BibitemOpen
  \bibfield  {author} {\bibinfo {author} {\bibfnamefont {D.~A.}\ \bibnamefont
  {Abanin}}, \bibinfo {author} {\bibfnamefont {E.}~\bibnamefont {Altman}},
  \bibinfo {author} {\bibfnamefont {I.}~\bibnamefont {Bloch}},\ and\ \bibinfo
  {author} {\bibfnamefont {M.}~\bibnamefont {Serbyn}},\ }\bibfield  {title}
  {\bibinfo {title} {Colloquium: {M}any-body localization, thermalization, and
  entanglement},\ }\href {https://doi.org/10.1103/RevModPhys.91.021001}
  {\bibfield  {journal} {\bibinfo  {journal} {{R}eviews of {M}odern {P}hysics}\
  }\textbf {\bibinfo {volume} {91}},\ \bibinfo {pages} {021001} (\bibinfo
  {year} {2019})}\BibitemShut {NoStop}%
\bibitem [{\citenamefont {Souza}\ \emph {et~al.}(2023)\citenamefont {Souza},
  \citenamefont {Pelster},\ and\ \citenamefont {dos Santos}}]{Souza2023}%
  \BibitemOpen
  \bibfield  {author} {\bibinfo {author} {\bibfnamefont {R.~S.}\ \bibnamefont
  {Souza}}, \bibinfo {author} {\bibfnamefont {A.}~\bibnamefont {Pelster}},\
  and\ \bibinfo {author} {\bibfnamefont {F.~E.}\ \bibnamefont {dos Santos}},\
  }\bibfield  {title} {\bibinfo {title} {Emergence of {Damped-Localized}
  excitations of the {Mott} state due to disorder},\ }\href
  {https://doi.org/10.1088/1367-2630/acdb92} {\bibfield  {journal} {\bibinfo
  {journal} {{N}ew {J}ournal of {P}hysics}\ }\textbf {\bibinfo {volume} {25}},\
  \bibinfo {pages} {063015} (\bibinfo {year} {2023})}\BibitemShut {NoStop}%
\bibitem [{\citenamefont {Islam}\ \emph {et~al.}(2015)\citenamefont {Islam},
  \citenamefont {Ma}, \citenamefont {Preiss}, \citenamefont {Tai},
  \citenamefont {Lukin}, \citenamefont {Rispoli},\ and\ \citenamefont
  {Greiner}}]{Greiner_QI_2015}%
  \BibitemOpen
  \bibfield  {author} {\bibinfo {author} {\bibfnamefont {R.}~\bibnamefont
  {Islam}}, \bibinfo {author} {\bibfnamefont {R.}~\bibnamefont {Ma}}, \bibinfo
  {author} {\bibfnamefont {P.~M.}\ \bibnamefont {Preiss}}, \bibinfo {author}
  {\bibfnamefont {M.~E.}\ \bibnamefont {Tai}}, \bibinfo {author} {\bibfnamefont
  {A.}~\bibnamefont {Lukin}}, \bibinfo {author} {\bibfnamefont
  {M.}~\bibnamefont {Rispoli}},\ and\ \bibinfo {author} {\bibfnamefont
  {M.}~\bibnamefont {Greiner}},\ }\bibfield  {title} {\bibinfo {title}
  {Measuring entanglement entropy in a quantum many-body system},\ }\href
  {https://doi.org/10.1038/nature15750} {\bibfield  {journal} {\bibinfo
  {journal} {{N}ature}\ }\textbf {\bibinfo {volume} {528}},\ \bibinfo {pages}
  {77} (\bibinfo {year} {2015})}\BibitemShut {NoStop}%
\bibitem [{\citenamefont {Pezz\`e}\ \emph {et~al.}(2018)\citenamefont
  {Pezz\`e}, \citenamefont {Smerzi}, \citenamefont {Oberthaler}, \citenamefont
  {Schmied},\ and\ \citenamefont {Treutlein}}]{RevModPhys.90.035005}%
  \BibitemOpen
  \bibfield  {author} {\bibinfo {author} {\bibfnamefont {L.}~\bibnamefont
  {Pezz\`e}}, \bibinfo {author} {\bibfnamefont {A.}~\bibnamefont {Smerzi}},
  \bibinfo {author} {\bibfnamefont {M.~K.}\ \bibnamefont {Oberthaler}},
  \bibinfo {author} {\bibfnamefont {R.}~\bibnamefont {Schmied}},\ and\ \bibinfo
  {author} {\bibfnamefont {P.}~\bibnamefont {Treutlein}},\ }\bibfield  {title}
  {\bibinfo {title} {Quantum metrology with nonclassical states of atomic
  ensembles},\ }\href {https://doi.org/10.1103/RevModPhys.90.035005} {\bibfield
   {journal} {\bibinfo  {journal} {Rev. Mod. Phys.}\ }\textbf {\bibinfo
  {volume} {90}},\ \bibinfo {pages} {035005} (\bibinfo {year}
  {2018})}\BibitemShut {NoStop}%
\bibitem [{\citenamefont {Pethick}\ and\ \citenamefont
  {Smith}(2008)}]{pethick_smith_2008}%
  \BibitemOpen
  \bibfield  {author} {\bibinfo {author} {\bibfnamefont {C.~J.}\ \bibnamefont
  {Pethick}}\ and\ \bibinfo {author} {\bibfnamefont {H.}~\bibnamefont
  {Smith}},\ }\href {https://doi.org/10.1017/CBO9780511802850} {\emph {\bibinfo
  {title} {Bose–Einstein Condensation in Dilute Gases}}},\ \bibinfo {edition}
  {2nd}\ ed.\ (\bibinfo  {publisher} {Cambridge University Press},\ \bibinfo
  {year} {2008})\BibitemShut {NoStop}%
\bibitem [{\citenamefont {Lewenstein}\ \emph {et~al.}(2012)\citenamefont
  {Lewenstein}, \citenamefont {Sanpera},\ and\ \citenamefont
  {Ahufinger}}]{optical}%
  \BibitemOpen
  \bibfield  {author} {\bibinfo {author} {\bibfnamefont {M.}~\bibnamefont
  {Lewenstein}}, \bibinfo {author} {\bibfnamefont {A.}~\bibnamefont
  {Sanpera}},\ and\ \bibinfo {author} {\bibfnamefont {V.}~\bibnamefont
  {Ahufinger}},\ }\href
  {https://doi.org/10.1093/acprof:oso/9780199573127.001.0001} {\emph {\bibinfo
  {title} {Ultracold Atoms in Optical Lattices: Simulating quantum many-body
  systems}}}\ (\bibinfo  {publisher} {Oxford University Press},\ \bibinfo
  {year} {2012})\BibitemShut {NoStop}%
\bibitem [{\citenamefont {Anderson}\ \emph {et~al.}(1995)\citenamefont
  {Anderson}, \citenamefont {Ensher}, \citenamefont {Matthews}, \citenamefont
  {Wieman},\ and\ \citenamefont {Cornell}}]{Wieman1995}%
  \BibitemOpen
  \bibfield  {author} {\bibinfo {author} {\bibfnamefont {M.}~\bibnamefont
  {Anderson}}, \bibinfo {author} {\bibfnamefont {J.}~\bibnamefont {Ensher}},
  \bibinfo {author} {\bibfnamefont {M.}~\bibnamefont {Matthews}}, \bibinfo
  {author} {\bibfnamefont {C.}~\bibnamefont {Wieman}},\ and\ \bibinfo {author}
  {\bibfnamefont {E.}~\bibnamefont {Cornell}},\ }\bibfield  {title} {\bibinfo
  {title} {{O}bservation of {B}ose-{E}instein condensation in a dilute atomic
  vapor},\ }\href {https://doi.org/10.1126/science.269.5221.198} {\bibfield
  {journal} {\bibinfo  {journal} {Science}\ }\textbf {\bibinfo {volume}
  {269}},\ \bibinfo {pages} {198} (\bibinfo {year} {1995})}\BibitemShut
  {NoStop}%
\bibitem [{\citenamefont {Davis}\ \emph {et~al.}(1995)\citenamefont {Davis},
  \citenamefont {Mewes}, \citenamefont {Andrews}, \citenamefont {van Druten},
  \citenamefont {Durfee}, \citenamefont {Kurn},\ and\ \citenamefont
  {Ketterle}}]{Ketterle1995}%
  \BibitemOpen
  \bibfield  {author} {\bibinfo {author} {\bibfnamefont {K.~B.}\ \bibnamefont
  {Davis}}, \bibinfo {author} {\bibfnamefont {M.~O.}\ \bibnamefont {Mewes}},
  \bibinfo {author} {\bibfnamefont {M.~R.}\ \bibnamefont {Andrews}}, \bibinfo
  {author} {\bibfnamefont {N.~J.}\ \bibnamefont {van Druten}}, \bibinfo
  {author} {\bibfnamefont {D.~S.}\ \bibnamefont {Durfee}}, \bibinfo {author}
  {\bibfnamefont {D.~M.}\ \bibnamefont {Kurn}},\ and\ \bibinfo {author}
  {\bibfnamefont {W.}~\bibnamefont {Ketterle}},\ }\bibfield  {title} {\bibinfo
  {title} {{B}ose-{E}instein condensation in a gas of sodium atoms},\ }\href
  {https://doi.org/10.1103/PhysRevLett.75.3969} {\bibfield  {journal} {\bibinfo
   {journal} {{P}hysical {R}eview {L}etters}\ }\textbf {\bibinfo {volume}
  {75}},\ \bibinfo {pages} {3969} (\bibinfo {year} {1995})}\BibitemShut
  {NoStop}%
\bibitem [{\citenamefont {van Es}\ \emph {et~al.}(2010)\citenamefont {van Es},
  \citenamefont {Wicke}, \citenamefont {van Amerongen}, \citenamefont {Rétif},
  \citenamefont {Whitlock},\ and\ \citenamefont {van Druten}}]{atom_chips}%
  \BibitemOpen
  \bibfield  {author} {\bibinfo {author} {\bibfnamefont {J.~J.~P.}\
  \bibnamefont {van Es}}, \bibinfo {author} {\bibfnamefont {P.}~\bibnamefont
  {Wicke}}, \bibinfo {author} {\bibfnamefont {A.~H.}\ \bibnamefont {van
  Amerongen}}, \bibinfo {author} {\bibfnamefont {C.}~\bibnamefont {Rétif}},
  \bibinfo {author} {\bibfnamefont {S.}~\bibnamefont {Whitlock}},\ and\
  \bibinfo {author} {\bibfnamefont {N.~J.}\ \bibnamefont {van Druten}},\
  }\bibfield  {title} {\bibinfo {title} {Box traps on an atom chip for
  one-dimensional quantum gases},\ }\href
  {https://doi.org/10.1088/0953-4075/43/15/155002} {\bibfield  {journal}
  {\bibinfo  {journal} {Journal of Physics B: Atomic, Molecular and Optical
  Physics}\ }\textbf {\bibinfo {volume} {43}},\ \bibinfo {pages} {155002}
  (\bibinfo {year} {2010})}\BibitemShut {NoStop}%
\bibitem [{\citenamefont {Al~Khawaja}\ and\ \citenamefont
  {Stoof}(2001)}]{Topolog_magn2001}%
  \BibitemOpen
  \bibfield  {author} {\bibinfo {author} {\bibfnamefont {U.}~\bibnamefont
  {Al~Khawaja}}\ and\ \bibinfo {author} {\bibfnamefont {H.}~\bibnamefont
  {Stoof}},\ }\bibfield  {title} {\bibinfo {title} {Skyrmions in a
  ferromagnetic {B}ose-{E}instein condensate},\ }\href
  {https://doi.org/10.1038/35082010} {\bibfield  {journal} {\bibinfo  {journal}
  {{N}ature}\ }\textbf {\bibinfo {volume} {411}},\ \bibinfo {pages} {918}
  (\bibinfo {year} {2001})}\BibitemShut {NoStop}%
\bibitem [{\citenamefont {Norcia}\ \emph {et~al.}(2021)\citenamefont {Norcia},
  \citenamefont {Politi}, \citenamefont {Klaus}, \citenamefont {Poli},
  \citenamefont {Sohmen}, \citenamefont {Mark}, \citenamefont {Bisset},
  \citenamefont {Santos},\ and\ \citenamefont {Ferlaino}}]{Ferlaino2021}%
  \BibitemOpen
  \bibfield  {author} {\bibinfo {author} {\bibfnamefont {M.~A.}\ \bibnamefont
  {Norcia}}, \bibinfo {author} {\bibfnamefont {C.}~\bibnamefont {Politi}},
  \bibinfo {author} {\bibfnamefont {L.}~\bibnamefont {Klaus}}, \bibinfo
  {author} {\bibfnamefont {E.}~\bibnamefont {Poli}}, \bibinfo {author}
  {\bibfnamefont {M.}~\bibnamefont {Sohmen}}, \bibinfo {author} {\bibfnamefont
  {M.~J.}\ \bibnamefont {Mark}}, \bibinfo {author} {\bibfnamefont {R.~N.}\
  \bibnamefont {Bisset}}, \bibinfo {author} {\bibfnamefont {L.}~\bibnamefont
  {Santos}},\ and\ \bibinfo {author} {\bibfnamefont {F.}~\bibnamefont
  {Ferlaino}},\ }\bibfield  {title} {\bibinfo {title} {Two-dimensional
  supersolidity in a dipolar quantum gas},\ }\href
  {https://doi.org/10.1038/s41586-021-03725-7} {\bibfield  {journal} {\bibinfo
  {journal} {{N}ature}\ }\textbf {\bibinfo {volume} {596}},\ \bibinfo {pages}
  {357} (\bibinfo {year} {2021})}\BibitemShut {NoStop}%
\bibitem [{\citenamefont {Eckel}\ \emph {et~al.}(2018)\citenamefont {Eckel},
  \citenamefont {Kumar}, \citenamefont {Jacobson}, \citenamefont {Spielman},\
  and\ \citenamefont {Campbell}}]{cosmic_inflation}%
  \BibitemOpen
  \bibfield  {author} {\bibinfo {author} {\bibfnamefont {S.}~\bibnamefont
  {Eckel}}, \bibinfo {author} {\bibfnamefont {A.}~\bibnamefont {Kumar}},
  \bibinfo {author} {\bibfnamefont {T.}~\bibnamefont {Jacobson}}, \bibinfo
  {author} {\bibfnamefont {I.~B.}\ \bibnamefont {Spielman}},\ and\ \bibinfo
  {author} {\bibfnamefont {G.~K.}\ \bibnamefont {Campbell}},\ }\bibfield
  {title} {\bibinfo {title} {A rapidly expanding {B}ose-{E}instein condensate:
  An expanding universe in the lab},\ }\href
  {https://doi.org/10.1103/PhysRevX.8.021021} {\bibfield  {journal} {\bibinfo
  {journal} {Physical Review X}\ }\textbf {\bibinfo {volume} {8}},\ \bibinfo
  {pages} {021021} (\bibinfo {year} {2018})}\BibitemShut {NoStop}%
\bibitem [{\citenamefont {Hall}\ \emph {et~al.}(1998)\citenamefont {Hall},
  \citenamefont {Matthews}, \citenamefont {Ensher}, \citenamefont {Wieman},\
  and\ \citenamefont {Cornell}}]{binary_mixture}%
  \BibitemOpen
  \bibfield  {author} {\bibinfo {author} {\bibfnamefont {D.~S.}\ \bibnamefont
  {Hall}}, \bibinfo {author} {\bibfnamefont {M.~R.}\ \bibnamefont {Matthews}},
  \bibinfo {author} {\bibfnamefont {J.~R.}\ \bibnamefont {Ensher}}, \bibinfo
  {author} {\bibfnamefont {C.~E.}\ \bibnamefont {Wieman}},\ and\ \bibinfo
  {author} {\bibfnamefont {E.~A.}\ \bibnamefont {Cornell}},\ }\bibfield
  {title} {\bibinfo {title} {Dynamics of component separation in a binary
  mixture of {B}ose-{E}instein condensates},\ }\href
  {https://doi.org/10.1103/PhysRevLett.81.1539} {\bibfield  {journal} {\bibinfo
   {journal} {{P}hysical {R}eview {L}etters}\ }\textbf {\bibinfo {volume}
  {81}},\ \bibinfo {pages} {1539} (\bibinfo {year} {1998})}\BibitemShut
  {NoStop}%
\bibitem [{\citenamefont {Javanainen}\ and\ \citenamefont
  {Yoo}(1996)}]{Javanainen1996}%
  \BibitemOpen
  \bibfield  {author} {\bibinfo {author} {\bibfnamefont {J.}~\bibnamefont
  {Javanainen}}\ and\ \bibinfo {author} {\bibfnamefont {S.~M.}\ \bibnamefont
  {Yoo}},\ }\bibfield  {title} {\bibinfo {title} {Quantum phase of a
  {B}ose-{E}instein condensate with an arbitrary number of atoms},\ }\href
  {https://doi.org/10.1103/PhysRevLett.76.161} {\bibfield  {journal} {\bibinfo
  {journal} {Physical Review Letters}\ }\textbf {\bibinfo {volume} {76}},\
  \bibinfo {pages} {161} (\bibinfo {year} {1996})}\BibitemShut {NoStop}%
\bibitem [{\citenamefont {Milburn}\ \emph {et~al.}(1997)\citenamefont
  {Milburn}, \citenamefont {Corney}, \citenamefont {Wright},\ and\
  \citenamefont {Walls}}]{Double_well1997}%
  \BibitemOpen
  \bibfield  {author} {\bibinfo {author} {\bibfnamefont {G.~J.}\ \bibnamefont
  {Milburn}}, \bibinfo {author} {\bibfnamefont {J.}~\bibnamefont {Corney}},
  \bibinfo {author} {\bibfnamefont {E.~M.}\ \bibnamefont {Wright}},\ and\
  \bibinfo {author} {\bibfnamefont {D.~F.}\ \bibnamefont {Walls}},\ }\bibfield
  {title} {\bibinfo {title} {Quantum dynamics of an atomic {B}ose-{E}instein
  condensate in a double-well potential},\ }\href
  {https://doi.org/10.1103/PhysRevA.55.4318} {\bibfield  {journal} {\bibinfo
  {journal} {Physical Review A}\ }\textbf {\bibinfo {volume} {55}},\ \bibinfo
  {pages} {4318} (\bibinfo {year} {1997})}\BibitemShut {NoStop}%
\bibitem [{\citenamefont {Dalton}\ and\ \citenamefont
  {Ghanbari}(2012)}]{Dalton}%
  \BibitemOpen
  \bibfield  {author} {\bibinfo {author} {\bibfnamefont {B.~J.}\ \bibnamefont
  {Dalton}}\ and\ \bibinfo {author} {\bibfnamefont {S.}~\bibnamefont
  {Ghanbari}},\ }\bibfield  {title} {\bibinfo {title} {Two mode theory of
  bose--einstein condensates: interferometry and the josephson model},\ }\href
  {https://doi.org/10.1080/09500340.2011.632100} {\bibfield  {journal}
  {\bibinfo  {journal} {Journal of Modern Optics}\ }\textbf {\bibinfo {volume}
  {59}},\ \bibinfo {pages} {287} (\bibinfo {year} {2012})}\BibitemShut
  {NoStop}%
\bibitem [{\citenamefont {Jaksch}\ \emph {et~al.}(1998)\citenamefont {Jaksch},
  \citenamefont {Bruder}, \citenamefont {Cirac}, \citenamefont {Gardiner},\
  and\ \citenamefont {Zoller}}]{Zoller1998}%
  \BibitemOpen
  \bibfield  {author} {\bibinfo {author} {\bibfnamefont {D.}~\bibnamefont
  {Jaksch}}, \bibinfo {author} {\bibfnamefont {C.}~\bibnamefont {Bruder}},
  \bibinfo {author} {\bibfnamefont {J.~I.}\ \bibnamefont {Cirac}}, \bibinfo
  {author} {\bibfnamefont {C.~W.}\ \bibnamefont {Gardiner}},\ and\ \bibinfo
  {author} {\bibfnamefont {P.}~\bibnamefont {Zoller}},\ }\bibfield  {title}
  {\bibinfo {title} {Cold bosonic atoms in optical lattices},\ }\href
  {https://doi.org/10.1103/PhysRevLett.81.3108} {\bibfield  {journal} {\bibinfo
   {journal} {{P}hysical {R}eview {L}etters}\ }\textbf {\bibinfo {volume}
  {81}},\ \bibinfo {pages} {3108} (\bibinfo {year} {1998})}\BibitemShut
  {NoStop}%
\bibitem [{\citenamefont {Bloch}(2005)}]{Bloch2005}%
  \BibitemOpen
  \bibfield  {author} {\bibinfo {author} {\bibfnamefont {I.}~\bibnamefont
  {Bloch}},\ }\bibfield  {title} {\bibinfo {title} {Ultracold quantum gases in
  optical lattices},\ }\href {https://doi.org/10.1038/nphys138} {\bibfield
  {journal} {\bibinfo  {journal} {Nature Physics}\ }\textbf {\bibinfo {volume}
  {1}},\ \bibinfo {pages} {23} (\bibinfo {year} {2005})}\BibitemShut {NoStop}%
\bibitem [{\citenamefont {Gross}\ and\ \citenamefont
  {Bloch}(2017)}]{Gross2017}%
  \BibitemOpen
  \bibfield  {author} {\bibinfo {author} {\bibfnamefont {C.}~\bibnamefont
  {Gross}}\ and\ \bibinfo {author} {\bibfnamefont {I.}~\bibnamefont {Bloch}},\
  }\bibfield  {title} {\bibinfo {title} {Quantum simulations with ultracold
  atoms in optical lattices},\ }\href {https://doi.org/10.1126/science.aal3837}
  {\bibfield  {journal} {\bibinfo  {journal} {Science}\ }\textbf {\bibinfo
  {volume} {357}},\ \bibinfo {pages} {995} (\bibinfo {year} {2017})},\ \Eprint
  {https://arxiv.org/abs/https://www.science.org/doi/pdf/10.1126/science.aal3837}
  {https://www.science.org/doi/pdf/10.1126/science.aal3837} \BibitemShut
  {NoStop}%
\bibitem [{\citenamefont {Sch{\"a}fer}\ \emph {et~al.}(2020)\citenamefont
  {Sch{\"a}fer}, \citenamefont {Fukuhara}, \citenamefont {Sugawa},
  \citenamefont {Takasu},\ and\ \citenamefont {Takahashi}}]{Schafer2020}%
  \BibitemOpen
  \bibfield  {author} {\bibinfo {author} {\bibfnamefont {F.}~\bibnamefont
  {Sch{\"a}fer}}, \bibinfo {author} {\bibfnamefont {T.}~\bibnamefont
  {Fukuhara}}, \bibinfo {author} {\bibfnamefont {S.}~\bibnamefont {Sugawa}},
  \bibinfo {author} {\bibfnamefont {Y.}~\bibnamefont {Takasu}},\ and\ \bibinfo
  {author} {\bibfnamefont {Y.}~\bibnamefont {Takahashi}},\ }\bibfield  {title}
  {\bibinfo {title} {Tools for quantum simulation with ultracold atoms in
  optical lattices},\ }\href {https://doi.org/10.1038/s42254-020-0195-3}
  {\bibfield  {journal} {\bibinfo  {journal} {Nature Reviews Physics}\ }\textbf
  {\bibinfo {volume} {2}},\ \bibinfo {pages} {411} (\bibinfo {year}
  {2020})}\BibitemShut {NoStop}%
\bibitem [{\citenamefont {Batrouni}\ \emph {et~al.}(2002)\citenamefont
  {Batrouni}, \citenamefont {Rousseau}, \citenamefont {Scalettar},
  \citenamefont {Rigol}, \citenamefont {Muramatsu}, \citenamefont {Denteneer},\
  and\ \citenamefont {Troyer}}]{Troyer2002}%
  \BibitemOpen
  \bibfield  {author} {\bibinfo {author} {\bibfnamefont {G.~G.}\ \bibnamefont
  {Batrouni}}, \bibinfo {author} {\bibfnamefont {V.}~\bibnamefont {Rousseau}},
  \bibinfo {author} {\bibfnamefont {R.~T.}\ \bibnamefont {Scalettar}}, \bibinfo
  {author} {\bibfnamefont {M.}~\bibnamefont {Rigol}}, \bibinfo {author}
  {\bibfnamefont {A.}~\bibnamefont {Muramatsu}}, \bibinfo {author}
  {\bibfnamefont {P.~J.~H.}\ \bibnamefont {Denteneer}},\ and\ \bibinfo {author}
  {\bibfnamefont {M.}~\bibnamefont {Troyer}},\ }\bibfield  {title} {\bibinfo
  {title} {Mott domains of bosons confined on optical lattices},\ }\href
  {https://doi.org/10.1103/PhysRevLett.89.117203} {\bibfield  {journal}
  {\bibinfo  {journal} {{P}hysical {R}eview {L}etters}\ }\textbf {\bibinfo
  {volume} {89}},\ \bibinfo {pages} {117203} (\bibinfo {year}
  {2002})}\BibitemShut {NoStop}%
\bibitem [{\citenamefont {Fromhold}\ \emph {et~al.}(2000)\citenamefont
  {Fromhold}, \citenamefont {Tench}, \citenamefont {Bujkiewicz}, \citenamefont
  {Wilkinson},\ and\ \citenamefont {Sheard}}]{Fromhold2000}%
  \BibitemOpen
  \bibfield  {author} {\bibinfo {author} {\bibfnamefont {T.~M.}\ \bibnamefont
  {Fromhold}}, \bibinfo {author} {\bibfnamefont {C.~R.}\ \bibnamefont {Tench}},
  \bibinfo {author} {\bibfnamefont {S.}~\bibnamefont {Bujkiewicz}}, \bibinfo
  {author} {\bibfnamefont {P.~B.}\ \bibnamefont {Wilkinson}},\ and\ \bibinfo
  {author} {\bibfnamefont {F.~W.}\ \bibnamefont {Sheard}},\ }\bibfield  {title}
  {\bibinfo {title} {Quantum chaos for cold atoms in an optical lattice with a
  tilted harmonic trap},\ }\href {https://doi.org/10.1088/1464-4266/2/5/310}
  {\bibfield  {journal} {\bibinfo  {journal} {Journal of Optics B: Quantum and
  Semiclassical Optics}\ }\textbf {\bibinfo {volume} {2}},\ \bibinfo {pages}
  {628} (\bibinfo {year} {2000})}\BibitemShut {NoStop}%
\bibitem [{\citenamefont {Pichler}\ \emph {et~al.}(2016)\citenamefont
  {Pichler}, \citenamefont {Zhu}, \citenamefont {Seif}, \citenamefont
  {Zoller},\ and\ \citenamefont {Hafezi}}]{Pichler2016}%
  \BibitemOpen
  \bibfield  {author} {\bibinfo {author} {\bibfnamefont {H.}~\bibnamefont
  {Pichler}}, \bibinfo {author} {\bibfnamefont {G.}~\bibnamefont {Zhu}},
  \bibinfo {author} {\bibfnamefont {A.}~\bibnamefont {Seif}}, \bibinfo {author}
  {\bibfnamefont {P.}~\bibnamefont {Zoller}},\ and\ \bibinfo {author}
  {\bibfnamefont {M.}~\bibnamefont {Hafezi}},\ }\bibfield  {title} {\bibinfo
  {title} {Measurement protocol for the entanglement spectrum of cold atoms},\
  }\href {https://doi.org/10.1103/PhysRevX.6.041033} {\bibfield  {journal}
  {\bibinfo  {journal} {Physical Review X}\ }\textbf {\bibinfo {volume} {6}},\
  \bibinfo {pages} {041033} (\bibinfo {year} {2016})}\BibitemShut {NoStop}%
\bibitem [{\citenamefont {Galitski}\ \emph {et~al.}(2019)\citenamefont
  {Galitski}, \citenamefont {Juzeliūnas},\ and\ \citenamefont
  {Spielman}}]{Galitski2019}%
  \BibitemOpen
  \bibfield  {author} {\bibinfo {author} {\bibfnamefont {V.}~\bibnamefont
  {Galitski}}, \bibinfo {author} {\bibfnamefont {G.}~\bibnamefont
  {Juzeliūnas}},\ and\ \bibinfo {author} {\bibfnamefont {I.~B.}\ \bibnamefont
  {Spielman}},\ }\bibfield  {title} {\bibinfo {title} {Artificial gauge fields
  with ultracold atoms},\ }\href {https://doi.org/10.1063/PT.3.4111} {\bibfield
   {journal} {\bibinfo  {journal} {Physics Today}\ }\textbf {\bibinfo {volume}
  {72}},\ \bibinfo {pages} {38} (\bibinfo {year} {2019})},\ \Eprint
  {https://arxiv.org/abs/https://pubs.aip.org/physicstoday/article-pdf/72/1/38/10121390/38\_1\_online.pdf}
  {https://pubs.aip.org/physicstoday/article-pdf/72/1/38/10121390/38\_1\_online.pdf}
  \BibitemShut {NoStop}%
\bibitem [{\citenamefont {Takamoto}\ \emph {et~al.}(2005)\citenamefont
  {Takamoto}, \citenamefont {Hong}, \citenamefont {Higashi},\ and\
  \citenamefont {Katori}}]{Takamoto2005}%
  \BibitemOpen
  \bibfield  {author} {\bibinfo {author} {\bibfnamefont {M.}~\bibnamefont
  {Takamoto}}, \bibinfo {author} {\bibfnamefont {F.-L.}\ \bibnamefont {Hong}},
  \bibinfo {author} {\bibfnamefont {R.}~\bibnamefont {Higashi}},\ and\ \bibinfo
  {author} {\bibfnamefont {H.}~\bibnamefont {Katori}},\ }\bibfield  {title}
  {\bibinfo {title} {An optical lattice clock},\ }\href
  {https://doi.org/10.1038/nature03541} {\bibfield  {journal} {\bibinfo
  {journal} {Nature}\ }\textbf {\bibinfo {volume} {435}},\ \bibinfo {pages}
  {321} (\bibinfo {year} {2005})}\BibitemShut {NoStop}%
\bibitem [{\citenamefont {Borkowski}(2018)}]{Borkowski2018}%
  \BibitemOpen
  \bibfield  {author} {\bibinfo {author} {\bibfnamefont {M.}~\bibnamefont
  {Borkowski}},\ }\bibfield  {title} {\bibinfo {title} {Optical lattice clocks
  with weakly bound molecules},\ }\href
  {https://doi.org/10.1103/PhysRevLett.120.083202} {\bibfield  {journal}
  {\bibinfo  {journal} {Physical Review Letters}\ }\textbf {\bibinfo {volume}
  {120}},\ \bibinfo {pages} {083202} (\bibinfo {year} {2018})}\BibitemShut
  {NoStop}%
\bibitem [{\citenamefont {Zoubi}\ and\ \citenamefont
  {Ritsch}(2009)}]{Zoubi2009}%
  \BibitemOpen
  \bibfield  {author} {\bibinfo {author} {\bibfnamefont {H.}~\bibnamefont
  {Zoubi}}\ and\ \bibinfo {author} {\bibfnamefont {H.}~\bibnamefont {Ritsch}},\
  }\bibfield  {title} {\bibinfo {title} {Quantum phases of bosonic atoms with
  two levels coupled by a cavity field in an optical lattice},\ }\href
  {https://doi.org/10.1103/PhysRevA.80.053608} {\bibfield  {journal} {\bibinfo
  {journal} {Physical Review A}\ }\textbf {\bibinfo {volume} {80}},\ \bibinfo
  {pages} {053608} (\bibinfo {year} {2009})}\BibitemShut {NoStop}%
\bibitem [{\citenamefont {Peyronel}\ \emph {et~al.}(2012)\citenamefont
  {Peyronel}, \citenamefont {Firstenberg}, \citenamefont {Liang}, \citenamefont
  {Hofferberth}, \citenamefont {Gorshkov}, \citenamefont {Pohl}, \citenamefont
  {Lukin},\ and\ \citenamefont {Vuleti{\'{c}}}}]{Pohl2012}%
  \BibitemOpen
  \bibfield  {author} {\bibinfo {author} {\bibfnamefont {T.}~\bibnamefont
  {Peyronel}}, \bibinfo {author} {\bibfnamefont {O.}~\bibnamefont
  {Firstenberg}}, \bibinfo {author} {\bibfnamefont {Q.-Y.}\ \bibnamefont
  {Liang}}, \bibinfo {author} {\bibfnamefont {S.}~\bibnamefont {Hofferberth}},
  \bibinfo {author} {\bibfnamefont {A.~V.}\ \bibnamefont {Gorshkov}}, \bibinfo
  {author} {\bibfnamefont {T.}~\bibnamefont {Pohl}}, \bibinfo {author}
  {\bibfnamefont {M.~D.}\ \bibnamefont {Lukin}},\ and\ \bibinfo {author}
  {\bibfnamefont {V.}~\bibnamefont {Vuleti{\'{c}}}},\ }\bibfield  {title}
  {\bibinfo {title} {Quantum nonlinear optics with single photons enabled by
  strongly interacting atoms},\ }\href {https://doi.org/10.1038/nature11361}
  {\bibfield  {journal} {\bibinfo  {journal} {Nature}\ }\textbf {\bibinfo
  {volume} {488}},\ \bibinfo {pages} {57} (\bibinfo {year} {2012})}\BibitemShut
  {NoStop}%
\bibitem [{\citenamefont {Ritsch}\ \emph {et~al.}(2013)\citenamefont {Ritsch},
  \citenamefont {Domokos}, \citenamefont {Brennecke},\ and\ \citenamefont
  {Esslinger}}]{Essingler2013}%
  \BibitemOpen
  \bibfield  {author} {\bibinfo {author} {\bibfnamefont {H.}~\bibnamefont
  {Ritsch}}, \bibinfo {author} {\bibfnamefont {P.}~\bibnamefont {Domokos}},
  \bibinfo {author} {\bibfnamefont {F.}~\bibnamefont {Brennecke}},\ and\
  \bibinfo {author} {\bibfnamefont {T.}~\bibnamefont {Esslinger}},\ }\bibfield
  {title} {\bibinfo {title} {Cold atoms in cavity-generated dynamical optical
  potentials},\ }\href {https://doi.org/10.1103/RevModPhys.85.553} {\bibfield
  {journal} {\bibinfo  {journal} {Rev. Mod. Phys.}\ }\textbf {\bibinfo {volume}
  {85}},\ \bibinfo {pages} {553} (\bibinfo {year} {2013})}\BibitemShut
  {NoStop}%
\bibitem [{\citenamefont {Schleich}(2001)}]{schleich01}%
  \BibitemOpen
  \bibfield  {author} {\bibinfo {author} {\bibfnamefont {W.~P.}\ \bibnamefont
  {Schleich}},\ }\href@noop {} {\emph {\bibinfo {title} {Quantum Optics in
  Phase Space}}}\ (\bibinfo  {publisher} {Wiley-VCH},\ \bibinfo {address}
  {Berlin},\ \bibinfo {year} {2001})\BibitemShut {NoStop}%
\bibitem [{\citenamefont {Li}\ \emph {et~al.}(2024)\citenamefont {Li},
  \citenamefont {Lu},\ and\ \citenamefont {Jiao}}]{photonics11030205}%
  \BibitemOpen
  \bibfield  {author} {\bibinfo {author} {\bibfnamefont {Z.}~\bibnamefont
  {Li}}, \bibinfo {author} {\bibfnamefont {W.-J.}\ \bibnamefont {Lu}},\ and\
  \bibinfo {author} {\bibfnamefont {Y.-F.}\ \bibnamefont {Jiao}},\ }\bibfield
  {title} {\bibinfo {title} {Quantum dynamics of cavity–bose–einstein
  condensates in a gravitational field},\ }\href
  {https://doi.org/10.3390/photonics11030205} {\bibfield  {journal} {\bibinfo
  {journal} {Photonics}\ }\textbf {\bibinfo {volume} {11}},\ \bibinfo {pages}
  {205} (\bibinfo {year} {2024})}\BibitemShut {NoStop}%
\bibitem [{\citenamefont {Farokh~Mivehvar}\ and\ \citenamefont
  {Ritsch}(2021)}]{Ritsch2021}%
  \BibitemOpen
  \bibfield  {author} {\bibinfo {author} {\bibfnamefont {T.~D.}\ \bibnamefont
  {Farokh~Mivehvar}, \bibfnamefont {Francesco~Piazza}}\ and\ \bibinfo {author}
  {\bibfnamefont {H.}~\bibnamefont {Ritsch}},\ }\bibfield  {title} {\bibinfo
  {title} {Cavity qed with quantum gases: new paradigms in many-body physics},\
  }\href {https://doi.org/10.1080/00018732.2021.1969727} {\bibfield  {journal}
  {\bibinfo  {journal} {Advances in Physics}\ }\textbf {\bibinfo {volume}
  {70}},\ \bibinfo {pages} {1} (\bibinfo {year} {2021})},\ \Eprint
  {https://arxiv.org/abs/https://doi.org/10.1080/00018732.2021.1969727}
  {https://doi.org/10.1080/00018732.2021.1969727} \BibitemShut {NoStop}%
\bibitem [{\citenamefont {Ghasemian}\ and\ \citenamefont
  {Tavassoly}(2017)}]{Ghasemian2017}%
  \BibitemOpen
  \bibfield  {author} {\bibinfo {author} {\bibfnamefont {E.}~\bibnamefont
  {Ghasemian}}\ and\ \bibinfo {author} {\bibfnamefont {M.~K.}\ \bibnamefont
  {Tavassoly}},\ }\bibfield  {title} {\bibinfo {title} {Quantum dynamics of a
  {B}{E}{C} interacting with a single-mode quantized field under the influence
  of a dissipation process: thermal and squeezed vacuum reservoirs},\ }\href
  {https://doi.org/10.1088/1555-6611/aa7dcf} {\bibfield  {journal} {\bibinfo
  {journal} {Laser Physics}\ }\textbf {\bibinfo {volume} {27}},\ \bibinfo
  {pages} {095202} (\bibinfo {year} {2017})}\BibitemShut {NoStop}%
\bibitem [{\citenamefont {Halati}\ \emph {et~al.}(2025)\citenamefont {Halati},
  \citenamefont {Sheikhan}, \citenamefont {Morigi},\ and\ \citenamefont
  {Kollath}}]{PhysRevLett.134.073604}%
  \BibitemOpen
  \bibfield  {author} {\bibinfo {author} {\bibfnamefont {C.-M.}\ \bibnamefont
  {Halati}}, \bibinfo {author} {\bibfnamefont {A.}~\bibnamefont {Sheikhan}},
  \bibinfo {author} {\bibfnamefont {G.}~\bibnamefont {Morigi}},\ and\ \bibinfo
  {author} {\bibfnamefont {C.}~\bibnamefont {Kollath}},\ }\bibfield  {title}
  {\bibinfo {title} {Controlling the dynamics of atomic correlations via the
  coupling to a dissipative cavity},\ }\href
  {https://doi.org/10.1103/PhysRevLett.134.073604} {\bibfield  {journal}
  {\bibinfo  {journal} {Phys. Rev. Lett.}\ }\textbf {\bibinfo {volume} {134}},\
  \bibinfo {pages} {073604} (\bibinfo {year} {2025})}\BibitemShut {NoStop}%
\bibitem [{\citenamefont {Ghasemian}\ and\ \citenamefont
  {Tavassoly}(2018)}]{Ghasemian18}%
  \BibitemOpen
  \bibfield  {author} {\bibinfo {author} {\bibfnamefont {E.}~\bibnamefont
  {Ghasemian}}\ and\ \bibinfo {author} {\bibfnamefont {M.~K.}\ \bibnamefont
  {Tavassoly}},\ }\bibfield  {title} {\bibinfo {title} {Spontaneous emission
  originating from atomic {B}{E}{C} interacting with a single-mode quantized
  field},\ }\href {https://doi.org/10.1088/0253-6102/69/6/711} {\bibfield
  {journal} {\bibinfo  {journal} {Communications in Theoretical Physics}\
  }\textbf {\bibinfo {volume} {69}},\ \bibinfo {pages} {711} (\bibinfo {year}
  {2018})}\BibitemShut {NoStop}%
\bibitem [{\citenamefont {Ghasemian}\ and\ \citenamefont
  {Tavassoly}(2021)}]{Ghasemian2021}%
  \BibitemOpen
  \bibfield  {author} {\bibinfo {author} {\bibfnamefont {E.}~\bibnamefont
  {Ghasemian}}\ and\ \bibinfo {author} {\bibfnamefont {M.}~\bibnamefont
  {Tavassoly}},\ }\bibfield  {title} {\bibinfo {title} {Dynamics of an atomic
  {B}ose–{E}instein condensate interacting with nonlinear quantized field
  under the influence of {S}tark effect},\ }\href
  {https://doi.org/https://doi.org/10.1016/j.physa.2020.125323} {\bibfield
  {journal} {\bibinfo  {journal} {Physica A: Statistical Mechanics and its
  Applications}\ }\textbf {\bibinfo {volume} {562}},\ \bibinfo {pages} {125323}
  (\bibinfo {year} {2021})}\BibitemShut {NoStop}%
\bibitem [{\citenamefont {Chanda}\ \emph {et~al.}(2025)\citenamefont {Chanda},
  \citenamefont {Barbiero}, \citenamefont {Lewenstein}, \citenamefont {Mark},\
  and\ \citenamefont {Zakrzewski}}]{Chanda_2025}%
  \BibitemOpen
  \bibfield  {author} {\bibinfo {author} {\bibfnamefont {T.}~\bibnamefont
  {Chanda}}, \bibinfo {author} {\bibfnamefont {L.}~\bibnamefont {Barbiero}},
  \bibinfo {author} {\bibfnamefont {M.}~\bibnamefont {Lewenstein}}, \bibinfo
  {author} {\bibfnamefont {M.~J.}\ \bibnamefont {Mark}},\ and\ \bibinfo
  {author} {\bibfnamefont {J.}~\bibnamefont {Zakrzewski}},\ }\bibfield  {title}
  {\bibinfo {title} {Recent progress on quantum simulations of non-standard
  bose–hubbard models},\ }\href {https://doi.org/10.1088/1361-6633/adc3a7}
  {\bibfield  {journal} {\bibinfo  {journal} {Reports on Progress in Physics}\
  }\textbf {\bibinfo {volume} {88}},\ \bibinfo {pages} {044501} (\bibinfo
  {year} {2025})}\BibitemShut {NoStop}%
\bibitem [{\citenamefont {Kumar}\ \emph {et~al.}(2011)\citenamefont {Kumar},
  \citenamefont {Bhattacherjee},\ and\ \citenamefont
  {{Man{M}ohan}}}]{TwophotonSpec2011}%
  \BibitemOpen
  \bibfield  {author} {\bibinfo {author} {\bibfnamefont {T.}~\bibnamefont
  {Kumar}}, \bibinfo {author} {\bibfnamefont {A.~B.}\ \bibnamefont
  {Bhattacherjee}},\ and\ \bibinfo {author} {\bibnamefont {{Man{M}ohan}}},\
  }\bibfield  {title} {\bibinfo {title} {{T}wo-photon nonlinear spectroscopy of
  periodically trapped ultracold atoms in a cavity},\ }\href
  {https://doi.org/10.1142/S0217979211100631} {\bibfield  {journal} {\bibinfo
  {journal} {International Journal of Modern Physics B}\ }\textbf {\bibinfo
  {volume} {25}},\ \bibinfo {pages} {1737} (\bibinfo {year}
  {2011})}\BibitemShut {NoStop}%
\bibitem [{\citenamefont {Maschler}\ and\ \citenamefont
  {Ritsch}(2005)}]{Maschler2005}%
  \BibitemOpen
  \bibfield  {author} {\bibinfo {author} {\bibfnamefont {C.}~\bibnamefont
  {Maschler}}\ and\ \bibinfo {author} {\bibfnamefont {H.}~\bibnamefont
  {Ritsch}},\ }\bibfield  {title} {\bibinfo {title} {Cold atom dynamics in a
  quantum optical lattice potential},\ }\href
  {https://doi.org/10.1103/PhysRevLett.95.260401} {\bibfield  {journal}
  {\bibinfo  {journal} {Physical Review Letters}\ }\textbf {\bibinfo {volume}
  {95}},\ \bibinfo {pages} {260401} (\bibinfo {year} {2005})}\BibitemShut
  {NoStop}%
\bibitem [{\citenamefont {Wu}\ \emph {et~al.}()\citenamefont {Wu},
  \citenamefont {Ray},\ and\ \citenamefont {Kroha}}]{DickBoseHubbard_bistable}%
  \BibitemOpen
  \bibfield  {author} {\bibinfo {author} {\bibfnamefont {T.}~\bibnamefont
  {Wu}}, \bibinfo {author} {\bibfnamefont {S.}~\bibnamefont {Ray}},\ and\
  \bibinfo {author} {\bibfnamefont {J.}~\bibnamefont {Kroha}},\ }\bibfield
  {title} {\bibinfo {title} {Temporal bistability in the dissipative
  {Dicke}-{Bose}-{Hubbard} system},\ }\href
  {https://doi.org/10.1002/andp.202300505} {\bibfield  {journal} {\bibinfo
  {journal} {Annalen der Physik}\ }\textbf {\bibinfo {volume} {536}},\ \bibinfo
  {pages} {2300505}}\BibitemShut {NoStop}%
\bibitem [{\citenamefont {Bloembergen}(1982)}]{BloeRMP}%
  \BibitemOpen
  \bibfield  {author} {\bibinfo {author} {\bibfnamefont {N.}~\bibnamefont
  {Bloembergen}},\ }\bibfield  {title} {\bibinfo {title} {Nonlinear optics and
  spectroscopy},\ }\href {https://doi.org/10.1103/RevModPhys.54.685} {\bibfield
   {journal} {\bibinfo  {journal} {Rev. Mod. Phys.}\ }\textbf {\bibinfo
  {volume} {54}},\ \bibinfo {pages} {685} (\bibinfo {year} {1982})}\BibitemShut
  {NoStop}%
\bibitem [{\citenamefont {Ciappina}\ \emph {et~al.}(2017)\citenamefont
  {Ciappina}, \citenamefont {Pérez-Hernández}, \citenamefont {Landsman},
  \citenamefont {Okell}, \citenamefont {Zherebtsov}, \citenamefont {Förg},
  \citenamefont {Schötz}, \citenamefont {Seiffert}, \citenamefont {Fennel},
  \citenamefont {Shaaran}, \citenamefont {Zimmermann}, \citenamefont {Chacón},
  \citenamefont {Guichard}, \citenamefont {Zaïr}, \citenamefont {Tisch},
  \citenamefont {Marangos}, \citenamefont {Witting}, \citenamefont {Braun},
  \citenamefont {Maier}, \citenamefont {Roso}, \citenamefont {Krüger},
  \citenamefont {Hommelhoff}, \citenamefont {Kling}, \citenamefont {Krausz},\
  and\ \citenamefont {Lewenstein}}]{physi}%
  \BibitemOpen
  \bibfield  {author} {\bibinfo {author} {\bibfnamefont {M.~F.}\ \bibnamefont
  {Ciappina}}, \bibinfo {author} {\bibfnamefont {J.~A.}\ \bibnamefont
  {Pérez-Hernández}}, \bibinfo {author} {\bibfnamefont {A.~S.}\ \bibnamefont
  {Landsman}}, \bibinfo {author} {\bibfnamefont {W.~A.}\ \bibnamefont {Okell}},
  \bibinfo {author} {\bibfnamefont {S.}~\bibnamefont {Zherebtsov}}, \bibinfo
  {author} {\bibfnamefont {B.}~\bibnamefont {Förg}}, \bibinfo {author}
  {\bibfnamefont {J.}~\bibnamefont {Schötz}}, \bibinfo {author} {\bibfnamefont
  {L.}~\bibnamefont {Seiffert}}, \bibinfo {author} {\bibfnamefont
  {T.}~\bibnamefont {Fennel}}, \bibinfo {author} {\bibfnamefont
  {T.}~\bibnamefont {Shaaran}}, \bibinfo {author} {\bibfnamefont
  {T.}~\bibnamefont {Zimmermann}}, \bibinfo {author} {\bibfnamefont
  {A.}~\bibnamefont {Chacón}}, \bibinfo {author} {\bibfnamefont
  {R.}~\bibnamefont {Guichard}}, \bibinfo {author} {\bibfnamefont
  {A.}~\bibnamefont {Zaïr}}, \bibinfo {author} {\bibfnamefont {J.~W.~G.}\
  \bibnamefont {Tisch}}, \bibinfo {author} {\bibfnamefont {J.~P.}\ \bibnamefont
  {Marangos}}, \bibinfo {author} {\bibfnamefont {T.}~\bibnamefont {Witting}},
  \bibinfo {author} {\bibfnamefont {A.}~\bibnamefont {Braun}}, \bibinfo
  {author} {\bibfnamefont {S.~A.}\ \bibnamefont {Maier}}, \bibinfo {author}
  {\bibfnamefont {L.}~\bibnamefont {Roso}}, \bibinfo {author} {\bibfnamefont
  {M.}~\bibnamefont {Krüger}}, \bibinfo {author} {\bibfnamefont
  {P.}~\bibnamefont {Hommelhoff}}, \bibinfo {author} {\bibfnamefont {M.~F.}\
  \bibnamefont {Kling}}, \bibinfo {author} {\bibfnamefont {F.}~\bibnamefont
  {Krausz}},\ and\ \bibinfo {author} {\bibfnamefont {M.}~\bibnamefont
  {Lewenstein}},\ }\bibfield  {title} {\bibinfo {title} {Attosecond physics at
  the nanoscale},\ }\href {https://doi.org/10.1088/1361-6633/aa574e} {\bibfield
   {journal} {\bibinfo  {journal} {Reports on Progress in Physics}\ }\textbf
  {\bibinfo {volume} {80}},\ \bibinfo {pages} {054401} (\bibinfo {year}
  {2017})}\BibitemShut {NoStop}%
\bibitem [{\citenamefont {Fu}\ and\ \citenamefont {Cui}(2019)}]{engine}%
  \BibitemOpen
  \bibfield  {author} {\bibinfo {author} {\bibfnamefont {X.}~\bibnamefont
  {Fu}}\ and\ \bibinfo {author} {\bibfnamefont {T.~J.}\ \bibnamefont {Cui}},\
  }\bibfield  {title} {\bibinfo {title} {Recent progress on metamaterials: From
  effective medium model to real-time information processing system},\ }\href
  {https://www.sciencedirect.com/science/article/pii/S0079672719300151}
  {\bibfield  {journal} {\bibinfo  {journal} {Progress in Quantum Electronics}\
  }\textbf {\bibinfo {volume} {67}},\ \bibinfo {pages} {100223} (\bibinfo
  {year} {2019})}\BibitemShut {NoStop}%
\bibitem [{\citenamefont {Andraud}\ and\ \citenamefont
  {Maury}(2009)}]{chemist}%
  \BibitemOpen
  \bibfield  {author} {\bibinfo {author} {\bibfnamefont {C.}~\bibnamefont
  {Andraud}}\ and\ \bibinfo {author} {\bibfnamefont {O.}~\bibnamefont
  {Maury}},\ }\bibfield  {title} {\bibinfo {title} {Lanthanide complexes for
  nonlinear optics: From fundamental aspects to applications},\ }\href
  {https://doi.org/https://doi.org/10.1002/ejic.200900534} {\bibfield
  {journal} {\bibinfo  {journal} {European Journal of Inorganic Chemistry}\
  }\textbf {\bibinfo {volume} {2009}},\ \bibinfo {pages} {4357} (\bibinfo
  {year} {2009})}\BibitemShut {NoStop}%
\bibitem [{\citenamefont {Yue}\ \emph {et~al.}(2011)\citenamefont {Yue},
  \citenamefont {Slipchenko},\ and\ \citenamefont {Cheng}}]{biolog}%
  \BibitemOpen
  \bibfield  {author} {\bibinfo {author} {\bibfnamefont {S.}~\bibnamefont
  {Yue}}, \bibinfo {author} {\bibfnamefont {M.}~\bibnamefont {Slipchenko}},\
  and\ \bibinfo {author} {\bibfnamefont {J.-X.}\ \bibnamefont {Cheng}},\
  }\bibfield  {title} {\bibinfo {title} {Multimodal nonlinear optical
  microscopy},\ }\href {https://doi.org/https://doi.org/10.1002/lpor.201000027}
  {\bibfield  {journal} {\bibinfo  {journal} {Laser \& Photonics Reviews}\
  }\textbf {\bibinfo {volume} {5}},\ \bibinfo {pages} {496} (\bibinfo {year}
  {2011})}\BibitemShut {NoStop}%
\bibitem [{\citenamefont {Combes}\ \emph {et~al.}(2021)\citenamefont {Combes},
  \citenamefont {Vučković}, \citenamefont {Perić~Bakulić}, \citenamefont
  {Antoine}, \citenamefont {Bonačić-Koutecky},\ and\ \citenamefont
  {Trajković}}]{medicine}%
  \BibitemOpen
  \bibfield  {author} {\bibinfo {author} {\bibfnamefont {G.~F.}\ \bibnamefont
  {Combes}}, \bibinfo {author} {\bibfnamefont {A.-M.}\ \bibnamefont
  {Vučković}}, \bibinfo {author} {\bibfnamefont {M.}~\bibnamefont
  {Perić~Bakulić}}, \bibinfo {author} {\bibfnamefont {R.}~\bibnamefont
  {Antoine}}, \bibinfo {author} {\bibfnamefont {V.}~\bibnamefont
  {Bonačić-Koutecky}},\ and\ \bibinfo {author} {\bibfnamefont
  {K.}~\bibnamefont {Trajković}},\ }\bibfield  {title} {\bibinfo {title}
  {Nanotechnology in tumor biomarker detection: The potential of liganded
  nanoclusters as nonlinear optical contrast agents for molecular diagnostics
  of cancer},\ }\href {https://doi.org/10.3390/cancers13164206} {\bibfield
  {journal} {\bibinfo  {journal} {Cancers}\ }\textbf {\bibinfo {volume} {13}},\
  \bibinfo {pages} {4206} (\bibinfo {year} {2021})}\BibitemShut {NoStop}%
\bibitem [{\citenamefont {Chen}\ \emph {et~al.}(2021)\citenamefont {Chen},
  \citenamefont {Hu}, \citenamefont {Kong},\ and\ \citenamefont
  {Mao}}]{SHGmaterials}%
  \BibitemOpen
  \bibfield  {author} {\bibinfo {author} {\bibfnamefont {J.}~\bibnamefont
  {Chen}}, \bibinfo {author} {\bibfnamefont {C.-L.}\ \bibnamefont {Hu}},
  \bibinfo {author} {\bibfnamefont {F.}~\bibnamefont {Kong}},\ and\ \bibinfo
  {author} {\bibfnamefont {J.-G.}\ \bibnamefont {Mao}},\ }\bibfield  {title}
  {\bibinfo {title} {High-performance second-harmonic-generation (shg)
  materials: New developments and new strategies},\ }\href
  {https://doi.org/10.1021/acs.accounts.1c00188} {\bibfield  {journal}
  {\bibinfo  {journal} {Accounts of Chemical Research}\ }\textbf {\bibinfo
  {volume} {54}},\ \bibinfo {pages} {2775} (\bibinfo {year}
  {2021})}\BibitemShut {NoStop}%
\bibitem [{\citenamefont {Cini}\ \emph {et~al.}(1993)\citenamefont {Cini},
  \citenamefont {D'Andrea},\ and\ \citenamefont {Verdozzi}}]{SHG93}%
  \BibitemOpen
  \bibfield  {author} {\bibinfo {author} {\bibfnamefont {M.}~\bibnamefont
  {Cini}}, \bibinfo {author} {\bibfnamefont {A.}~\bibnamefont {D'Andrea}},\
  and\ \bibinfo {author} {\bibfnamefont {C.}~\bibnamefont {Verdozzi}},\
  }\bibfield  {title} {\bibinfo {title} {Many-photon effects in inelastic
  light-scattering},\ }\href {https://doi.org/10.1016/0375-9601(93)90294-A}
  {\bibfield  {journal} {\bibinfo  {journal} {Physics Letters A}\ }\textbf
  {\bibinfo {volume} {180}},\ \bibinfo {pages} {430} (\bibinfo {year}
  {1993})}\BibitemShut {NoStop}%
\bibitem [{\citenamefont {Cini}\ \emph {et~al.}(1995)\citenamefont {Cini},
  \citenamefont {D'Andrea},\ and\ \citenamefont {Verdozzi}}]{SHG95}%
  \BibitemOpen
  \bibfield  {author} {\bibinfo {author} {\bibfnamefont {M.}~\bibnamefont
  {Cini}}, \bibinfo {author} {\bibfnamefont {A.}~\bibnamefont {D'Andrea}},\
  and\ \bibinfo {author} {\bibfnamefont {C.}~\bibnamefont {Verdozzi}},\
  }\bibfield  {title} {\bibinfo {title} {Many-photon effects in inelastic
  light-scattering - theory and model applications},\ }\href
  {https://doi.org/10.1142/S0217979295000501} {\bibfield  {journal} {\bibinfo
  {journal} {International Journal of Modern Physics B}\ }\textbf {\bibinfo
  {volume} {9}},\ \bibinfo {pages} {1185} (\bibinfo {year} {1995})}\BibitemShut
  {NoStop}%
\bibitem [{\citenamefont {Vi\~nas Bostr\"om}\ \emph {et~al.}(2020)\citenamefont
  {Vi\~nas Bostr\"om}, \citenamefont {D'Andrea}, \citenamefont {Cini},\ and\
  \citenamefont {Verdozzi}}]{Emil2020}%
  \BibitemOpen
  \bibfield  {author} {\bibinfo {author} {\bibfnamefont {E.}~\bibnamefont
  {Vi\~nas Bostr\"om}}, \bibinfo {author} {\bibfnamefont {A.}~\bibnamefont
  {D'Andrea}}, \bibinfo {author} {\bibfnamefont {M.}~\bibnamefont {Cini}},\
  and\ \bibinfo {author} {\bibfnamefont {C.}~\bibnamefont {Verdozzi}},\
  }\bibfield  {title} {\bibinfo {title} {Time-resolved multiphoton effects in
  the fluorescence spectra of two-level systems at rest and in motion},\ }\href
  {https://doi.org/10.1103/PhysRevA.102.013719} {\bibfield  {journal} {\bibinfo
   {journal} {Physical Review A}\ }\textbf {\bibinfo {volume} {102}},\ \bibinfo
  {pages} {013719} (\bibinfo {year} {2020})}\BibitemShut {NoStop}%
\bibitem [{\citenamefont {Landig}\ \emph {et~al.}(2016)\citenamefont {Landig},
  \citenamefont {Hruby}, \citenamefont {Dogra}, \citenamefont {Landini},
  \citenamefont {Mottl}, \citenamefont {Donner},\ and\ \citenamefont
  {Esslinger}}]{Landig2016}%
  \BibitemOpen
  \bibfield  {author} {\bibinfo {author} {\bibfnamefont {R.}~\bibnamefont
  {Landig}}, \bibinfo {author} {\bibfnamefont {L.}~\bibnamefont {Hruby}},
  \bibinfo {author} {\bibfnamefont {N.}~\bibnamefont {Dogra}}, \bibinfo
  {author} {\bibfnamefont {M.}~\bibnamefont {Landini}}, \bibinfo {author}
  {\bibfnamefont {R.}~\bibnamefont {Mottl}}, \bibinfo {author} {\bibfnamefont
  {T.}~\bibnamefont {Donner}},\ and\ \bibinfo {author} {\bibfnamefont
  {T.}~\bibnamefont {Esslinger}},\ }\bibfield  {title} {\bibinfo {title}
  {Quantum phases from competing short- and long-range interactions in an
  optical lattice},\ }\href {https://doi.org/10.1038/nature17409} {\bibfield
  {journal} {\bibinfo  {journal} {Nature}\ }\textbf {\bibinfo {volume} {532}},\
  \bibinfo {pages} {476} (\bibinfo {year} {2016})}\BibitemShut {NoStop}%
\bibitem [{\citenamefont {Nagy}\ \emph {et~al.}(2018)\citenamefont {Nagy},
  \citenamefont {K\'onya}, \citenamefont {Domokos},\ and\ \citenamefont
  {Szirmai}}]{Nagy2018}%
  \BibitemOpen
  \bibfield  {author} {\bibinfo {author} {\bibfnamefont {D.}~\bibnamefont
  {Nagy}}, \bibinfo {author} {\bibfnamefont {G.}~\bibnamefont {K\'onya}},
  \bibinfo {author} {\bibfnamefont {P.}~\bibnamefont {Domokos}},\ and\ \bibinfo
  {author} {\bibfnamefont {G.}~\bibnamefont {Szirmai}},\ }\bibfield  {title}
  {\bibinfo {title} {Quantum noise in a transversely-pumped-cavity bose-hubbard
  model},\ }\href {https://doi.org/10.1103/PhysRevA.97.063602} {\bibfield
  {journal} {\bibinfo  {journal} {Physical Review A}\ }\textbf {\bibinfo
  {volume} {97}},\ \bibinfo {pages} {063602} (\bibinfo {year}
  {2018})}\BibitemShut {NoStop}%
\bibitem [{\citenamefont {Gopalakrishna}\ \emph {et~al.}(2023)\citenamefont
  {Gopalakrishna}, \citenamefont {Vi\~{n}as Bostr\"om},\ and\ \citenamefont
  {Verdozzi}}]{Scipost}%
  \BibitemOpen
  \bibfield  {author} {\bibinfo {author} {\bibfnamefont {M.}~\bibnamefont
  {Gopalakrishna}}, \bibinfo {author} {\bibfnamefont {E.}~\bibnamefont
  {Vi\~{n}as Bostr\"om}},\ and\ \bibinfo {author} {\bibfnamefont
  {C.}~\bibnamefont {Verdozzi}},\ }\bibfield  {title} {\bibinfo {title}
  {{Photon pumping, photodissociation and dissipation at interplay for the
  fluorescence of a molecule in a cavity}},\ }\href
  {https://doi.org/10.21468/SciPostPhys.15.4.138} {\bibfield  {journal}
  {\bibinfo  {journal} {SciPost Phys.}\ }\textbf {\bibinfo {volume} {15}},\
  \bibinfo {pages} {138} (\bibinfo {year} {2023})}\BibitemShut {NoStop}%
\bibitem [{\citenamefont {Bauch}\ \emph {et~al.}(2009)\citenamefont {Bauch},
  \citenamefont {Balzer}, \citenamefont {Henning},\ and\ \citenamefont
  {Bonitz}}]{BreathingBonitz2009}%
  \BibitemOpen
  \bibfield  {author} {\bibinfo {author} {\bibfnamefont {S.}~\bibnamefont
  {Bauch}}, \bibinfo {author} {\bibfnamefont {K.}~\bibnamefont {Balzer}},
  \bibinfo {author} {\bibfnamefont {C.}~\bibnamefont {Henning}},\ and\ \bibinfo
  {author} {\bibfnamefont {M.}~\bibnamefont {Bonitz}},\ }\bibfield  {title}
  {\bibinfo {title} {Quantum breathing mode of trapped bosons and fermions at
  arbitrary coupling},\ }\href {https://doi.org/10.1103/PhysRevB.80.054515}
  {\bibfield  {journal} {\bibinfo  {journal} {Phys. Rev. B}\ }\textbf {\bibinfo
  {volume} {80}},\ \bibinfo {pages} {054515} (\bibinfo {year}
  {2009})}\BibitemShut {NoStop}%
\bibitem [{\citenamefont {Heimsoth}\ and\ \citenamefont
  {Bonitz}(2010)}]{BeyondGP2010}%
  \BibitemOpen
  \bibfield  {author} {\bibinfo {author} {\bibfnamefont {M.}~\bibnamefont
  {Heimsoth}}\ and\ \bibinfo {author} {\bibfnamefont {M.}~\bibnamefont
  {Bonitz}},\ }\bibfield  {title} {\bibinfo {title} {Interacting bosons beyond
  the gross-pitaevskii mean field},\ }\href
  {https://doi.org/10.1016/j.physe.2009.06.040} {\bibfield  {journal} {\bibinfo
   {journal} {Physica E: Low-dimensional Systems and Nanostructures}\ }\textbf
  {\bibinfo {volume} {42}},\ \bibinfo {pages} {420} (\bibinfo {year} {2010})},\
  \bibinfo {note} {proceedings of the international conference Frontiers of
  Quantum and Mesoscopic Thermodynamics FQMT '08}\BibitemShut {NoStop}%
\bibitem [{\citenamefont {Balzer}\ \emph {et~al.}(2010)\citenamefont {Balzer},
  \citenamefont {Bauch},\ and\ \citenamefont {Bonitz}}]{LaserBonitz2010}%
  \BibitemOpen
  \bibfield  {author} {\bibinfo {author} {\bibfnamefont {K.}~\bibnamefont
  {Balzer}}, \bibinfo {author} {\bibfnamefont {S.}~\bibnamefont {Bauch}},\ and\
  \bibinfo {author} {\bibfnamefont {M.}~\bibnamefont {Bonitz}},\ }\bibfield
  {title} {\bibinfo {title} {Time-dependent second-order born calculations for
  model atoms and molecules in strong laser fields},\ }\href
  {https://doi.org/10.1103/PhysRevA.82.033427} {\bibfield  {journal} {\bibinfo
  {journal} {Phys. Rev. A}\ }\textbf {\bibinfo {volume} {82}},\ \bibinfo
  {pages} {033427} (\bibinfo {year} {2010})}\BibitemShut {NoStop}%
\bibitem [{\citenamefont {Joost}\ and\ \citenamefont
  {Bonitz}(2025)}]{Laser_Bonitz_Graphene2025}%
  \BibitemOpen
  \bibfield  {author} {\bibinfo {author} {\bibfnamefont {J.-P.}\ \bibnamefont
  {Joost}}\ and\ \bibinfo {author} {\bibfnamefont {M.}~\bibnamefont {Bonitz}},\
  }\bibfield  {title} {\bibinfo {title} {Ultrafast charge separation induced by
  a uniform field in graphene nanoribbons},\ }\href
  {https://doi.org/10.1103/dtk9-xv6n} {\bibfield  {journal} {\bibinfo
  {journal} {Phys. Rev. Res.}\ }\textbf {\bibinfo {volume} {7}},\ \bibinfo
  {pages} {L032068} (\bibinfo {year} {2025})}\BibitemShut {NoStop}%
\bibitem [{\citenamefont {Bonitz}\ \emph {et~al.}(1996)\citenamefont {Bonitz},
  \citenamefont {Kremp}, \citenamefont {Scott}, \citenamefont {Binder},
  \citenamefont {Kraeft},\ and\ \citenamefont {Köhler}}]{MBonitz_1996}%
  \BibitemOpen
  \bibfield  {author} {\bibinfo {author} {\bibfnamefont {M.}~\bibnamefont
  {Bonitz}}, \bibinfo {author} {\bibfnamefont {D.}~\bibnamefont {Kremp}},
  \bibinfo {author} {\bibfnamefont {D.~C.}\ \bibnamefont {Scott}}, \bibinfo
  {author} {\bibfnamefont {R.}~\bibnamefont {Binder}}, \bibinfo {author}
  {\bibfnamefont {W.~D.}\ \bibnamefont {Kraeft}},\ and\ \bibinfo {author}
  {\bibfnamefont {H.~S.}\ \bibnamefont {Köhler}},\ }\bibfield  {title}
  {\bibinfo {title} {Numerical analysis of non-markovian effects in
  charge-carrier scattering: one-time versus two-time kinetic equations},\
  }\href {https://doi.org/10.1088/0953-8984/8/33/012} {\bibfield  {journal}
  {\bibinfo  {journal} {Journal of Physics: Condensed Matter}\ }\textbf
  {\bibinfo {volume} {8}},\ \bibinfo {pages} {6057} (\bibinfo {year}
  {1996})}\BibitemShut {NoStop}%
\bibitem [{\citenamefont {Schl\"unzen}\ \emph {et~al.}(2016)\citenamefont
  {Schl\"unzen}, \citenamefont {Hermanns}, \citenamefont {Bonitz},\ and\
  \citenamefont {Verdozzi}}]{PhysRevB.93.035107}%
  \BibitemOpen
  \bibfield  {author} {\bibinfo {author} {\bibfnamefont {N.}~\bibnamefont
  {Schl\"unzen}}, \bibinfo {author} {\bibfnamefont {S.}~\bibnamefont
  {Hermanns}}, \bibinfo {author} {\bibfnamefont {M.}~\bibnamefont {Bonitz}},\
  and\ \bibinfo {author} {\bibfnamefont {C.}~\bibnamefont {Verdozzi}},\
  }\bibfield  {title} {\bibinfo {title} {Dynamics of strongly correlated
  fermions: Ab initio results for two and three dimensions},\ }\href
  {https://doi.org/10.1103/PhysRevB.93.035107} {\bibfield  {journal} {\bibinfo
  {journal} {Phys. Rev. B}\ }\textbf {\bibinfo {volume} {93}},\ \bibinfo
  {pages} {035107} (\bibinfo {year} {2016})}\BibitemShut {NoStop}%
\bibitem [{\citenamefont {Schl\"unzen}\ \emph {et~al.}(2020)\citenamefont
  {Schl\"unzen}, \citenamefont {Joost},\ and\ \citenamefont
  {Bonitz}}]{PhysRevLett.124.076601}%
  \BibitemOpen
  \bibfield  {author} {\bibinfo {author} {\bibfnamefont {N.}~\bibnamefont
  {Schl\"unzen}}, \bibinfo {author} {\bibfnamefont {J.-P.}\ \bibnamefont
  {Joost}},\ and\ \bibinfo {author} {\bibfnamefont {M.}~\bibnamefont
  {Bonitz}},\ }\bibfield  {title} {\bibinfo {title} {Achieving the scaling
  limit for nonequilibrium green functions simulations},\ }\href
  {https://doi.org/10.1103/PhysRevLett.124.076601} {\bibfield  {journal}
  {\bibinfo  {journal} {Phys. Rev. Lett.}\ }\textbf {\bibinfo {volume} {124}},\
  \bibinfo {pages} {076601} (\bibinfo {year} {2020})}\BibitemShut {NoStop}%
\bibitem [{\citenamefont {Zoubi}\ and\ \citenamefont
  {Ritsch}(2007)}]{Zoubi2007}%
  \BibitemOpen
  \bibfield  {author} {\bibinfo {author} {\bibfnamefont {H.}~\bibnamefont
  {Zoubi}}\ and\ \bibinfo {author} {\bibfnamefont {H.}~\bibnamefont {Ritsch}},\
  }\bibfield  {title} {\bibinfo {title} {Excitons and cavity polaritons for
  ultracold atoms in an optical lattice},\ }\href
  {https://doi.org/10.1103/PhysRevA.76.013817} {\bibfield  {journal} {\bibinfo
  {journal} {Physical Review A}\ }\textbf {\bibinfo {volume} {76}},\ \bibinfo
  {pages} {013817} (\bibinfo {year} {2007})}\BibitemShut {NoStop}%
\bibitem [{\citenamefont {Maschler}\ \emph {et~al.}(2008)\citenamefont
  {Maschler}, \citenamefont {Mekhov},\ and\ \citenamefont
  {Ritsch}}]{Maschler2008}%
  \BibitemOpen
  \bibfield  {author} {\bibinfo {author} {\bibfnamefont {C.}~\bibnamefont
  {Maschler}}, \bibinfo {author} {\bibfnamefont {I.~B.}\ \bibnamefont
  {Mekhov}},\ and\ \bibinfo {author} {\bibfnamefont {H.}~\bibnamefont
  {Ritsch}},\ }\bibfield  {title} {\bibinfo {title} {Ultracold atoms in optical
  lattices generated by quantized light fields},\ }\href
  {https://doi.org/10.1140/epjd/e2008-00016-4} {\bibfield  {journal} {\bibinfo
  {journal} {The European Physical Journal D}\ }\textbf {\bibinfo {volume}
  {46}},\ \bibinfo {pages} {545} (\bibinfo {year} {2008})}\BibitemShut
  {NoStop}%
\bibitem [{\citenamefont {Zoubi}\ and\ \citenamefont
  {Ritsch}(2013)}]{Zoubi2013}%
  \BibitemOpen
  \bibfield  {author} {\bibinfo {author} {\bibfnamefont {H.}~\bibnamefont
  {Zoubi}}\ and\ \bibinfo {author} {\bibfnamefont {H.}~\bibnamefont {Ritsch}},\
  }\bibinfo {title} {Chapter 3 - {Excitons} and cavity polaritons for optical
  lattice ultracold atoms},\ in\ \href
  {https://doi.org/https://doi.org/10.1016/B978-0-12-408090-4.00003-7} {\emph
  {\bibinfo {booktitle} {Advances In Atomic, Molecular, and Optical
  Physics}}},\ Vol.~\bibinfo {volume} {62},\ \bibinfo {editor} {edited by\
  \bibinfo {editor} {\bibfnamefont {E.}~\bibnamefont {Arimondo}}, \bibinfo
  {editor} {\bibfnamefont {P.~R.}\ \bibnamefont {Berman}},\ and\ \bibinfo
  {editor} {\bibfnamefont {C.~C.}\ \bibnamefont {Lin}}}\ (\bibinfo  {publisher}
  {Academic Press},\ \bibinfo {year} {2013})\ pp.\ \bibinfo {pages}
  {171--229}\BibitemShut {NoStop}%
\bibitem [{\citenamefont {Gopalakrishna}\ \emph {et~al.}(2024)\citenamefont
  {Gopalakrishna}, \citenamefont {Pavlyukh},\ and\ \citenamefont
  {Verdozzi}}]{Dicke24}%
  \BibitemOpen
  \bibfield  {author} {\bibinfo {author} {\bibfnamefont {M.}~\bibnamefont
  {Gopalakrishna}}, \bibinfo {author} {\bibfnamefont {Y.}~\bibnamefont
  {Pavlyukh}},\ and\ \bibinfo {author} {\bibfnamefont {C.}~\bibnamefont
  {Verdozzi}},\ }\bibfield  {title} {\bibinfo {title} {Time-resolved optical
  response of the dicke's model via the nonequilibrium green's function
  approach},\ }\href {https://doi.org/https://doi.org/10.1002/pssb.202300576}
  {\bibfield  {journal} {\bibinfo  {journal} {physica status solidi (b)}\
  }\textbf {\bibinfo {volume} {261}},\ \bibinfo {pages} {2300576} (\bibinfo
  {year} {2024})},\ \Eprint
  {https://arxiv.org/abs/https://onlinelibrary.wiley.com/doi/pdf/10.1002/
  pssb.202300576} {https://onlinelibrary.wiley.com/doi/pdf/10.1002/
  pssb.202300576} \BibitemShut {NoStop}%
\bibitem [{cur()}]{currently}%
  \BibitemOpen
  \href@noop {} {\ }\bibinfo {note} {A direct way to go beyond finite
  Bose-Hubbard systems would be to consider an optical lattice described by the
  Bose Hubbard model at large $L$ and, for example, to focus on the ideal/non
  ideal superfluid phase in the presence of quantum light modes. This is a
  rather interesting problem, which however requires additional developments;
  work in along these lines is currently in process.}\BibitemShut {Stop}%
\bibitem [{\citenamefont {Kuang}\ and\ \citenamefont {Zhou}(2003)}]{Kuang}%
  \BibitemOpen
  \bibfield  {author} {\bibinfo {author} {\bibfnamefont {L.-M.}\ \bibnamefont
  {Kuang}}\ and\ \bibinfo {author} {\bibfnamefont {L.}~\bibnamefont {Zhou}},\
  }\bibfield  {title} {\bibinfo {title} {Generation of atom-photon entangled
  states in atomic {B}ose-{E}instein condensate via electromagnetically induced
  transparency},\ }\href {https://doi.org/10.1103/PhysRevA.68.043606}
  {\bibfield  {journal} {\bibinfo  {journal} {Physical Review A}\ }\textbf
  {\bibinfo {volume} {68}},\ \bibinfo {pages} {043606} (\bibinfo {year}
  {2003})}\BibitemShut {NoStop}%
\bibitem [{\citenamefont {Huang}\ \emph {et~al.}(2008)\citenamefont {Huang},
  \citenamefont {Fang}, \citenamefont {He}, \citenamefont {Kong},\ and\
  \citenamefont {Zhou}}]{Huang08}%
  \BibitemOpen
  \bibfield  {author} {\bibinfo {author} {\bibfnamefont {C.}~\bibnamefont
  {Huang}}, \bibinfo {author} {\bibfnamefont {J.}~\bibnamefont {Fang}},
  \bibinfo {author} {\bibfnamefont {H.}~\bibnamefont {He}}, \bibinfo {author}
  {\bibfnamefont {F.}~\bibnamefont {Kong}},\ and\ \bibinfo {author}
  {\bibfnamefont {M.}~\bibnamefont {Zhou}},\ }\bibfield  {title} {\bibinfo
  {title} {Squeezing of an atom laser originating from atomic {B}ose-{E}instein
  condensate interacting with light field},\ }\href
  {https://www.sciencedirect.com/science/article/pii/S0378437108001969}
  {\bibfield  {journal} {\bibinfo  {journal} {Physica A: Statistical Mechanics
  and its Applications}\ }\textbf {\bibinfo {volume} {387}},\ \bibinfo {pages}
  {3449} (\bibinfo {year} {2008})}\BibitemShut {NoStop}%
\bibitem [{\citenamefont {Stefanucci}(2024)}]{Stefanucci_dissip}%
  \BibitemOpen
  \bibfield  {author} {\bibinfo {author} {\bibfnamefont {G.}~\bibnamefont
  {Stefanucci}},\ }\bibfield  {title} {\bibinfo {title} {Kadanoff-baym
  equations for interacting systems with dissipative lindbladian dynamics},\
  }\href {https://doi.org/10.1103/PhysRevLett.133.066901} {\bibfield  {journal}
  {\bibinfo  {journal} {Phys. Rev. Lett.}\ }\textbf {\bibinfo {volume} {133}},\
  \bibinfo {pages} {066901} (\bibinfo {year} {2024})}\BibitemShut {NoStop}%
\bibitem [{\citenamefont {Caldeira}\ and\ \citenamefont
  {Leggett}(1983)}]{Caldeira1983}%
  \BibitemOpen
  \bibfield  {author} {\bibinfo {author} {\bibfnamefont {A.~O.}\ \bibnamefont
  {Caldeira}}\ and\ \bibinfo {author} {\bibfnamefont {A.~J.}\ \bibnamefont
  {Leggett}},\ }\bibfield  {title} {\bibinfo {title} {Quantum tunnelling in a
  dissipative system},\ }\href {https://doi.org/10.1016/0003-4916(83)90202-6}
  {\bibfield  {journal} {\bibinfo  {journal} {Annals of Physics}\ }\textbf
  {\bibinfo {volume} {149}},\ \bibinfo {pages} {374} (\bibinfo {year}
  {1983})}\BibitemShut {NoStop}%
\bibitem [{\citenamefont {Venkataraman}\ \emph {et~al.}(2013)\citenamefont
  {Venkataraman}, \citenamefont {Plato}, \citenamefont {Tufarelli},\ and\
  \citenamefont {Kim}}]{Venkataraman2014}%
  \BibitemOpen
  \bibfield  {author} {\bibinfo {author} {\bibfnamefont {V.}~\bibnamefont
  {Venkataraman}}, \bibinfo {author} {\bibfnamefont {A.~D.~K.}\ \bibnamefont
  {Plato}}, \bibinfo {author} {\bibfnamefont {T.}~\bibnamefont {Tufarelli}},\
  and\ \bibinfo {author} {\bibfnamefont {M.~S.}\ \bibnamefont {Kim}},\
  }\bibfield  {title} {\bibinfo {title} {Affecting non-markovian behaviour by
  changing bath structures},\ }\href
  {https://doi.org/10.1088/0953-4075/47/1/015501} {\bibfield  {journal}
  {\bibinfo  {journal} {Journal of Physics B: Atomic, Molecular and Optical
  Physics}\ }\textbf {\bibinfo {volume} {47}},\ \bibinfo {pages} {015501}
  (\bibinfo {year} {2013})}\BibitemShut {NoStop}%
\bibitem [{\citenamefont {Horsfield}\ \emph {et~al.}(2004)\citenamefont
  {Horsfield}, \citenamefont {Bowler}, \citenamefont {Fisher}, \citenamefont
  {Todorov},\ and\ \citenamefont {Montgomery}}]{DetBal1}%
  \BibitemOpen
  \bibfield  {author} {\bibinfo {author} {\bibfnamefont {A.~P.}\ \bibnamefont
  {Horsfield}}, \bibinfo {author} {\bibfnamefont {D.~R.}\ \bibnamefont
  {Bowler}}, \bibinfo {author} {\bibfnamefont {A.~J.}\ \bibnamefont {Fisher}},
  \bibinfo {author} {\bibfnamefont {T.~N.}\ \bibnamefont {Todorov}},\ and\
  \bibinfo {author} {\bibfnamefont {M.~J.}\ \bibnamefont {Montgomery}},\
  }\bibfield  {title} {\bibinfo {title} {Power dissipation in nanoscale
  conductors: classical, semi-classical and quantum dynamics},\ }\href
  {https://doi.org/10.1088/0953-8984/16/21/010} {\bibfield  {journal} {\bibinfo
   {journal} {Journal of Physics: Condensed Matter}\ }\textbf {\bibinfo
  {volume} {16}},\ \bibinfo {pages} {3609} (\bibinfo {year}
  {2004})}\BibitemShut {NoStop}%
\bibitem [{\citenamefont {Stahl}\ and\ \citenamefont
  {Potthoff}(2017)}]{DetBal2}%
  \BibitemOpen
  \bibfield  {author} {\bibinfo {author} {\bibfnamefont {C.}~\bibnamefont
  {Stahl}}\ and\ \bibinfo {author} {\bibfnamefont {M.}~\bibnamefont
  {Potthoff}},\ }\bibfield  {title} {\bibinfo {title} {Anomalous spin
  precession under a geometrical torque},\ }\href
  {https://doi.org/10.1103/PhysRevLett.119.227203} {\bibfield  {journal}
  {\bibinfo  {journal} {Physical Review Letters}\ }\textbf {\bibinfo {volume}
  {119}},\ \bibinfo {pages} {227203} (\bibinfo {year} {2017})}\BibitemShut
  {NoStop}%
\bibitem [{\citenamefont {Bai}\ \emph {et~al.}(2022)\citenamefont {Bai},
  \citenamefont {Han}, \citenamefont {Feng}, \citenamefont {Zhou},
  \citenamefont {Su}, \citenamefont {Wang}, \citenamefont {Liao}, \citenamefont
  {Zhu}, \citenamefont {Chen}, \citenamefont {Pan}, \citenamefont {Fan},\ and\
  \citenamefont {Song}}]{DetBal3}%
  \BibitemOpen
  \bibfield  {author} {\bibinfo {author} {\bibfnamefont {H.}~\bibnamefont
  {Bai}}, \bibinfo {author} {\bibfnamefont {L.}~\bibnamefont {Han}}, \bibinfo
  {author} {\bibfnamefont {X.~Y.}\ \bibnamefont {Feng}}, \bibinfo {author}
  {\bibfnamefont {Y.~J.}\ \bibnamefont {Zhou}}, \bibinfo {author}
  {\bibfnamefont {R.~X.}\ \bibnamefont {Su}}, \bibinfo {author} {\bibfnamefont
  {Q.}~\bibnamefont {Wang}}, \bibinfo {author} {\bibfnamefont {L.~Y.}\
  \bibnamefont {Liao}}, \bibinfo {author} {\bibfnamefont {W.~X.}\ \bibnamefont
  {Zhu}}, \bibinfo {author} {\bibfnamefont {X.~Z.}\ \bibnamefont {Chen}},
  \bibinfo {author} {\bibfnamefont {F.}~\bibnamefont {Pan}}, \bibinfo {author}
  {\bibfnamefont {X.~L.}\ \bibnamefont {Fan}},\ and\ \bibinfo {author}
  {\bibfnamefont {C.}~\bibnamefont {Song}},\ }\bibfield  {title} {\bibinfo
  {title} {Observation of spin splitting torque in a collinear antiferromagnet
  ${\mathrm{ruo}}_{2}$},\ }\href
  {https://doi.org/10.1103/PhysRevLett.128.197202} {\bibfield  {journal}
  {\bibinfo  {journal} {Physical Review Letters}\ }\textbf {\bibinfo {volume}
  {128}},\ \bibinfo {pages} {197202} (\bibinfo {year} {2022})}\BibitemShut
  {NoStop}%
\bibitem [{\citenamefont {Gay-Balmaz}\ and\ \citenamefont
  {Tronci}(2023)}]{DetBal4}%
  \BibitemOpen
  \bibfield  {author} {\bibinfo {author} {\bibfnamefont {F.}~\bibnamefont
  {Gay-Balmaz}}\ and\ \bibinfo {author} {\bibfnamefont {C.}~\bibnamefont
  {Tronci}},\ }\bibfield  {title} {\bibinfo {title} {Dynamics of mixed
  quantum–classical spin systems},\ }\href
  {https://doi.org/10.1088/1751-8121/acc145} {\bibfield  {journal} {\bibinfo
  {journal} {Journal of Physics A: Mathematical and Theoretical}\ }\textbf
  {\bibinfo {volume} {56}},\ \bibinfo {pages} {144002} (\bibinfo {year}
  {2023})}\BibitemShut {NoStop}%
\bibitem [{\citenamefont {Andrade}\ \emph {et~al.}(2009)\citenamefont
  {Andrade}, \citenamefont {Castro}, \citenamefont {Zueco}, \citenamefont
  {Alonso}, \citenamefont {Echenique}, \citenamefont {Falceto},\ and\
  \citenamefont {Rubio}}]{DetBal5}%
  \BibitemOpen
  \bibfield  {author} {\bibinfo {author} {\bibfnamefont {X.}~\bibnamefont
  {Andrade}}, \bibinfo {author} {\bibfnamefont {A.}~\bibnamefont {Castro}},
  \bibinfo {author} {\bibfnamefont {D.}~\bibnamefont {Zueco}}, \bibinfo
  {author} {\bibfnamefont {J.~L.}\ \bibnamefont {Alonso}}, \bibinfo {author}
  {\bibfnamefont {P.}~\bibnamefont {Echenique}}, \bibinfo {author}
  {\bibfnamefont {F.}~\bibnamefont {Falceto}},\ and\ \bibinfo {author}
  {\bibfnamefont {A.}~\bibnamefont {Rubio}},\ }\bibfield  {title} {\bibinfo
  {title} {Modified {E}hrenfest formalism for efficient large-scale ab initio
  molecular dynamics},\ }\href {https://doi.org/10.1021/ct800518j} {\bibfield
  {journal} {\bibinfo  {journal} {Journal of Chemical Theory and Computation}\
  }\textbf {\bibinfo {volume} {5}},\ \bibinfo {pages} {728} (\bibinfo {year}
  {2009})}\BibitemShut {NoStop}%
\bibitem [{\citenamefont {Stenquist}\ \emph {et~al.}(2024)\citenamefont
  {Stenquist}, \citenamefont {Zapata}, \citenamefont {Olofsson}, \citenamefont
  {Liao}, \citenamefont {Svegborn}, \citenamefont {Bruhnke}, \citenamefont
  {Verdozzi},\ and\ \citenamefont {Dahlstr\"om}}]{PhysRevLett.133.063202}%
  \BibitemOpen
  \bibfield  {author} {\bibinfo {author} {\bibfnamefont {A.}~\bibnamefont
  {Stenquist}}, \bibinfo {author} {\bibfnamefont {F.}~\bibnamefont {Zapata}},
  \bibinfo {author} {\bibfnamefont {E.}~\bibnamefont {Olofsson}}, \bibinfo
  {author} {\bibfnamefont {Y.}~\bibnamefont {Liao}}, \bibinfo {author}
  {\bibfnamefont {E.}~\bibnamefont {Svegborn}}, \bibinfo {author}
  {\bibfnamefont {J.~N.}\ \bibnamefont {Bruhnke}}, \bibinfo {author}
  {\bibfnamefont {C.}~\bibnamefont {Verdozzi}},\ and\ \bibinfo {author}
  {\bibfnamefont {J.~M.}\ \bibnamefont {Dahlstr\"om}},\ }\bibfield  {title}
  {\bibinfo {title} {Mollow-like triplets in ultrafast resonant absorption},\
  }\href {https://doi.org/10.1103/PhysRevLett.133.063202} {\bibfield  {journal}
  {\bibinfo  {journal} {Phys. Rev. Lett.}\ }\textbf {\bibinfo {volume} {133}},\
  \bibinfo {pages} {063202} (\bibinfo {year} {2024})}\BibitemShut {NoStop}%
\bibitem [{\citenamefont {Dicke}(1954)}]{Dicke1954}%
  \BibitemOpen
  \bibfield  {author} {\bibinfo {author} {\bibfnamefont {R.~H.}\ \bibnamefont
  {Dicke}},\ }\bibfield  {title} {\bibinfo {title} {Coherence in spontaneous
  radiation processes},\ }\href {https://doi.org/10.1103/PhysRev.93.99}
  {\bibfield  {journal} {\bibinfo  {journal} {Phys. Rev.}\ }\textbf {\bibinfo
  {volume} {93}},\ \bibinfo {pages} {99} (\bibinfo {year} {1954})}\BibitemShut
  {NoStop}%
\bibitem [{\citenamefont {Kirton}\ \emph {et~al.}(2019)\citenamefont {Kirton},
  \citenamefont {Roses}, \citenamefont {Keeling},\ and\ \citenamefont
  {Dalla~Torre}}]{Kirton2019}%
  \BibitemOpen
  \bibfield  {author} {\bibinfo {author} {\bibfnamefont {P.}~\bibnamefont
  {Kirton}}, \bibinfo {author} {\bibfnamefont {M.~M.}\ \bibnamefont {Roses}},
  \bibinfo {author} {\bibfnamefont {J.}~\bibnamefont {Keeling}},\ and\ \bibinfo
  {author} {\bibfnamefont {E.~G.}\ \bibnamefont {Dalla~Torre}},\ }\bibfield
  {title} {\bibinfo {title} {Introduction to the dicke model: From equilibrium
  to nonequilibrium, and vice versa},\ }\href
  {https://doi.org/https://doi.org/10.1002/qute.201800043} {\bibfield
  {journal} {\bibinfo  {journal} {Advanced Quantum Technologies}\ }\textbf
  {\bibinfo {volume} {2}},\ \bibinfo {pages} {1800043} (\bibinfo {year}
  {2019})}\BibitemShut {NoStop}%
\bibitem [{\citenamefont {Nagy}\ \emph {et~al.}(2010)\citenamefont {Nagy},
  \citenamefont {K\'onya}, \citenamefont {Szirmai},\ and\ \citenamefont
  {Domokos}}]{Nagy2010}%
  \BibitemOpen
  \bibfield  {author} {\bibinfo {author} {\bibfnamefont {D.}~\bibnamefont
  {Nagy}}, \bibinfo {author} {\bibfnamefont {G.}~\bibnamefont {K\'onya}},
  \bibinfo {author} {\bibfnamefont {G.}~\bibnamefont {Szirmai}},\ and\ \bibinfo
  {author} {\bibfnamefont {P.}~\bibnamefont {Domokos}},\ }\bibfield  {title}
  {\bibinfo {title} {Dicke-model phase transition in the quantum motion of a
  {B}ose-{E}instein condensate in an optical cavity},\ }\href
  {https://doi.org/10.1103/PhysRevLett.104.130401} {\bibfield  {journal}
  {\bibinfo  {journal} {Physical Review Letters}\ }\textbf {\bibinfo {volume}
  {104}},\ \bibinfo {pages} {130401} (\bibinfo {year} {2010})}\BibitemShut
  {NoStop}%
\end{thebibliography}%

 \FloatBarrier
 \appendix  
 
\begin{figure}
{\includegraphics[width=0.9\columnwidth]{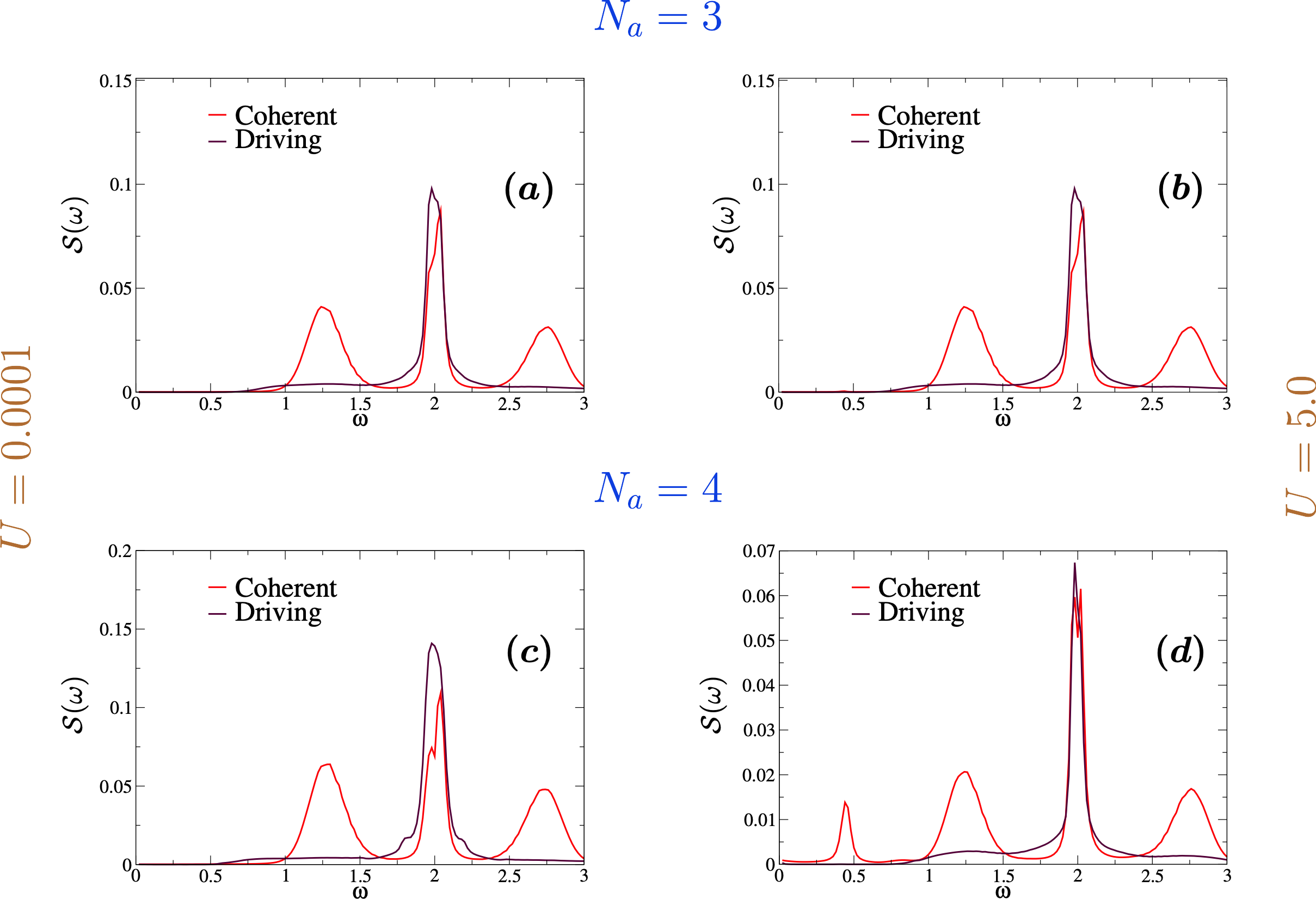}}
	\caption{Resonance calculations ($\omega_0=2.0$) with $L=3$. In all panels,
	$U_{eg}=U/2$.  
	 The calculations with photon driving are with $\tau_1=\tau_2=2.0$, $t_1=\frac{6\pi}{\omega_0}$, $t_2=\frac{31\pi}{\omega_0}$, the initial state is $|\Phi_0^{Res}\rangle$ and the coupling $\gamma_d$ is chosen such that $\langle \hat{b}^\dagger \hat{b}\rangle\approx16$.  For consistency, the calculation 
	 with the initial coherent state $\hat{b} |\eta\rangle= \eta|\eta\rangle$ 
	are performed with $\eta=4$. } 
	\label{A2fig1}
\end{figure}

\begin{figure}
		\centering
		\includegraphics[width=0.95\columnwidth]{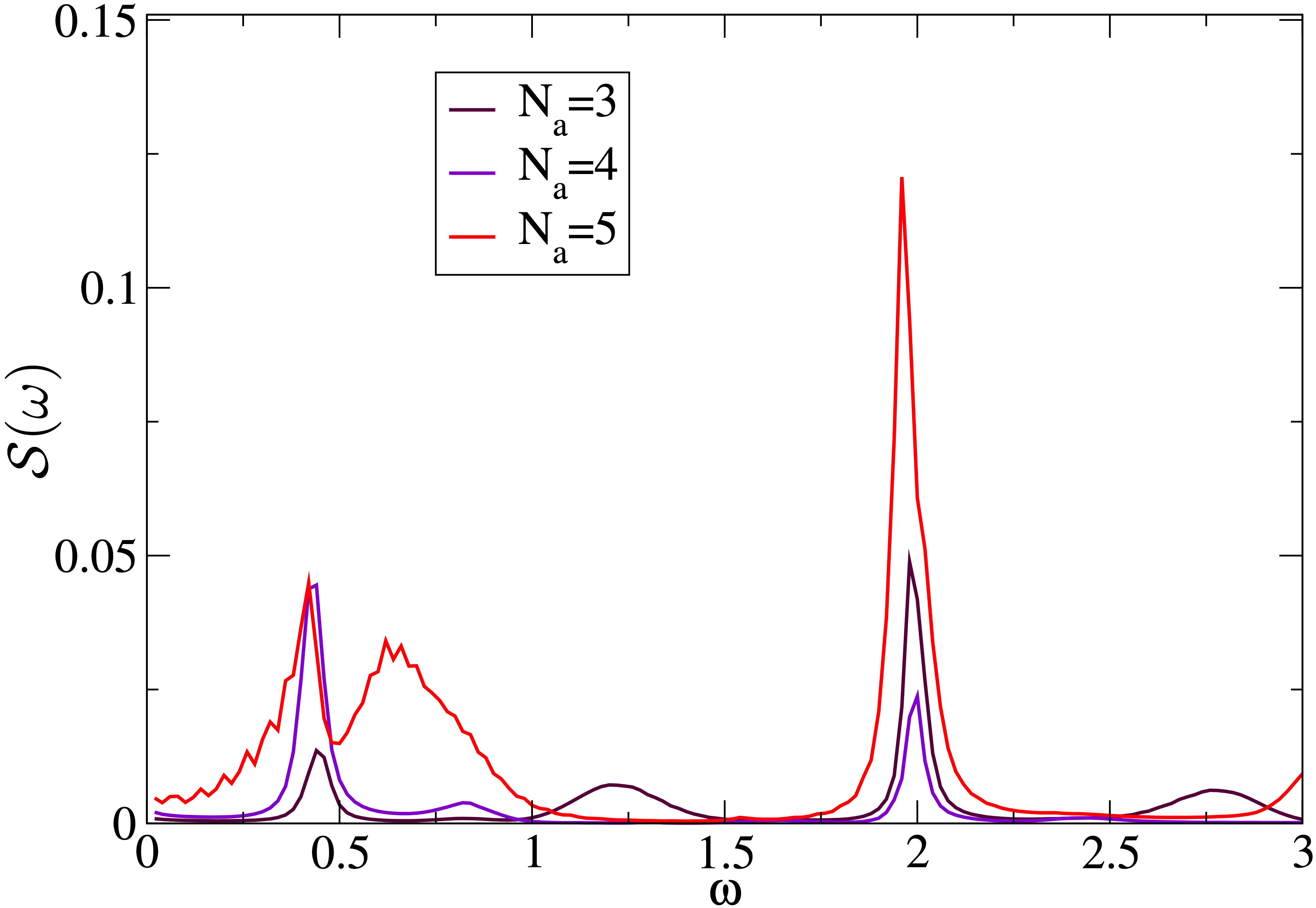}
		\caption{Results are of resonance calculations with $U=5.0$ and $U_{eg}=2.5$ (high interaction). In all calculations $L=2$, $\omega_0=2.0$ and $\eta=4$.  } 
	\label{A2fig2}
\end{figure}

\begin{figure}
 \centering
 	\includegraphics*[scale=0.75]{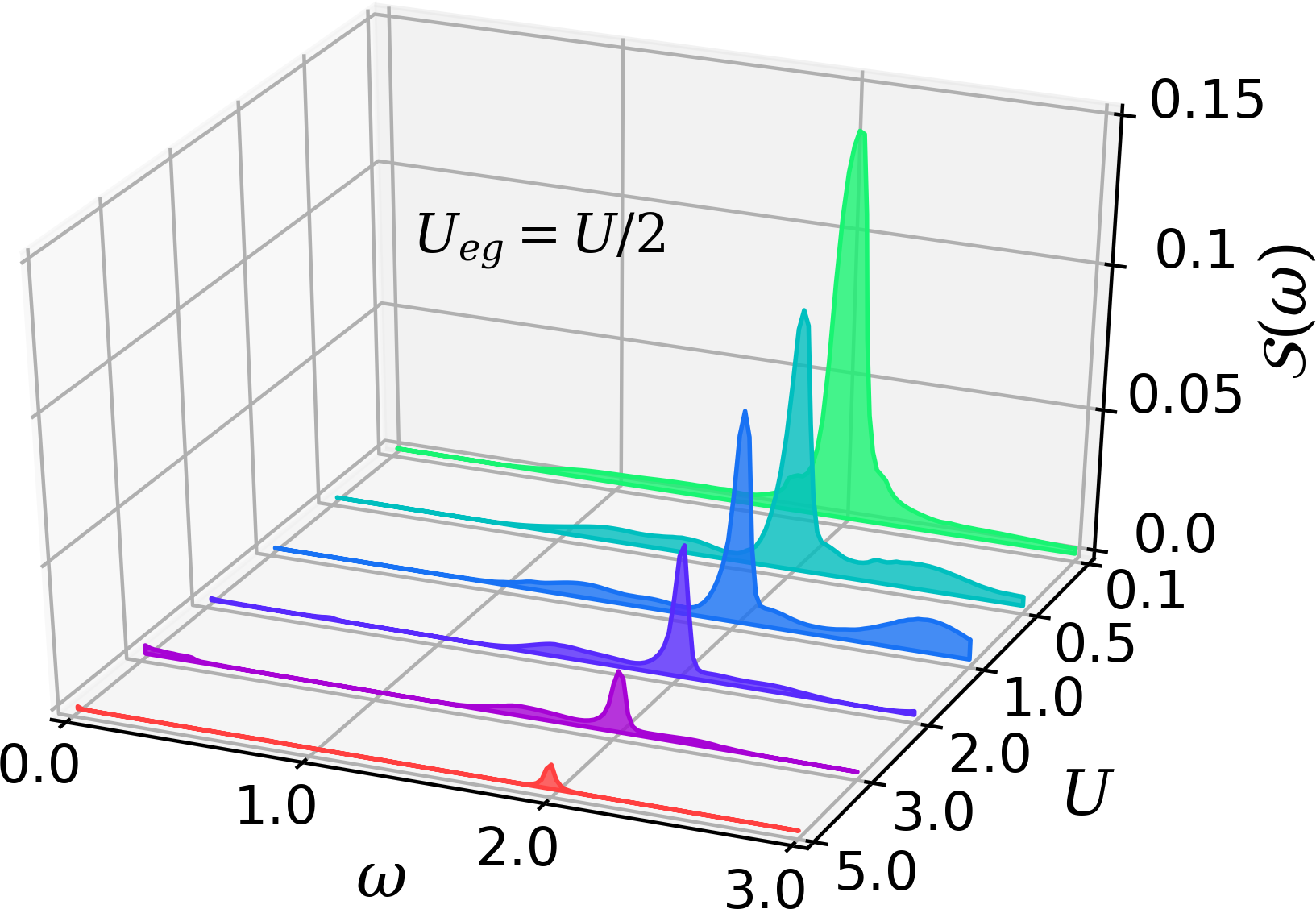}
 	\caption{ Resonant fluorescent spectrum ($\omega_0=2$) in a dimer ($L=2$) 
	with $N_a=4$ atoms, and its dependence on the atom-atom interaction strength 
	$U$ (with $U_{gg}=U/2$).}	
 	\label{figAPPdiffU}
\end{figure}

\begin{figure}
		\centering
		\includegraphics[width=0.9\columnwidth]{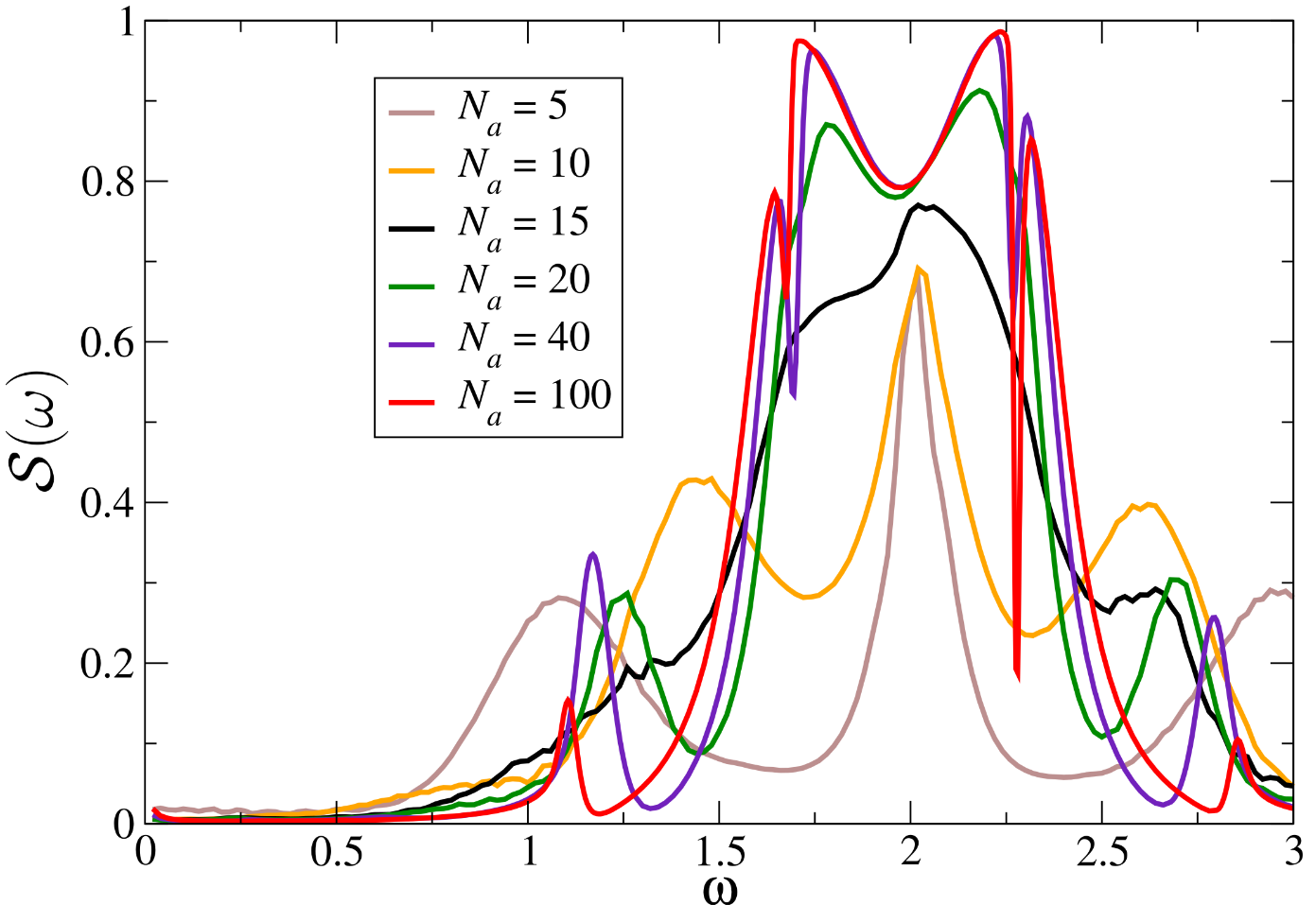}
		\caption{2BEC results at resonance as function of $N_a$ obtained with an initial coherent photon state satisfying $\langle \hat{b}^\dagger \hat{b}\rangle = 16$ and cavity frequency $\omega_0=\Omega_R=2.0$.  
Calculations use $U_{gg} = U_{ee} = U_{eg}=0$, $\gamma_0 = 0.3$, $\gamma_f = 0.1$, $\Gamma = 0.02$, $E_g = 0.0$, and $E_e = 2.0$.} 
	\label{Mollow_BEC_App}
\end{figure}

 \section{Additional Results at Resonance (Fig.~\ref{A2fig1} - Fig.~\ref{Mollow_BEC_App}) }\label{AdditionalReso} 

We further illustrate the impact of finite-rate laser driving. 
In Fig.~\ref{A2fig1} we compare fluorescence spectra obtained by 
(i) starting from the cavity--system state and injecting photons through 
time-dependent driving, and 
(ii) starting from the system's ground state with an initial coherent photon state. 
In the latter case the Mollow sidebands are clearly visible, whereas finite-rate 
driving suppresses them. 
The example shown corresponds to $L=3$ with $N_a=3,4$; all other parameters are 
given in the figure caption. 
Similar behavior is observed for other choices of $L$ and $N_a$.

Starting from a coherent state rather than populating the cavity via laser driving 
has an additional consequence. 
In Fig.~\ref{A2fig2} we show resonance spectra for $L=2$ and $N_a=3,4,5$, with 
$U=5.0$ and $U_{eg}=U/2$, using coherent-state initialisation. 
In this case the spectrum displays lower-frequency structures that are absent when 
photons enter the cavity through finite-rate driving---the situation examined in 
the main text.

In Fig.~\ref{figAPPdiffU} we consider the fluorescent response at 
resonant incident frequency $\omega_0=2.0$, for different interaction regimes,
as determined by the value of $U$ ( $0 \le U \le 5$, with $U_{eg} = U/2$ in all cases). 
The results shown are for a dimer ($L=2$) with $N_a=4$ atoms. The system
is initially in the state $|\Phi_0^{Res}\rangle$, as specified in the main text. The incident
pulse parameters are, $\tau_1=\tau_2=2.0$, $t_1=\frac{6\pi}{\omega_0}$, $t_2=\frac{31\pi}{\omega_0}$. The strength of the driving $\gamma_d$ is chosen such that $\langle \hat{b}^\dagger \hat{b}\rangle\approx 16.0$ at the end of the drive.
As observed for $N_a=3$ in the main text, the intensity of the spectrum decreases with increasing the interaction $U$.

We also show additional resonance results for the two-mode BEC (2BEC) system, with the purpose of
illustrating how the spectrum depends on the number of atoms $N_a$ in the system. As reported in 
Fig.~\ref{Mollow_BEC_App}, at low $N_a$,  $\mathcal{S}(\omega)$ exhibits a clear three-peaked Mollow structure,
which however disappears at larger number of particles,
with the spectrum approaching multiple broad peaks that eventually merge into a large, almost plateau-like
structure, as discussed in Fig.~\ref{figBEC2} of the main text.

\begin{figure}[H]
	\begin{tikzpicture}
	\node at (-2.3, 6) {\includegraphics[width=0.9\columnwidth]{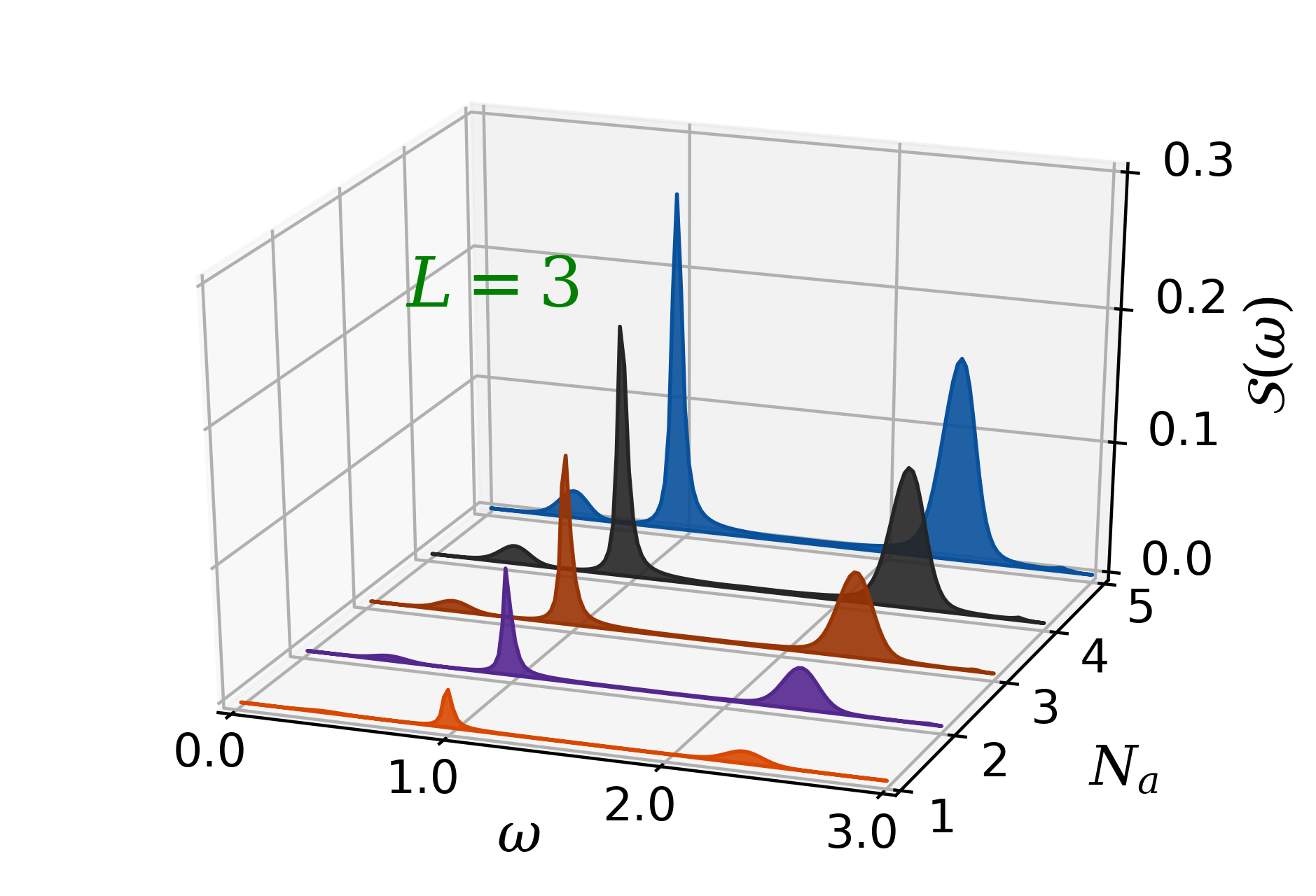}};
	\node at (-2.3, 1) {\includegraphics[width=0.9\columnwidth]{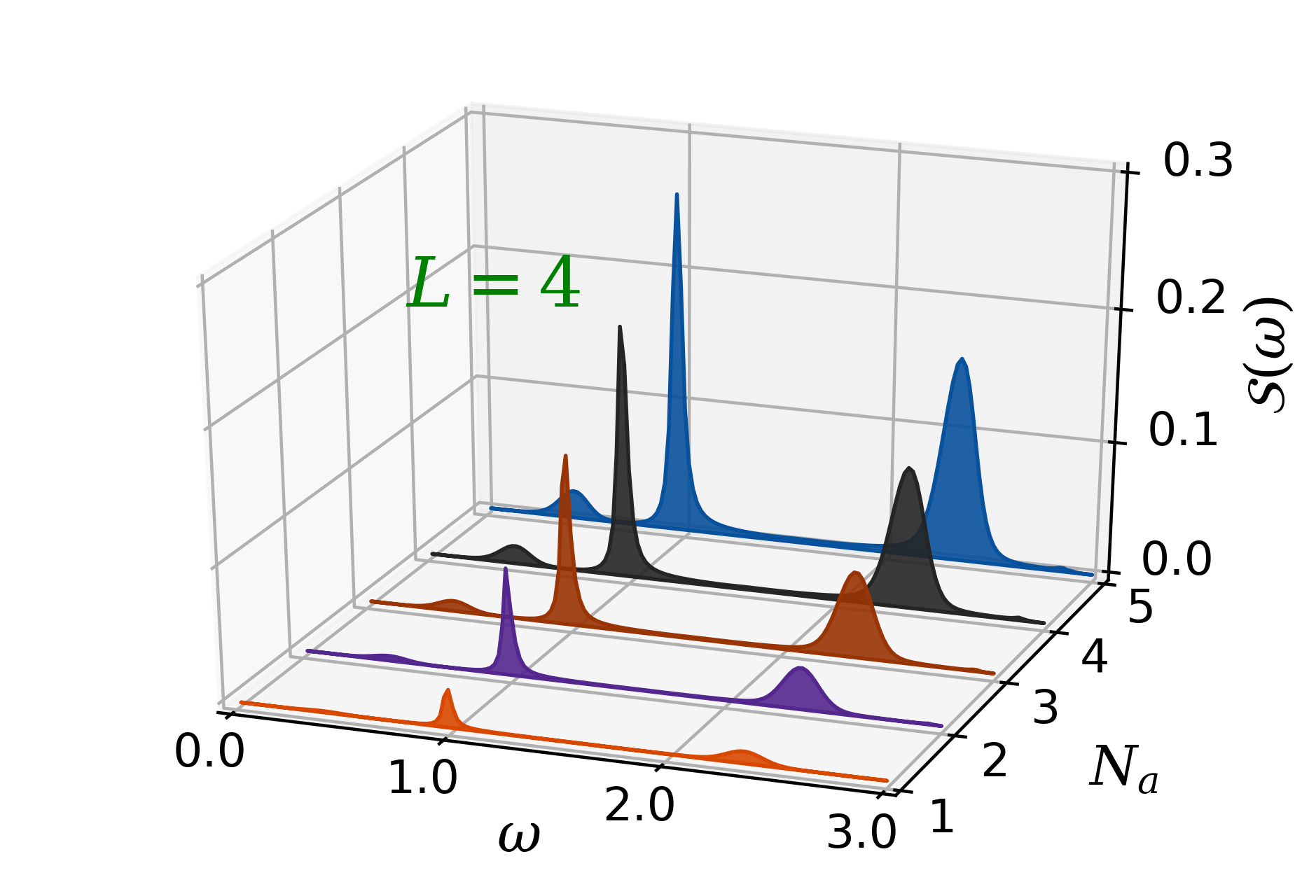}};
	\node[black] at (-4.50,6.5) {$\bm{(a)}$};
	\node[black] at (-4.50,1.5)  {$\bm{(b)}$};
	\end{tikzpicture}
	\caption{SHG for $U = 0.0001$, $U_{eg} = U/2$ when $L=3$ and $L=4$. In all calculations $\eta=4.0$, $\omega_0=1.0$ and the initial state is $|\Phi_0^{SHG}\rangle$. } 
	\label{figAPPL34}
\end{figure}

\section{Additional SHG Results (see Figure~\ref{figAPPL34})}\label{AdditionalSHG}
Fig~\ref{figAPPL34} displays the SHG results in the low interaction regime with $L=3$ and $L=4$. We observe that increasing the number of sites from $3$ to $4$ brings negligible effect to the fluorescent spectra. Furthermore, the results are very similar to those for $L=2$ in the main text.

\end{document}